\documentclass[final,5p,times,twocolumn]{elsarticle}

\usepackage{amssymb}
\usepackage{amsmath}

\usepackage[caption=false,font=footnotesize]{subfig}
\usepackage{booktabs}				
\usepackage{array}					
\usepackage{multirow}				
\usepackage{siunitx}				
\usepackage{flafter}				
\usepackage{comment}				
\usepackage[usenames,dvipsnames]{xcolor}	
\usepackage{flafter}				
\usepackage{lipsum} 				
\usepackage{setspace}               

\usepackage{multicol}

\usepackage{amssymb}
\usepackage{soul}
\usepackage{url}
\usepackage{epstopdf}
\usepackage{rotating}
\usepackage{placeins}
\usepackage{pifont}
\usepackage{threeparttable}
\usepackage{makecell}
\usepackage{float}

\usepackage{hyperref}	
\usepackage{cleveref}	

\usepackage{amsmath, colortbl}
\usepackage{algorithm}
\usepackage[noend]{algpseudocode}
\usepackage{eqparbox}
\newdimen{\algindent}
\algnewcommand\LeftComment[2]{%
\hspace{#1\algindent}$\triangleright$ \eqparbox{COMMENT}{\textit{#2}} \hfill %
}
\algnewcommand{\IIf}[1]{\State\algorithmicif\ #1\ \algorithmicthen}
\algnewcommand{\ElseIIf}[1]{\algorithmicelse\ #1} 
\algnewcommand{\EndIIf}{\unskip\ \algorithmicend\ \algorithmicif}
\usepackage{svg}
\usepackage{dsfont}
\usepackage{textcomp}
\usepackage{tabularray}
\usepackage{enumitem}
\usepackage{subfig}
\DeclareMathOperator*{\argmin}{argmin}
\DeclareMathOperator*{\argmax}{argmax}
\usepackage{nicematrix}
\usepackage{pdfpages}

\begin{document}

\begin{frontmatter}


\title{Multi-level informed optimization via decomposed Kriging for large design problems under uncertainty}

\author[RRE]{Enrico~Ampellio\corref{cor1}}
\cortext[cor1]{Corresponding author. Address: Leonhardstrasse 21, 8092 Zurich, Switzerland. E-mail: eampellio@ethz.ch}
\author[RRE]{Blazhe~Gjorgiev}
\author[RRE]{Giovanni~Sansavini}

\affiliation[RRE]{organization={Reliability and Risk Engineering Laboratory, Institute of Energy and Process Engineering, Department of Mechanical and Process Engineering, ETH Zurich},
            country={Switzerland}}


\begin{abstract}

Engineering design involves demanding models encompassing many decision variables and uncontrollable parameters. In addition, unavoidable aleatoric and epistemic uncertainties can be very impactful and add further complexity.  
The state-of-the-art adopts two steps, uncertainty quantification and design optimization, to optimize systems under uncertainty by means of robust or stochastic metrics. However, conventional scenario-based, surrogate-assisted, and mathematical programming methods are not sufficiently scalable to be affordable and precise in large and complex cases.
Here, a multi-level approach is proposed to accurately optimize resource-intensive, high-dimensional, and complex engineering problems under uncertainty with minimal resources. A non-intrusive, fast-scaling, Kriging-based surrogate is developed to map the combined design/parameter domain efficiently. Multiple surrogates are adaptively updated by hierarchical and orthogonal decomposition to leverage the fewer and most uncertainty-informed data.
The proposed method is statistically compared to the state-of-the-art via an analytical testbed and is shown to be concurrently faster and more accurate by orders of magnitude.
  
\end{abstract}


\begin{graphicalabstract}
\begin{figure}[htbp]
  \centering
    \includegraphics[width=1\textwidth]{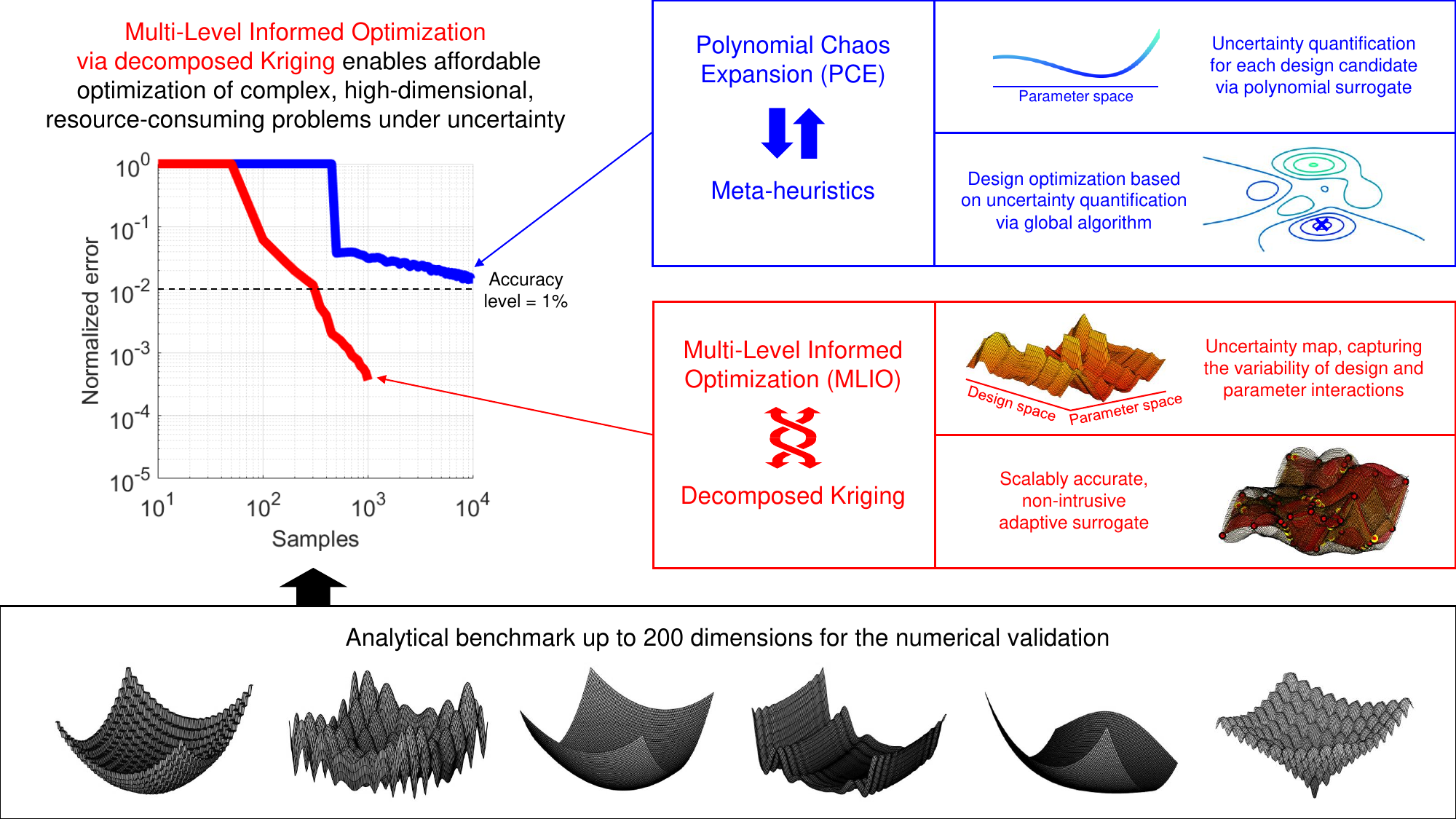}
\end{figure}
\end{graphicalabstract}



\begin{highlights}
\section*{highlights}
    \item Optimizing large and complex problems under uncertainty requires scalable methods.
    \item An adaptive decomposed Kriging surrogate maps parametric effects over design options.
    \item A multi-level informed optimization updates the map aiming for the best design.
    \item Numerically validated versus the state-of-the-art on a heterogenous analytical testbed.
    \item Complex, high-dimensional, resource-consuming problems become tractable.
\end{highlights}


\begin{keyword}
design under uncertainty, large complex systems, multi-level optimization, adaptive Kriging surrogate
\end{keyword}

\end{frontmatter}

\section{Introduction}
\label{sec_intro}


 Engineering and applied sciences, driven by the new paradigm of sustainability, deal with challenging design problems, whether regarding structures, machines, or systems. Mathematical models to capture the underlying physics are essential, featuring many decision variables and uncontrollable parameters, and eventually involving complex patterns and constraints. Gradient-based, meta-heuristics, or data-driven optimization is adopted to find the best compromise design, utilizing quality metrics and simulation results.
 A deterministic approach is short-sighted since epistemic and aleatoric uncertainties affect the model and its assumed parameters, respectively, and may significantly impact the nominal results. Therefore, decision-making in the presence of uncertainties is paramount but demanding~\cite{kochenderfer2015decision}, and it may be intractable when considering complex, high-dimensional, and resource-consuming problems, from efficient engines to sustainable energy networks.


 High-dimensional and complex conditions are common in engineering. Real-world cases can be very challenging to tackle~\cite{crespo2020nasa} and require sophisticated methods to quantify uncertainty with limited information \cite{gray2022inference}. Many problems involve Ordinary or Partial Differential Equations (ODE/PDE) and minimize a loss function, herein referred to as $COST(\mathbf{u},\mathbf{p})$. They may be irregular and affected by noise, but the dimensionality is usually limited to the order of ten~\cite{beck2012comparison}, counting design variables $\mathbf{u}$, uncertain parameters $\mathbf{p}$ (bold for multi-variate vectors), and constraints. Instead, system and network problems are usually more regular (linearized) but large, including several hundred up to billions of dimensions~\cite{pfenninger2014energy}, even after aggregation. They can also involve non-convexities such as optimal flows in transports and power grids~\cite{gjorgiev2022nexus}. 

 A remarkable example is operations research, a branch where choices have important political and social implications. It is relevant to mention the ongoing transition toward net-zero energy systems \cite{international2021net} which involves many aspects, such as seasonal energy storage, international collaboration, the role of hydrogen, gas, and carbon capturing. The related models are finely resolved in space and time while embedding dozens of social, technological, economic, and strategic indicators. To reach tractability, they are commonly linearized and solved as large and complex Mixed Integer Linear Programming (MILP) problems. The goal is to minimize a total expenditure, over multi-year investments $\mathbf{u}$ and finely resolved operations $\mathbf{o}$ considering several parameters $\mathbf{p}$ and constraints. 
 Despite typically aggregated on the time domain, the problem is very resource-intense and turns inherently non-linear and non-convex as a function of $\mathbf{p}$. Moreover, time correlations (e.g., storage) and non-linear effects related to impactful parameters, like climate and weather fluctuations \cite{panteli2015influence} including extreme events, add to the complexity.
 Optimizing this system under uncertainty is arduous \cite{liu2020energy}, and a certain level of compression is mandatory to manage the curse of dimensionality. A sensitivity can eliminate secondary parameters \cite{moret2017strategic} and one could focus on investments $\mathbf{u}$ only as a function of $\mathbf{p}$, embedding optimal operations $\mathbf{o}$. Still, the problem remains complex and large, hence potentially untreatable. 

 This demands scalable methods that are accurate enough to efficiently support the optimization of realistic, resource-consuming systems under uncertainty.
 Two components are essential to this task, defining the so-called two-step approach:
 \begin{itemize}[noitemsep,topsep=0pt]
    \item Uncertainty Quantification ($UQ$): to assess the impact on $COST$ of parametric uncertainties for a given design $\mathbf{u}$, according to a $UQ_p$ operator over $\mathbf{p}$, $UQ(\mathbf{u}) = UQ_p(COST(\mathbf{u},\mathbf{p}))$. Robust \cite{gabrel2014recent} or stochastic \cite{shapiro2021lectures} criteria define the operator as either the maximum, $UQ_p \equiv max_p$, independent of probability, or a statistical moment, such as $UQ_p \equiv \mathbb{E}_p$, reliant on a probability distribution.
    \item Design optimization ($OPT$): to find the configuration minimizing $COST$ and the uncertainty impact on it together, $OPT_u = min_u UQ_p(COST(\mathbf{u},\mathbf{p}))$.
 \end{itemize}
 Both are difficult tasks that require advanced problem-specific methods. Uncertainty quantification using a few relevant scenarios per engineering judgment \cite{schlachtberger2018cost} is scalable with dimensionality but not accurate. A statistical number of scenarios, like in Monte Carlo (MC) \cite{karmellos2019comparison}, is accurate but not scalable, hence potentially unaffordable. Surrogates \cite{jin2003use} are efficient to compute but can be too inaccurate, and their training scales poorly \cite{barton2006metamodel}. On the other hand, optimizing via mathematical programming is fast and scalable, but accuracy is guaranteed only on convex problems. Generalized algorithms for global optimization, like meta-heuristics, data-driven, and surrogate-assisted, work on any problem, but scalable training is difficult to achieve, and convergence cannot be guaranteed. 
 In conclusion, such methods lack scalable accuracy with the number of dimensions in complex problems. This often forces oversimplification or partitioning to attempt any design-under-uncertainty task \cite{moret2020decision}, which is therefore incomplete or unrealistic.
 
 In the case of MILPs, mathematical programming is enabled through scenario-based approaches \cite{schlachtberger2018cost}. However, inefficient out-of-sampling is needed to quantify uncertainty for either robust or stochastic optimization. Sensitivities and near-optimal approaches \cite{neumann2023broad} address exploration and epistemic uncertainty, but explode the number of observations. 
 Widespread robust optimization \cite{gabrielli2019robust} is cheaper and more intuitive than stochastic optimization \cite{mavromatidis2018comparison}, but over-conservative. Distributionally robust, stochastic-robust, and chance-constrained optimization try to balance the cheapness of robust and the thoroughness of stochastic methods, but are subject to the cons of both. Similar techniques also apply to non-convex formulations with non-linear parametric effects, but they are expensive and inaccurate.

 Surrogates are generally suitable for any problem formulation, whether concerning $UQ$ or $OPT$ tasks, and robust or stochastic criteria. Close to order reduction techniques, they are machine learning approximations valuable when complex and/or resource-consuming processes are involved. 
 Among many options, Support Vector Machines (SVM) as a form of generalized kernel-based regression are popular for reliability analysis \cite{roy2023support}, but feature limited interpretation, difficult setting and tuning, long training times, large datasets, and are unsuitable for very high-dimensionality. Polynomial Chaos Expansion (PCE), thanks to the built-in principles of orthogonal decomposition and stochasticity, is widely adopted for sensitivities and uncertainty quantification \cite{sudret2008global}, in structures, thermoacoustics, computational fluid dynamics, power systems, and many others. Advanced versions using sparsity for low-rank truncations \cite{luthen2021sparse} and adaptivity via regularized regression \cite{blatman2011adaptive} are efficient and return analytical variance indexes. However, they are undermined by irregular landscapes due to their polynomial nature, and become inaccurate or computationally prohibitive for a dimensionality around 100 or higher. 
 Kriging \cite{chiles2018fifty} is also extensively applied to complex design \cite{raponi2019kriging}, especially for global optimization in crashworthiness, structures, aerodynamics, electromagnetics, and more. Surrogates in general and Kriging in particular have recently raised interest in the context of risk and safety as stochastic emulators \cite{peng2025efficient}, for global sensitivities \cite{shang2023efficient}, for reliability analyses \cite{li2025stacking,chen2025reliability,wan2025new}, for multi-objective optimization under uncertainty \cite{rivier2022surrogate}, and to support decision-making for resilient systems \cite{hu2025surrogate}. As a form of Bayesian regression, Kriging predicts the behavior of any process in unexplored locations as weighted average of known observations, and provides confidence intervals. However, it suffers from computational complexity especially when the number of observations grows. 
 
 This work tackles the challenge of scalable accuracy on complex problems, characterized within min/max ranges of variables ad parameters. A Multi-Level Informed Optimization (MLIO) scheme based on decomposed surrogates is proposed to map the uncertainty impact on design choices, providing affordable optimization in realistic conditions.
 The method is non-intrusive and assumption-free, except for $\mathcal{C}_0$ continuity. It encompasses three levels:
 \begin{enumerate}[noitemsep,topsep=0pt]
     \item Solve: physically informed solution of a deterministic realization, $COST(\mathbf{u},\mathbf{p})$, given both design and parameter sets. $COST$ evaluation includes eventual operations, constraints, and penalizations, and is treated as a black-box.
     \item Explore: adaptive surrogate to map $COST(\mathbf{u},\mathbf{p})$ through a fast-scaling yet accurate ensemble of Kriging layers, incorporating hierarchical and orthogonal decomposition principles and called from now on decomposed Kriging.
     \item Exploit: design optimization (OPT) leveraging decomposed Kriging, to refine the best regions of the uncertainty map while the surrogate is being trained.
 \end{enumerate} 
 The innovative contribution of this work is two-fold:
 \begin{itemize}[noitemsep,topsep=0pt]
     \item Uncertainty map: the multi-level informed scheme goes beyond the traditional two-step approach for optimization under uncertainty. It is a problem-learning perspective that maps the interactions between decision variables and uncertain parameters.
     \item Decomposed Kriging algorithm: a multi-layer ensemble of surrogates developed to be accurate, scalable, inexpensive, self-adaptive, informative, and non-intrusive, and to minimize hyperparameters and assumptions. As a piece of fundamental research, it is indeed generalizable to a wide range of applications per se.
 \end{itemize}

 The multi-level informed method and the decomposed Kriging algorithm are described in Sections~\ref{sec_method}~and~\ref{sec_algorithm}, respectively, supported by \ref{app_KRG} and \ref{app_SmrKRG}. Section~\ref{sec_castudy} introduces the analytical benchmark for numerical validation and a two-step state-of-the-art method for comparison. Results are reported in Section \ref{sec_results} and discussed in Section~\ref{sec_discussion}. Finally, Section \ref{sec_conclusions} summarizes the most relevant insights of the study. 
 


 \section{Multi-Level Informed Optimization}
 \label{sec_method}

 Here, we first provide an overview of the new mapping perspective among design choices and uncertain parameters (Section~\ref{subsec_perspective}). Then, we describe the three levels of the proposed Multi-Level Informed Optimization method (MILO) for design under uncertainty and iterations between them (Section~\ref{subsec_scheme}).

 \subsection{The importance of uncertainty mapping} 
 \label{subsec_perspective}

\begin{figure}
    \begin{flushleft}
    \includegraphics[width=.45\textwidth]{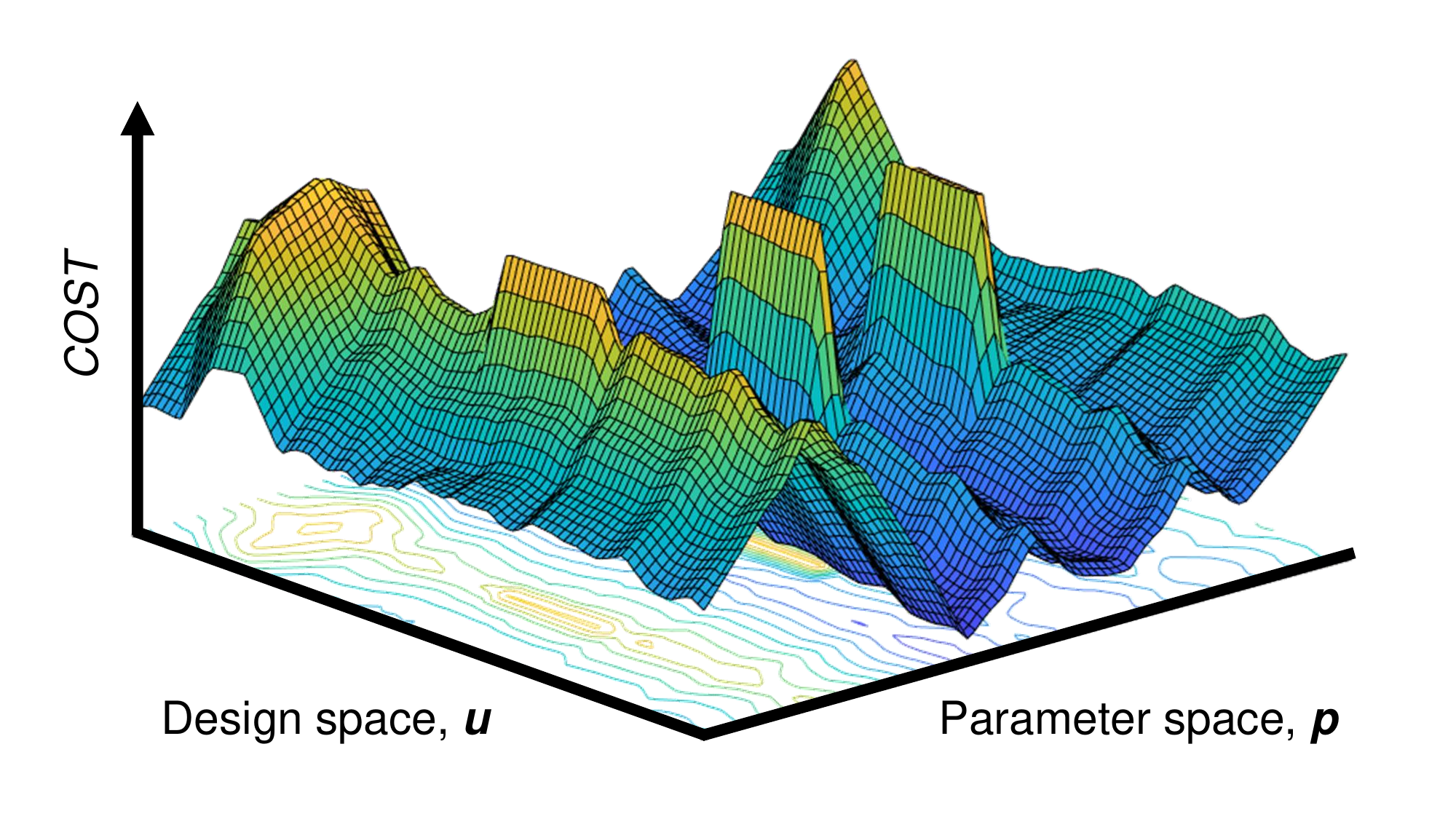}\\
     \caption{Graphical representation of the uncertainty map, projected from a multi-variate energy system on parameter and design spaces}\label{fig_UncertaintyMAP}
    \end{flushleft}
 \end{figure}

 Two conflicting needs are manifested in optimization under uncertainty: i) limiting the number of $COST(\mathbf{u},\mathbf{p})$ evaluations to maintain tractability and ii) ensuring accuracy in both $UQ$ and $OPT$ phases. 
 Separating $OPT_u(\mathbf{p})$ acting on $\mathbf{u}$ as a function of $\mathbf{p}$ and $UQ_p(\mathbf{u})$ acting on $\mathbf{p}$ as a function of $\mathbf{u}$ is inefficient because it requires a factorial out-of-sampling grid. Instead, interactions among design choices and uncertain parameters could be exploited with multiple advantages: i) minimize the number of samples to capture the overall correlated map; ii) pursue uncertainty quantification and design optimization concurrently; iii) focus on the variability of the $COST$ function, regardless of assumptions on both parameter (probability distribution) or design (convexity) spaces.
 This changes the perspective of design under uncertainty: drawing the full uncertainty map of $COST(\mathbf{u},\mathbf{p})$ (Fig.\ref{fig_UncertaintyMAP}) will drive decisions in the best-informed way possible. Moreover, exploring the two multi-variate spaces at once leads to important insights about epistemic uncertainty, similarly to near-optimal methods like Model for Generating Alternatives (MGA) \cite{decarolis2011using}, but in a comprehensive way.
 
 Building a surrogate for the uncertainty map is a natural choice, widely adopted in engineering from energy systems to chemical processes, especially when resource-intense evaluations are involved. Three major challenges, however, limit the progression of the state-of-the-art: i) $COST_p(\mathbf{u})$ landscape is expected to be irregular and/or multi-modal, mainly due to design-related constraints; ii) the overall dimensionality $D=D_u+D_p$ can easily reach several hundreds or thousands for realistic complex problems, albeit aggregated; iii) unlike $OPT_u$ that is a single-point optimization, $UQ_p$ is a characterization and requires ubiquitous accuracy. Available surrogates find typical applications to regular ($\mathcal{C}_1$ or higher) and/or low-dimensional, $D$$\sim$$\mathcal{O}(10)$, problems \cite{moustapha2018comparative} and so are relegated to uncertainty quantification over a limited number of parameters. Design optimization is separately achieved through mathematical programming when possible and global optimization when not. Surrogating the entire map over $[\mathbf{u},\mathbf{p}]$ is instead an efficient but ambitious enabler for optimization under uncertainty. It is independent of the specific problem's properties, but requires surrogates to be scalably accurate on complex landscapes.
 
 \subsection{MLIO logical scheme} 
 \label{subsec_scheme}

 \begin{figure*}
   \begin{center}
         \includegraphics[trim={2.5cm 0 2.5cm 0},clip,width=0.95\textwidth]{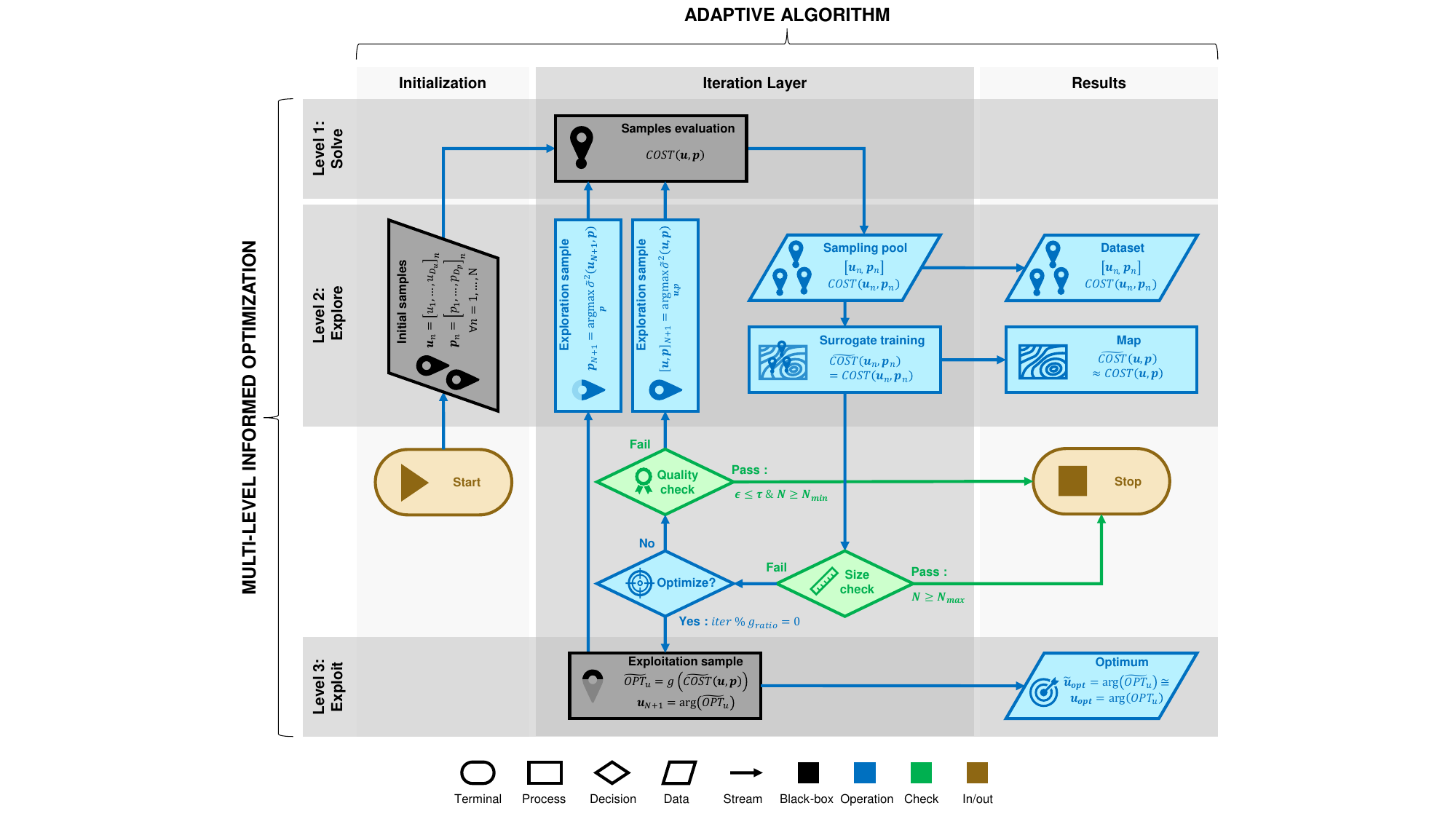}\\
         \caption{Logical flow chart of MLIO approach with the three levels, solution, exploration, and exploitation, and the iterative loops of the adaptive algorithm }\label{fig_MLIOflow}
   \end{center}
 \end{figure*} 
 To overcome the scalability limitations of state-of-the-art, we develop a multi-level informed optimization. It empowers a change of perspective in design under uncertainty, from the direct two-step to interlaced map capturing.
 The method aims to accurately approximate the whole uncertainty map with a minimal number of $COST(\mathbf{u},\mathbf{p})$ evaluations. They are accessed as a black-box so that the $COST$ function can be freely defined, and the MLIO is non-intrusive and problem-independent.
 The flow chart in Fig.\ref{fig_MLIOflow} illustrates the general form of tri-leveled MLIO. The levels are represented horizontally and decouple physical information (treated in "Level 1: Solve"), variability exploration (treated in "Level 2: Explore"), and optimality exploitation (treated in "Level 3: Exploit"). Different adaptive algorithms can be used under the same MLIO arrangement, all characterized by at least three phases depicted as columns in Fig.\ref{fig_MLIOflow}, namely, input ("Initialization"), output ("Results"), and the adaptive feedback loops (the "Iteration Layer"). This structure is flexible and generalizable thanks to the decoupling of the three levels from the iterative layer. The three levels and the iterations between them are described in detail hereafter in Subsections \ref{subsubsec_solve},\ref{subsubsec_explore}, and \ref{subsubsec_exploit}, while the initialization and results Subsections, \ref{subsubsec_init} and \ref{subsubsec_res}, wrap the entire procedure. 

 \subsubsection{Initialization}  
 \label{subsubsec_init}
 
 MLIO is initialized with an explorative set of $N$ parametric scenarios defined as $\mathbf{p}_n = [p_1,..,p_{D_p}]_n ~\forall n=1,..,N$, and corresponding design options $\mathbf{u}_n = [u_1,..,u_{D_u}]_n$. A minimum of two distinct $[\mathbf{u},\mathbf{p}]$ sets must be provided, $N \geq 2$, to capture differences on the map. It can be a random pair of two $\mathbf{u}$ and $\mathbf{p}$ sets without any knowledge of the problem being solved, but usually a baseline and some other relevant scenarios are known. For instance, in operations research one can define $COST(\mathbf{u},\mathbf{p})$ with embedded operations $\mathbf{o}$ as a MILP optimization step, $COST(\mathbf{u},\mathbf{p})=\min_{\mathbf{o}} cost(\mathbf{u},\mathbf{o},\mathbf{p})$, where the total cost as a function of design, parameters, and operations is defined as $cost(\mathbf{u},\mathbf{o},\mathbf{p})$. In this case, the initial $\mathbf{u_n}$ are the optima corresponding to each parametric set. In general, design sets can be initialized solving $\mathbf{u_n} = \argmin_{\mathbf{u}} COST(\mathbf{u},\mathbf{p_n})$ for any $COST$ formulation. If $COST$ is difficult to solve, $u$ sets can alternatively be determined by engineering judgment or sampled randomly. Thanks to MLIO self-adaptiveness, initialization size is meant to be a small fraction (ideally $\sim 1-10\%$, smaller on larger problems) of the total sampling budget allowed, $N_{max}$ and this phase is conceived as a black-box from the MLIO perspective. Essentially, $[\mathbf{u}_n,\mathbf{p}_n]$ can be independently provided in any suitable manner for the problem being addressed.

 \subsubsection{Level 1: Solve}  
 \label{subsubsec_solve} 
 
 The $[\mathbf{u},\mathbf{p}]$ sets are processed at "Level 1: Solve" entering the "Iteration Layer", which deterministically solves the physics of the problem evaluating the samples, $COST(\mathbf{u},\mathbf{p})$. This can be an analytical evaluation, a simulation, or an optimization, and it is treated as a black-box exogenous process by MLIO. From here on, the multi-level informed optimization chooses every subsequent $[\mathbf{u},\mathbf{p}]$ sample to evaluate, thanks to the transversal "Iteration Layer".

 \subsubsection{Level 2: Explore} 
 \label{subsubsec_explore}
 
 After the evaluation is performed at Level 1, the first surrogate $\widetilde{COST}$ is built at "Level 2: Explore" of the "Iteration Layer". A sampling pool, updated with all the $[\mathbf{u_n},\mathbf{p_n}]$ sets and corresponding $COST(\mathbf{u_n},\mathbf{p_n})$ collected so far, feeds the surrogate training. The surrogate is progressively updated as new samples are added, following the exploration (on "Level 2: Explore") and the exploitation (on "Level 3: Exploit") feedbacks of the "Iteration Layer", which will enlarge the sampling pool one point at a time, from $N$ to $N+1$. The surrogate $\widetilde{COST}$ interpolates the original loss function, i.e. it matches $COST$ in the sampled points, $\widetilde{COST}(\mathbf{u_n},\mathbf{p_n})=COST(\mathbf{u_n},\mathbf{p_n}) ~\forall n$, and approximates it elsewhere, $\widetilde{COST}(\mathbf{u},\mathbf{p}) \approx COST(\mathbf{u},\mathbf{p})$.
 
 As in any adaptive algorithm, MLIO loops back to Level 1 after the surrogate training, concluding one iteration. Additional exploration samples will be employed to maximize the surrogate confidence along the exploration feedback line at Level 2, according to $[\mathbf{u}_{N+1},\mathbf{p}_{N+1}] = argmin_{\mathbf{u},\mathbf{p}} \tilde{\sigma}^2(\mathbf{u},\mathbf{p})$. The additional samples are selected in correspondence to the surrogate's maximum expected variance $\tilde{\sigma}^2$ with respect to the actual $COST$. Iterations break when the surrogate`s approximation error $\epsilon$ complies to a satisfactory threshold $\tau$, $\epsilon \leq \tau$. Error calculation involves a certain number of validation samples over training samples, $v_{ratio}$, and Kriging confidence interval, as described in Section \ref{sec_algorithm}. Overall, compliance to min/max samples, $N_{min} \leq N \leq N_{max}$, is also performed as a size check before the quality check, to impose a hard containment of premature/late convergence. However, up to Level 2, the resulting surrogate $\widetilde{COST}$ is purely explorative and, as such, not efficient in searching for any preferred optimum design. 

 \subsubsection{Level 3: Exploit}
 \label{subsubsec_exploit}
 
 The value in drawing the problem's map is allowing the identification of desired configurations, referred to as optimal. Consequently, "Level 3: Exploit" is introduced to leverage the surrogate already during the training (i.e., immediately after Level 2) and refine its quality in correspondence to the expected most interesting regions. These regions are defined through a greedy operator $g$ of $\widetilde{COST}$ over $\mathbf{u}$, $\widetilde{OPT}_{u}=g(\widetilde{COST}(\mathbf{u},\mathbf{p}))$. $g$ is treated as black-box, and can be any process acting on the cost surrogate, $\widetilde{COST}$. The specific case of optimization under uncertainty entails the solution of the problem $\widetilde{OPT}_{u}=min_{\mathbf{u}} UQ_p(\widetilde{COST}(\mathbf{u},\mathbf{p}))$, where $g=min_{\mathbf{u}} UQ_p$ and $\widetilde{COST}(\mathbf{u},\mathbf{p})$ approximates the uncertainty map.
 The next exploitation sample $[\mathbf{u}_{N+1},\mathbf{p}_{N+1}]$ is selected as $\mathbf{u}_{N+1}=\arg(\widetilde{OPT}_{u})$, defined at Level 3, and $\mathbf{p}_{N+1} = \argmin_{\mathbf{p}} \tilde{\sigma}^2(\mathbf{u}_{N+1},\mathbf{p})$ defined at Level 2, to maximize the surrogate confidence for the selected design. In general, no quality check can be applied for the convergence to the global optimum, so the stopping conditions of the exploration and exploitation loops remain tied to the overall approximation quality of the surrogate. In specific cases, an optimality criterion could be added, like tolerance on the error or its convergence, if the expected value of the global optimum is known or predictable but its location is unknown. Otherwise, Level 3 still improves the map's confidence in the best areas.

 \subsubsection{Results and overarching rationale}  
 \label{subsubsec_res}
 
 The collected dataset, $[\mathbf{u}_n,\mathbf{p}_n]$ and $COST(\mathbf{u}_n,\mathbf{p}_n) ~\forall n=1,...,N$, the current optimum, $\mathbf{u_{opt}} = arg(\widetilde{OPT}_{u})$, and the current surrogate, $\widetilde{COST}(\mathbf{u},\mathbf{p})$, found are returned at the end of the procedure. While the optimum is already decision-oriented, the surrogate generally holds as a valid representation of the entire map. $\widetilde{COST}$ can be used afterward for any task, even outside the original intent, without re-running the expensive training. This makes the method flexible and attractive. 
 
 Overall, Level 2 and Level 3 loop back according to explorative and explotative rewards in the "Iteration Layer" as in reinforcement learning, to boost the surrogate quality driven by strategic information collected from the physical $COST$ observations. Moreover, they follow the acquisition functions of Bayesian optimization, but explicitly split exploration and exploitation phases. On the one side, this allows for pursuing any optimization or characterization task through an appropriate definition of Level's 3 greedy operator $g$ (e.g., quantile estimation) with the very same overall structure. On the other side, explorative and exploitative attitudes can be directly balanced as in meta-heuristics \cite{hussain2019exploration}, by alternating them through the iterations, indexed by $iter$, with frequency governed via a dedicated hyperparameter, $g_{ratio}$. In Fig.\ref{fig_MLIOflow}, this is represented by the "Optimize?" decision block, asking if an exploitation feedback loop should be pursued in place of an exploration feedback loop at each iteration, in order to maintain the prescribed $g_{ratio}$ between the two. Indeed, unlike the majority of machine learning and heuristic algorithms, the number of hyperparameters and their impact on results is minimized in MLIO, thanks to its architectural self-adaptiveness. In total, there are three hyperparameters, namely, $N$ initial samples (initialization), $\epsilon$ calculation (quality check, involving $v_{ratio}$), and $g_{ratio}$ (balancing exploration and exploitation), all falling within predefined ranges (see \ref{subapp_hyper}). However, this study will show how the most influential of these, namely $N$ initial samples, can actually be standardized (Section \ref{sec_results}); hence, only two hyperparameters remain for fine-tuning. $\tau$, $N_{min}$, $N_{max}$ are operative parameters for exit conditions, not altering MLIO logic.  
 
 \section{Decomposed Kriging algorithm}
 \label{sec_algorithm}

 Section \ref{sec_algorithm} details the surrogate training and the feedback iterations of MLIO. First, we provide an overview of the decomposed algorithm (Section~\ref{subsec_flow}); then, the key mathematics behind the training of decomposed Kriging surrogates are presented (Section \ref{subsec_SurrMath}); and lastly, the mathematical details about the adaptive iterative loops are presented (Section~\ref{subsec_IterMath}).

 \subsection{Overview and flowchart} 
 \label{subsec_flow}

  \begin{figure*}
   \begin{center}
         \includegraphics[width=0.9\textwidth]{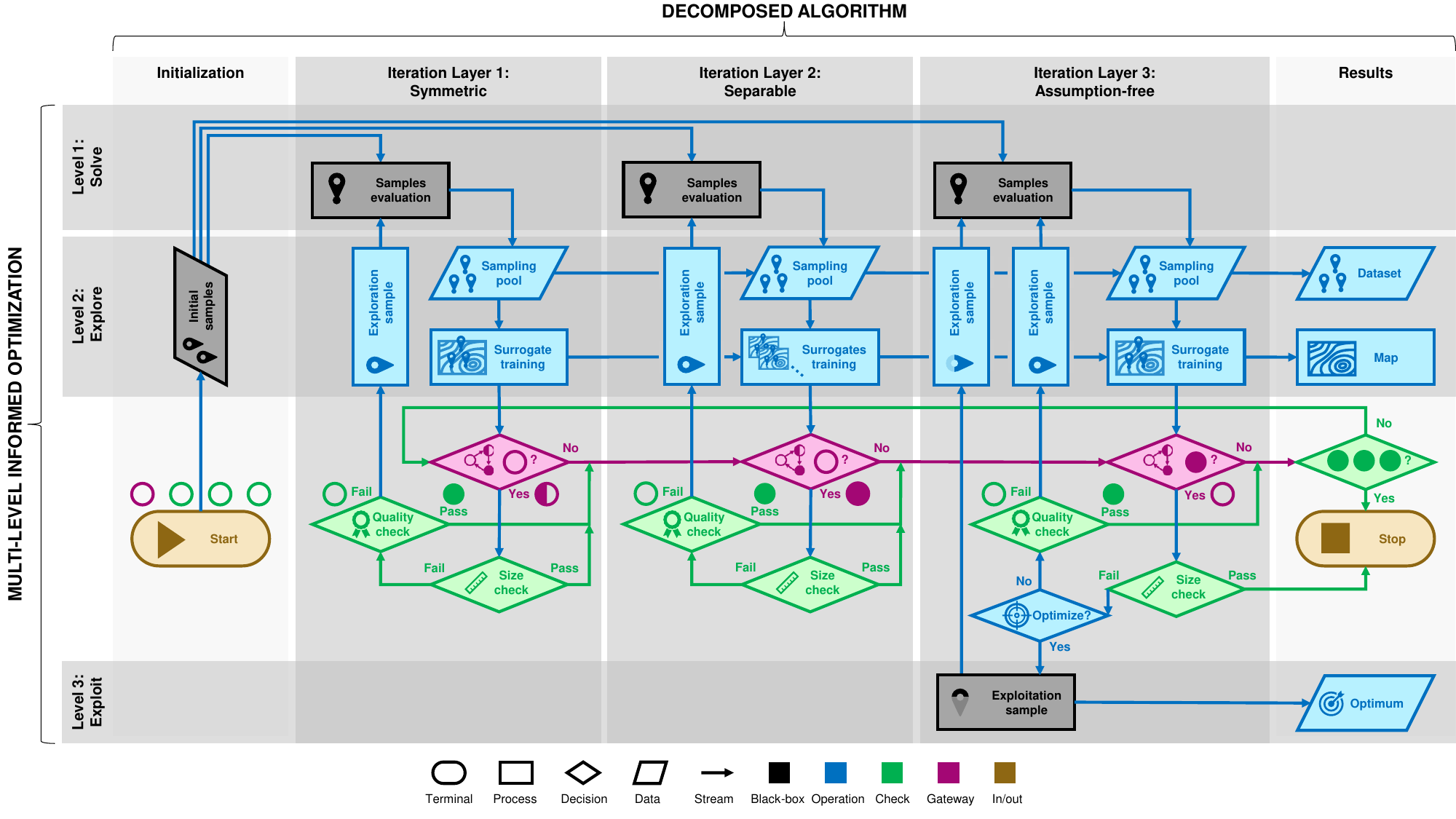}\\
         \caption{Decomposed Kriging algorithm within the MLIO scheme and its three iteration layers.}\label{fig_SmrKRGflow}
   \end{center}
 \end{figure*}  
 The surrogate is the central element of the proposed MLIO.
 It needs to absorb potential irregularities without destabilization or over-fitting, fast-scale with multi-dimensional problems, and return a confidence estimate. 
 Kriging is the best option thanks to its statistical nature, suitability for complex functions, strong adaptiveness via Bayesian optimization, and scalability through the distance-based radial kernel. 
 Originally developed in geostatistics, Kriging is the Best Linear Unbiased Prediction (BLUP) based on Gaussian processes. It assumes that nearby samples are similar, which holds at least partially for any problem with a certain regularity.
 The Kriging surrogate $\tilde{z}(\mathbf{x})$ approximates a function $z(\mathbf{x})$ of a multi-variate variable $\mathbf{x} \in \mathbb{R}^D$, confided within min/max box-bounds, $\mathbf{B}$. Predictions in unobserved locations $\mathbf{x_0}$ are calculated as the weighted sum of already measured observations $\mathbf{x_n}$, $ \tilde{z}(\mathbf{x_0}) = \sum_{n=1}^{N}{w_{n,0} z(\mathbf{x_n})}$, through a set of weights $w_{n,0}$ depending on the auto-correlation model (kernel) fitted on observations. \ref{app_KRG} offers a compact summary of fundamental Kriging mathematics. 
 
 Many Kriging variants are adopted in engineering applications, mainly for sensitivity and optimization \cite{kleijnen2017regression}. One eminent example is the renowned Efficient Global Optimization algorithm (EGO) \cite{jones1998efficient}. Kriging's ideal dimensionality is $\mathcal{O}(10)$, alongside Bayesian optimization and similar surrogates. Universal Kriging \cite{wackernagel2003multivariate} generalizes the ordinary Kriging by replacing the constant mean term with a deterministic trend or drift composed of basis functions. Nevertheless, scalability remains an issue even when efficient surrogates are adopted as a trend, like advanced PCE \cite{schobi2015polynomial}. Further efforts have been spent to boost scalability, including gradient-enhancements (GEK) \cite{ulaganathan2014use}, Co-Kriging for multiple correlated datasets \cite{le2014recursive}, Bayesian Kriging where model parameters are considered in turn random variables \cite{ranftl2021bayesian}, and latent Kriging based on low-dimensional underlying patterns \cite{lee2022bayesian} and preliminary aggregation/reduction \cite{song2024improved}. Multi-fidelity \cite{peherstorfer2018survey} is often combined with Kriging, especially hierarchical Kriging \cite{abdallah2019parametric} using low-fidelity ordinary Kriging as a drift for high-fidelity universal Kriging. Nevertheless, all these variants still struggle with approximation quality on high-dimensional complex functions because they lack adaptivity and/or effective pattern recognition. Therefore, they are not suitable as surrogates for Level 2 in MLIO.
 
 A multi-purpose generalization of the Efficient Global Optimization is needed, where an ensemble of Kriging surrogates \cite{viana2013efficient} can boost efficiency. To unlock superior efficiency and scalability, this paper introduces the orthogonal and hierarchical decomposition of the original problem, $COST(\mathbf{u},\mathbf{p}) \equiv z(\mathbf{x})$, by decomposing the adaptive algorithm of Fig.\ref{fig_MLIOflow} into an ensemble of symmetric, sum separable, and assumption-free layers in Fig.\ref{fig_SmrKRGflow}. The orthogonal components are carried by the symmetric and separable layers, while the final information is hierarchically reconstructed through the composition of the three layers. Ordinary Kriging is used for Level 2 in the MLIO scheme, but scalability is greatly improved thanks to the specific hypotheses on simplification patterns active for $z$ on Layers 1 and 2. Namely, the symmetric layer assumes that a single dimensional function $f_0$ governs the original problem, $z(\mathbf{x})=\sum_{d=1}^{D} f_0(x_d)$, while the separable layer assumes that $z$ is the sum of different functions, $f_d$, on per each dimensional component, $d$, i.e., $z(\mathbf{x})=\sum_{d=1}^{D} f_d(x_d)$. This means that multiple surrogates with orthogonal properties approximating the $f$ functions are needed on these two layers, one per each dimension $d$. The third layer is assumption-free and recovers the rest of the multi-variate interactions to reconstruct the final Kriging prediction. Such surrogates must be fed by observations with different properties hence belonging to different sampling pools. This implies that the single iterative loop in the "Iteration Layer" of Fig.\ref{fig_MLIOflow} is expanded into three iterative layers, namely, "Layer 1: Symmetric", "Layer 2: Separable" and "Layer 3: Assumption-free" (Fig.\ref{fig_SmrKRGflow}). Essentially, the single Iteration Layer of Fig.\ref{fig_MLIOflow} is transposed to Layer 3 of the decomposed algorithm, acting on the multi-variate space without assumptions. "Level 3: Exploit" is active solely on Layer 3, and Layers 1 and 2 provide intermediate samples and approximations to Layer 3, with a number of samples that scale constantly and linearly with $D$, respectively. The three sampling pools and Kriging surrogates in "Level 2: Explore" are continuously and simultaneously updated throughout the adaptive training process, from "Initialization" to "Results". New samples are called in "Level 1: Solve" one at a time in a cyclic sequence, to inform directly Layer 3 and speed up the overall approximation. This sequence is regulated by one gateway per layer (blocks with switchlight diagrams in Fig.\ref{fig_SmrKRGflow}) and it ends when the quality criteria are met at all layers simultaneously ("Quality check" blocks in Fig.\ref{fig_SmrKRGflow}), or when the maximum number of allowed observations is exceeded ("Size check" blocks in Fig.\ref{fig_SmrKRGflow}).
 
 The stepwise adaptive training process ensures the efficient use of computational resources. Thus, only the fewest, most informative, physics-driven data are strategically added to the three layers in order to maximize the confidence of each surrogate. 
 In addition to hierarchical Kriging, the decomposition process is inspired by the separable interleaved solver in \cite{baudivs2015global}, which shows remarkable scalability and can effectively hybridize with other strategies \cite{ampellio2016hybrid}.
 The decomposition scheme directly imposes symmetry and separability in Layers 1 and 2, in contrast to pricey and potentially deceptive projections searching for an orthogonal basis, as in Principal Component Analysis (PCA), Proper Orthogonal Decomposition (POD), and Polynomial Chaos Expansion (PCE). A great interpretation advantage of high-dimensional complex problems is derived when symmetrical or separable traits align with the problem's natural coordinates. Indeed, this is not uncommon in engineering sciences, although perhaps only partially or locally. Even in case of misalignment or strong correlations, Layers 1 and 2 act as a computationally efficient trend (quote from universal Kriging) for Layer 3, promoting stability and scalability. Furthermore, a preliminary step of PCA or active subspaces \cite{constantine2014active} can provide the separable surrogates with principal latent directions.
 
 \subsection{Surrogate training at Level 2} 
 \label{subsec_SurrMath}
 
 Mathematically, the final prediction of decomposed Kriging $\tilde{z}^{DKG}(\mathbf{x_0}) \approx z(\mathbf{x_0})$ at an unobserved location $\mathbf{x_0}$ is defined as the summation of the four contributions:
 \begin{equation}\label{eq_SmrKRG}
     \begin{gathered}
         \tilde{z}^{DKG}(\mathbf{x_0}) = z^{REF} + \tilde{z}^{SYM\Delta}(\mathbf{x_0}) + \tilde{z}^{SEP\Delta}(\mathbf{x_0}) + \tilde{z}^{FRE\Delta}(\mathbf{x_0}) \\   
     \end{gathered}
 \end{equation}
 where $z^{REF}=z(\mathbf{x^{REF}})$ is a constant part corresponding to a reference configuration $\mathbf{x^{REF}}$, $\tilde{z}^{SYM\Delta}(\mathbf{x_0})$ is a symmetric part, $\tilde{z}^{SEP\Delta}(\mathbf{x_0})$ is a separable part, and $\tilde{z}^{FRE\Delta}(\mathbf{x_0})$ is an assumption-free part.
 They are calculated as sequential differences one after the other. As a consequence, Eq.\ref{eq_OrdKRGweight} for ordinary Kriging splits into the three delta predictors, $\tilde{z}^{SYM\Delta}$, $\tilde{z}^{SEP\Delta}$, and $\tilde{z}^{FRE\Delta}$:
 \begin{equation}\label{eq_SmrKRGsymsplit}
     \begin{gathered}
        \tilde{z}^{SYM\Delta}(\mathbf{x_0}) = \sum^{D}_{d=1} \sum^{N^{SYM}}_{n=1} w^{SYM}_{d,n,0}\left(z(\mathbf{x^{SYM}_{n}})-z^{REF}\right) \\
        \tilde{z}^{SYM}(\mathbf{x_0}) = z^{REF} + \tilde{z}^{SYM\Delta}(\mathbf{x_0})
     \end{gathered}
 \end{equation}  
 \begin{equation}\label{eq_SmrKRGsepsplit}
     \begin{gathered}
        \tilde{z}^{SEP\Delta}(\mathbf{x_0}) = \sum^{D}_{d=2} \sum^{N^{SEP}_d}_{n=1} w^{SEP}_{d,n,0}\left(z(\mathbf{x^{SEP}_{d,n}})-\tilde{z}^{SYM}(\mathbf{x^{SEP}_{d,n}})\right) \\
        \tilde{z}^{SEP}(\mathbf{x_0}) = \tilde{z}^{SYM}(\mathbf{x_0}) + \tilde{z}^{SEP\Delta}(\mathbf{x_0})
     \end{gathered}
 \end{equation}  
 \begin{equation}\label{eq_SmrKRGvarsplit}
     \begin{gathered}
        \tilde{z}^{FRE\Delta}(\mathbf{x_0}) = \sum^{N^{DKG}}_{n=1} w^{FRE}_{n,0}\left(z(\mathbf{x^{DKG}_{n}})-\tilde{z}^{SEP}(\mathbf{x^{DKG}_{n}})\right) \\
        \tilde{z}^{FRE}(\mathbf{x_0}) = \tilde{z}^{SEP}(\mathbf{x_0}) + \tilde{z}^{FRE\Delta}(\mathbf{x_0}) = \tilde{z}^{DKG}(\mathbf{x_0})
     \end{gathered}
 \end{equation} 
 For each of them is possible to reconstruct a partial prediction, $\tilde{z}^{SYM}$, $\tilde{z}^{SEP}$, and $\tilde{z}^{FRE}$, by summing the current delta to the previous layer. Note that the prediction reconstructed at Layer 3 is complete and, therefore, equivalent to that returned by the whole decomposed Kriging. 
 Symmetric and separable surrogates are trained on symmetric and separable sampling pools, made of $N^{SYM}$ and $\sum_{d=2}^{D} N^{SEP}_d$ samples. $\mathbf{x^{SYM}_{n}} = [x^{SYM}_{n},x^{REF}_2,...,x^{REF}_D] ~\forall n=1,...,N^{SYM}$ and $\mathbf{x^{SEP}_{d,n}} = [x^{REF}_1,x^{SEP}_{d,n},...,x^{REF}_D]  ~\forall n=1,...,N^{SEP}_d ~\forall d=2,...,D$ apply per dimensional component, while $\mathbf{x^{DKG}_{n}}$ are the union of all symmetric, separable, and assumption-free samples, $\mathbf{x^{FRE}_{n}} ~\forall n=1,...,N^{FRE}$. 
 Both $\mathbf{x^{SYM}_{n}}$ and $\mathbf{x^{SEP}_{d,n}}$ differ along one dimension at a time with respect to the reference $\mathbf{x^{REF}}$ vector. The reference point is arbitrarily chosen as part of the initialization (e.g., nominal condition), then fixed and pivotal among symmetric and separable sampling pools. In practice, Kriging surrogates belonging to Layers 1 and 2 are built along orthogonal cuts centered in $\mathbf{x^{REF}}$ and extend their prediction to the entire space for any $\mathbf{x_0}$.
 The symmetric surrogate is an extremization of the separable one, pretending that the whole multi-variate space can be traced back to a single-dimensional investigation. Any component can be chosen as a basis for the symmetric surrogate; without problem-related preferences, the first dimension in the problem, $d=1$, is used. At least one more observation on each layer (i.e., minimum 2 in total) completes the initialization, one per dimension in the separable pool. 
 
 The weights $w$ on each layer are calculated by solving the linear system of ordinary Kriging (Eq.\ref{eq_OrdKRGsys}) in terms of differences with respect to the previous layer. This extends to the residual auto-correlation function $\gamma$ as in universal Kriging (Eq.\ref{eq_UnvKRGsys}). The matrix form of the symmetric weights reads as follows:
  \begin{equation}\label{eq_SmrKRGsymweights}
     \begin{gathered}   
        \begin{bmatrix}
            \mathbf{\Gamma^{SYM}} & \mathds{1} \\
            \mathds{1}^T & 0 \\
        \end{bmatrix}
        \begin{bmatrix}
            \mathbf{w^{SYM}_{d,0}} \\
            \lambda^{SYM}_{d,0} \\
        \end{bmatrix}
        = 
        \begin{bmatrix}
            \mathbf{\gamma^{SYM}_{d,0}} \\
            1 \\
        \end{bmatrix}
        \forall d = 1,...,D
     \end{gathered}
 \end{equation} 
 \begin{equation}\label{eq_SmrKRGsymweightG}
     \begin{gathered}   
        \mathbf{\Gamma^{SYM}} = 
        \mathop{                     
            \begin{bNiceMatrix}
                \gamma^{SYM}(|x^{SYM}_{1}-x^{SYM}_{1}|) & \hdots & \gamma^{SYM}(|x^{SYM}_{1}-x^{SYM}_{ N^{SYM}}|) \\
                \vdots & \Ddots & \vdots \\
                \gamma^{SYM}(|x^{SYM}_{N^{SYM}}-x^{SYM}_{1}|) & \hdots  & \gamma^{SYM}(|x^{SYM}_{N^{SYM}}-x^{SYM}_{N^{SYM}}|) \\
            \end{bNiceMatrix}
        }\limits_{\in \mathbb{R}^{N^{SYM} \times N^{SYM}}} \\
     \end{gathered}
 \end{equation} 
  \begin{equation}\label{eq_SmrKRGsymweightw}
     \begin{gathered}   
        \mathbf{w^{SYM}_{d,0}} \in \mathbb{R}^{N^{SYM}} = [w^{SYM}_{d,1,0},..., w^{SYM}_{d,N^{SYM},0}]^T, \lambda^{SYM}_{d,0} \in \mathbb{R} \\
     \end{gathered}
 \end{equation} 
  \begin{equation}\label{eq_SmrKRGsymweightg}
     \begin{gathered}   
        \mathbf{\gamma^{SYM}_{d,0}} \in \mathbb{R}^{N^{SYM}} =
        \begin{bmatrix}
            \gamma^{SYM}(|x^{SYM}_{1}-x_{d,0}|) \\
            \vdots\\
            \gamma^{SYM}(|x^{SYM}_{N^{SYM}}-x_{d,0}|) \\
        \end{bmatrix}
        \\
        \gamma^{SYM}(|x_{i}-x_{j}|) \approx \frac{1}{2}\left(\left(z(\mathbf{x_{1,i}}) - z^{REF}\right) - \left(z(\mathbf{x_{1,j}}) - z^{REF}\right)\right)^2
     \end{gathered}
 \end{equation} 
 where the Euclidean distance operator $||\mathbf{x_i}-\mathbf{x_j}||$ between $i$ and $j$ points collapse to the absolute value $|x_i-x_j|$ since the symmetric surrogate considers only one dimension. $\mathds{1}$ is the unitary vector and $\lambda$ is the Lagrangian multiplier for ordinary Kriging (see \ref{app_KRG}).
 $|x^{SYM}_{n}-x_{d,0}|$ terms represent the projected distance of the new sample $\mathbf{x_{0}}$ from its $d$-th component to $d=1$, i.e., on the symmetric space. The symmetric auto-correlation matrix $\mathbf{\Gamma^{SYM}}$ is unique and can be calculated only once, regardless of the number of dimensions.
 
 Similarly to Eq.\ref{eq_SmrKRGsymweights}-\ref{eq_SmrKRGsymweightg}, the weights for the separable surrogate are obtained as:
 \begin{equation}\label{eq_SmrKRGsepweights}
     \begin{gathered}   
        \begin{bmatrix}
            \mathbf{\Gamma^{SEP}_d} & \mathds{1} \\
            \mathds{1}^T & 0 \\
        \end{bmatrix}
        \begin{bmatrix}
            \mathbf{w^{SEP}_{d,0}} \\
            \lambda^{SEP}_{d,0} \\
        \end{bmatrix}
        = 
        \begin{bmatrix}
            \mathbf{\gamma^{SEP}_{d,0}} \\
            1 \\
        \end{bmatrix}
        \forall d = 2,...,D
     \end{gathered}
 \end{equation} 
 In this case, the auto-correlation matrix $\mathbf{\Gamma^{SEP}_d}$ varies as a function of the considered dimensional component and must be recalculated $D-1$ times, solving the following linear system:
 \begin{equation}\label{eq_SmrKRGsepweightG}
     \begin{gathered}   
        \mathbf{\Gamma^{SEP}_d} = 
        \mathop{                     
            \begin{bNiceMatrix}
                \gamma^{SEP}(|x^{SEP}_{d,1}-x^{SEP}_{d,1}|) & \hdots & \gamma^{SEP}(|x^{SEP}_{d,1}-x^{SEP}_{d,N^{SEP}_d}|) \\
                \vdots & \Ddots & \vdots \\
                \gamma^{SEP}(|x^{SEP}_{d,N^{SEP}_d}-x^{SEP}_{d,1}|) & \hdots & \gamma^{SEP}(|x^{SEP}_{d,N^{SEP}_d}-x^{SEP}_{d,N^{SEP}_d}|) \\
            \end{bNiceMatrix}
        }\limits_{\in \mathbb{R}^{N^{SEP}_d} \times\mathbb{R}^{N^{SEP}_d}} \\
     \end{gathered}
 \end{equation}  
  \begin{equation}\label{eq_SmrKRGsepweightw}
     \begin{gathered}   
        \mathbf{w^{SEP}_{d,0}} \in \mathbb{R}^{N^{SEP}_d} = [w^{SEP}_{d,1,0},..., w^{SEP}_{d,N^{SEP}_d,0}]^T, \lambda^{SEP}_{d,0} \in \mathbb{R} \\
     \end{gathered}
 \end{equation}  
  \begin{equation}\label{eq_SmrKRGsepweightg}
     \begin{gathered}   
        \mathbf{\gamma^{SEP}_{d,0}} \in \mathbb{R}^{N^{SEP}_d} =
        \begin{bmatrix}
            \gamma^{SEP}(|x^{SEP}_{d,1}-x_{d,0}|) \\
            \vdots\\
            \gamma^{SEP}(|x^{SEP}_{d,N^{SEP}_d}-x_{d,0}|) \\
        \end{bmatrix}
        \\
        \gamma^{SEP}(|x_{d,i}-x_{d,j}|) \approx \\ \approx \frac{1}{2}\left(\left(z(\mathbf{x_{d,i}}) - \tilde{z}^{SYM}(\mathbf{x_{d,i}})\right) - \left(z(\mathbf{x_{d,j}}) - \tilde{z}^{SYM}(\mathbf{x_{d,j}})\right)\right)^2
     \end{gathered}
 \end{equation}  
 Symmetric and separable Kriging layers calculate the weights and add new samples only in the symmetric pool, along the first dimension, $\mathbf{x^{SYM}_{n}}$, or excluding it to avoid double-accounting in the separable pool, $\mathbf{x^{SEP}_{n}}$, respectively.
 
 The assumption-free system for the weights is instead:
  \begin{equation}\label{eq_SmrKRGvarweights}
     \begin{gathered}   
        \begin{bmatrix}
            \mathbf{\Gamma^{FRE}} & \mathds{1} \\
            \mathds{1}^T & 0 \\
        \end{bmatrix}
        \begin{bmatrix}
            \mathbf{w^{FRE}_{0}} \\
            \lambda^{FRE}_{0} \\
        \end{bmatrix}
        = 
        \begin{bmatrix}
            \mathbf{\gamma^{FRE}_{0}} \\
            1 \\
        \end{bmatrix}
     \end{gathered}
 \end{equation} 
 \begin{equation}\label{eq_SmrKRGvarweightG}
     \begin{gathered}   
        \mathbf{\Gamma^{FRE}} = 
        \mathop{                     
            \begin{bNiceMatrix}
                \gamma^{FRE}(||\mathbf{x^{DKG}_{1}}-\mathbf{x^{DKG}_{1}}||) & \hdots & \gamma^{FRE}(||\mathbf{x^{DKG}_{1}}-\mathbf{x^{DKG}_{N^{DKG}}}||) \\
                \vdots & \Ddots & \vdots \\
                \gamma^{FRE}(||\mathbf{x^{DKG}_{N^{DKG}}}-\mathbf{x^{DKG}_{1}}||) & \hdots & \gamma^{FRE}(||\mathbf{x^{DKG}_{N^{DKG}}}-\mathbf{x^{DKG}_{N^{DKG}}}||) \\
            \end{bNiceMatrix}
        }\limits_{\in \mathbb{R}^{N^{DKG}} \times\mathbb{R}^{N^{DKG}}} \\
     \end{gathered}
 \end{equation}
  \begin{equation}\label{eq_SmrKRGvarweightw}
     \begin{gathered}   
        \mathbf{w^{FRE}_{0}} \in \mathbb{R}^{N^{DKG}} = [w^{FRE}_{1,0},..., w^{FRE}_{N^{DKG},0}]^T, \lambda^{FRE}_{0} \in \mathbb{R} \\
     \end{gathered}
 \end{equation}
  \begin{equation}\label{eq_SmrKRGvarweightg}
     \begin{gathered}   
        \mathbf{\gamma^{FRE}_{0}} \in \mathbb{R}^{N^{DKG}} =
        \begin{bmatrix}
            \gamma^{FRE}(||\mathbf{x^{DKG}_{1}}-\mathbf{x_{0}}||) \\
            \vdots\\
            \gamma^{FRE}(||\mathbf{x^{DKG}_{N^{DKG}}}-\mathbf{x_{0}}||) \\
        \end{bmatrix}
        \\
        \gamma^{FRE}(||\mathbf{x_{i}}-\mathbf{x_{j}}||) \approx \frac{1}{2}\left(\left(z(\mathbf{x_{i}}) - \tilde{z}^{SEP}(\mathbf{x_{i}})\right) - \left(z(\mathbf{x_{j}}) - \tilde{z}^{SEP}(\mathbf{x_{j}})\right)\right)^2
     \end{gathered}
 \end{equation}
 Layer 3 adds new samples into the assumption-free pool only, and assumption-free weights are calculated once regardless of dimensionality. However, computation is heavy due to the large sample basis and multi-dimensional distances ($||\cdot||$ operator). 
 
 The variance $\tilde{\sigma}^2(\mathbf{x_0})$ of the prediction at each layer in Eqs.\ref{eq_SmrKRGsymsigmasys}-\ref{eq_SmrKRGvarsigmasys} is derived from Eq.\ref{eq_OrdKRGsigmasys}: 
 \begin{equation}\label{eq_SmrKRGsymsigmasys}
     \begin{gathered}     
         \tilde{\sigma}^{{SYM}^2}(\mathbf{x_0}) = \sum^{D}_{d=1} 
         \begin{bmatrix}
            \mathbf{w^{SYM}_{d,0}}^T & \lambda^{SYM}_{d,0}\\
         \end{bmatrix}
         \begin{bmatrix}
            \mathbf{\gamma^{SYM}_{d,0}} \\
            1 \\
         \end{bmatrix} 
     \end{gathered}
  \end{equation} 
  \begin{equation}\label{eq_SmrKRGsepsigmasys}
     \begin{gathered}    
        \tilde{\sigma}^{{SEP}^2}(\mathbf{x_0}) = \sum^{D}_{d=2} 
        \begin{bmatrix}
           \mathbf{w^{SEP}_{d,0}}^T & \lambda^{SEP}_{d,0}\\
        \end{bmatrix}
        \begin{bmatrix}
           \mathbf{\gamma^{SEP}_{d,0}} \\
           1 \\
        \end{bmatrix}  
     \end{gathered}
 \end{equation} 
 \begin{equation}\label{eq_SmrKRGvarsigmasys}
     \begin{gathered}    
        \tilde{\sigma}^{{FRE}^2}(\mathbf{x_0}) = 
        \begin{bmatrix}
           \mathbf{w^{FRE}_{0}}^T & \lambda^{FRE}_{0}\\
        \end{bmatrix}
        \begin{bmatrix}
           \mathbf{\gamma^{FRE}_{0}} \\
           1 \\
        \end{bmatrix}    
     \end{gathered}
 \end{equation}   
 Variance along different directions is summed since orthogonal models are independent. 
 A Confidence Interval ($CI$), centered around the expected value of the prediction, $\tilde{z}(\mathbf{x_{0}})$, can be defined via $\tilde{\sigma}^2$. $CI$ is expected to contain the actual value $z(\mathbf{x_{0}})$ according to a probability $P=[0,1]$, which determines the confidence level. For each new sample $\mathbf{x_0}$, $CI_{\mathbf{x_0}}(P)$ is calculated as the inverse of the cumulative distribution function $\Phi$ of the Kriging normal distribution with mean $\mu=\tilde{z}(\mathbf{x_{0}})$ and variance $\sigma^2={\tilde{\sigma}}^2(\mathbf{x_0})$, for the symmetric probability interval subtended by $P$, $[0.5(1-P),0.5(1+P)]$: 
  \begin{equation}\label{eq_SmrKRGCI}
     \begin{gathered}     
         CI_{\mathbf{x_0}}(P) = \Phi^{-1}_{\tilde{z}(\mathbf{x_{0}}),{\tilde{\sigma}}^2(\mathbf{x_0})}\left( \left[ \frac{1-P}{2},\frac{1+P}{2} \right] \right) \\
     \end{gathered}
 \end{equation}  
 
 Symmetric and separable layers are as powerful in extrapolating orthogonal properties as they are in propagating eventual errors. Preventing misinformation from escalating up to Layer 3 is essential for decomposed Kriging accuracy. 
 For this reason, separable and assumption-free layers are computed twice, namely, the first in terms of delta with respect to the previous layer, as presented so far, and the second directly from $z^{REF}$, using previous samples but not previous predictions. The following direct forms are alternatives to Eq.\ref{eq_SmrKRGsepsplit} and Eq.\ref{eq_SmrKRGvarsplit}:
 \begin{equation}\label{eq_SmrKRGsepsplitdirect}
     \begin{gathered}
         \tilde{z}^{SEP}(\mathbf{x_0}) = z^{REF} + \sum^{D}_{d=2} \sum^{N^{SEP}_d}_{n=1} w^{SEP}_{d,n,0}\left(z(\mathbf{x^{SEP}_{d,n}})-z^{REF}\right) \\
     \end{gathered}
 \end{equation} 
 \begin{equation}\label{eq_SmrKRGvarsplitdirect}
     \begin{gathered}
         \tilde{z}^{FRE}(\mathbf{x_0}) = z^{REF} + \sum^{N^{DKG}}_{n=1} w^{FRE}_{n,0}\left(z(\mathbf{x^{DKG}_{n}})-z^{REF}\right) \\
     \end{gathered}
 \end{equation} 
 Weights and semivariogram calculations are adapted to reflect the $z-z^{REF}$ nature of direct surrogates, and decomposed Kriging predictor in Eq.\ref{eq_SmrKRG} becomes either $\tilde{z}^{DKG}(\mathbf{x_0}) = \tilde{z}^{SEP}(\mathbf{x_0}) + \tilde{z}^{FRE\Delta}(\mathbf{x_0})$ or simply $\tilde{z}^{DKG}(\mathbf{x_0}) = \tilde{z}^{FRE}(\mathbf{x_0})$ if Eq.\ref{eq_SmrKRGsepsplitdirect} or Eq.\ref{eq_SmrKRGvarsplitdirect} are respectively active.
 The smallest validation error (Eq.\ref{eq_SmrKRGsymexitVAL}, \ref{eq_SmrKRGsepexitVAL}, \ref{eq_SmrKRGvarexitVAL}) is used to control whether the delta or the direct Kriging surrogate is activated in each layer. As a result, a total number of $[1]^{SYM} + [2(D-1)]^{SEP}+[2]^{FRE}=2D+1$ surrogates are calculated at each iteration. The final decomposed Kriging surrogate is made of delta or direct form at Layer 2 and delta at Layer 3 or direct at Level 3. This nearly doubles the algorithm's computational complexity, but it also significantly alleviates the risk of producing inaccurate surrogates, thus stabilizing fast-scaling properties. Furthermore, only one new sample is used at each iteration, following the sequence [SYM, SEP, FRE, SYM,...] and skipping layers that pass the quality check, until the training is completed. This is critical to contain and correct divergent behaviors thanks to the one-by-one sample progression, if a surrogate momentarily misinterprets the $z$ properties.

 Kriging can approximate multi-modality and higher-order discontinuities, including sudden steps and slope changes. This demands well-placed samples; thus, selecting an appropriate parametric semivariance kernel function $\gamma$ is crucial for the quality of the surrogate. Decomposed Kriging automatically fits the hyperparameters of different auto-correlation $\gamma$ models, for each layer at each iteration. Every time, the model presenting the smallest validation error is chosen. Variograms and $\gamma$ models are expressed in residual terms of values, $z-\tilde{z}^{SEP}$ at Layer 3, $z-\tilde{z}^{SYM}$ at Layer 2, and $z-z^{REF}$ at Layer 1, according to delta equations, Eq.\ref{eq_SmrKRGsymsplit}-\ref{eq_SmrKRGvarsplit}. Residuals at Layer 2 or Layer 3 become simply $z-z^{REF}$ if direct equations are instead activated Eq.\ref{eq_SmrKRGsymsplit}-\ref{eq_SmrKRGvarsplit}. The details of the fitting process are reported in \ref{subapp_variogram} and represent a non-trivial meta-optimization to be solved several times during the decomposed Kriging training. 
 It is important to mention that $\gamma$ models are fit on the semivariogram, instead of directly using any cross-validation error, validation error, or maximum likelihood estimation \cite{moustapha2016quantile}. The reason lies in the very structure and purpose of decomposed Kriging, which minimizes the number of observations on each layer, even for high-dimensional and complex problems. Indeed, error-based fitting relies on a smaller dataset and a more complicated least squares problem than variogram fitting. This can mislead the surrogates, especially on Layers 1 and 2 and/or in the early phases, thus compromising scalability. 

 \subsection{Exploration and validation samples ar Levels 2 and 3} 
 \label{subsec_IterMath}
 
 The new explorative sample at Level 2 is selected in correspondence to the current Kriging's maximum variance at each surrogate layer. The sampling pools are enlarged one point at a time ($N+1$ index):  
 \begin{equation}\label{eq_SmrKRGsymnext}
     \begin{gathered}     
        \mathbf{x^{SYM}_{N^{SYM}+1}} = \argmax_{x^{SYM}_0} \tilde{\sigma}^{{SYM}^2}(x^{SYM}_0)
     \end{gathered}
 \end{equation}   
  \begin{equation}\label{eq_SmrKRGsepnext}
     \begin{gathered}     
        \mathbf{x^{SEP}_{d,N^{SEP}_d+1}} = \argmax_{d,x^{SEP}_{d,0}}\left[\max_{x^{SEP}_{d,0}} \tilde{\sigma}^{{SEP}^2}(x^{SEP}_{d,0})\right]_{d=2,...,D} \\
     \end{gathered}
 \end{equation}   
  \begin{equation}\label{eq_SmrKRGvarnext}
     \begin{gathered}     
        \mathbf{x^{FRE}_{N^{FRE}+1}} = \argmax_{\mathbf{x^{FRE}_0}} \tilde{\sigma}^{{FRE}^2}(\mathbf{x^{FRE}_0})
     \end{gathered}
 \end{equation}    
 As in Bayesian optimization, the acquisition function \cite{snoek2012practical} is straightforward to define but challenging to solve, since the associated box-bounded meta-optimization problem is non-linear, non-convex, multi-modal, and increasingly large from Layer 1 to Layer 3. Moreover, it must be solved numerous times throughout the adaptive training process. Fortunately, evaluating Kriging is fast (see Section \ref{sec_discussion}), so meta-heuristics becomes an accessible global meta-optimizer. Refer to \ref{subapp_hyper} for the meta-optimization settings adopted in decomposed Kriging.
  
 Error estimation is fundamental to stop the demanding iterative process for new samples, as soon as the surrogate's approximation is deemed good. In this regard, the confidence interval returned by Kriging (Eq.\ref{eq_SmrKRGCI}) is already a quality estimator. If $\max_{\mathbf{x_0}} (CI_{\mathbf{x_0}}(0.95)-\tilde{z}(\mathbf{x_0}))$, corresponding to Eq.\ref{eq_SmrKRGsymnext}-\ref{eq_SmrKRGvarnext} conditions, is below a reference threshold, $\tau_{CI}$, there is a 95\% chance the actual value will fall within the $CI$ range. Since this quality metric is bound to the Kriging assumptions, an unbiased validation error, $\epsilon_{VAL}$, to be lowered below a threshold, $\tau_{VAL}$, is also needed. The error is computed on validation samples at each layer, $\mathbf{v^{DKG}}=[\mathbf{v^{SYM}},\mathbf{v^{SEP}},\mathbf{v^{FRE}}]$, that are never used for the training and are included since initialization. Details on error formulations are provided in \ref{subapp_error}.
 Exit criteria on tolerance thresholds are complemented by a minimum amount of overall validation points, $V^{DKG} = V^{SYM} + \sum^{D}_{d=2} V^{SEP}_d + V^{FRE} \geq V^{DKG}_{min}$, to counteract premature arrest. Furthermore, a maximum amount of total samples, $N^{TOT} = N^{DKG}+V^{DKG} \leq N^{TOT}_{max}$, and samples per dimension, $ N^{SYM}+V^{SYM} \leq N^{SS}_{1,max}, N^{SEP}_d+V^{SEP}_d \leq N^{SS}_{d,max}$, impose an hard stop. If the process is terminated this way, the quality of the surrogate is judged by the error level achieved so far.

 For a trustworthy error estimate, the validation samples are progressively increased from initialization using a constant ratio to training points, $v_{ratio} = N^{DKG}/V^{DKG}$. Validation samples are selected for maximum diversity with respect to all the previous observations:
 \begin{equation}\label{eq_SmrKRGsymnextVAL}
     \begin{gathered}    
        \mathbf{v^{SYM}_{V^{SYM}+1}} = \argmax_{v^{SYM}_0} 
        \begin{bmatrix}
            |x^{SYM}_n - v^{SYM}_0| ~\forall n=1,..,N^{SYM} \\
            |v^{SYM}_n - v^{SYM}_0| ~\forall n=1,..,V^{SYM} \\
        \end{bmatrix}        
     \end{gathered}
 \end{equation}  
 \begin{equation}\label{eq_SmrKRGsepnextVAL}
     \begin{gathered}    
        \mathbf{v^{SEP}_{d,V^{SEP}+1}} = \argmax_{d} \left[ \max_{v^{SEP}_{d,0}} 
        \begin{bmatrix}
            |x^{SEP}_{d,n} - v^{SEP}_{d,0}| ~\forall n=1,..,N^{SEP} \\
            |v^{SEP}_{d,n} - v^{SEP}_{d,0}| ~\forall n=1,..,V^{SEP} \\
        \end{bmatrix}
        \right]_{d=2,...,D} \\             
     \end{gathered}
 \end{equation}  
 \begin{equation}\label{eq_SmrKRGvarnextVAL}
     \begin{gathered}    
        \mathbf{v^{FRE}_{V^{FRE}+1}} = \argmax_{\mathbf{v^{FRE}_0}} 
        \begin{bmatrix}
            ||\mathbf{x^{DKG}_n} - \mathbf{v^{FRE}_0}|| ~\forall n=1,..,N^{DKG} \\
            ||\mathbf{v^{DKG}_n} - \mathbf{v^{FRE}_0}|| ~\forall n=1,..,V^{DKG} \\
        \end{bmatrix}
     \end{gathered}
 \end{equation}  
 This prevents clustering and abates the risk of compromising the validation error, while entailing another complex meta-optimization (Section \ref{sec_castudy} for settings). The use of validation points is preferred over cross-validation for two reasons: i) avoid recomputing an increasing number of $N^{SYM} + 2\sum_{d=2}^{D} N^{SEP}_d + 2N^{FRE}$ surrogates at each iteration; and ii) account for information exogenous to the training process, to mitigate biases. Hence, the algorithm will additionally ask for a new validation sample (Eq.\ref{eq_SmrKRGsymnextVAL}-\ref{eq_SmrKRGvarnextVAL}) after a new training sample (Eq.\ref{eq_SmrKRGsymnext}-\ref{eq_SmrKRGvarnext}) to respect $v_{ratio}$ in all the sampling pools. 

 According to the flow scheme in Fig.\ref{fig_SmrKRGflow} and the appended pseudo-code  in~\ref{subapp_pseudocode}, the MLIO exploitation phase at Level 3 takes place only in correspondence to assumption-free decomposed Kriging at Layer 3. The exploitative feedback loop pursues a greedy action foreign to the explorative Kriging infill. Potentially whatever task, even completely different from design under uncertainty, can be described in the greedy phase, exploiting the surrogate under construction. Any acquisition operator $g$ over the decomposed Kriging prediction $\tilde{z}^{DKG}(\mathbf{x})$ and/or the original problem $z(\mathbf{x})$ serves the purpose, as long as it returns a subset of possible new samples $\mathbf{X^{FRE}_0}$, of $N^g$ size, to maximize the Kriging confidence upon $\mathbf{x^{FRE}_{N^{FRE+1}}}$ as a modification to Eq.\ref{eq_SmrKRGvarnext}:
 \begin{equation}\label{eq_SmrKRGvargreedy}
     \begin{gathered}    
        \mathbf{X^{FRE}_0} \in \mathbb{R}^{N^g \times D} = g(\tilde{z}^{DKG}(\mathbf{x}),z(\mathbf{x})) \\
        \mathbf{x^{FRE}_{N^{FRE}+1}} = \argmax_{\mathbf{X^{FRE}_0}} \tilde{\sigma}^{{FRE}^2}(\mathbf{X^{FRE}_0})
     \end{gathered}
 \end{equation} 
 Greedy exploitation is performed alternatively to exploration in Kriging Layer 3, respecting a constant $g_{ratio}$ between the number of iterations in which the two are activated. The greedy phase is skipped if $g_{ratio}$ is not provided; in this case, the decomposed Kriging remains a fast-scaling, highly exploratory surrogate, without incorporating any decision intent while training.
 
 Key hyperparameters associated with decompositions (Eq.\ref{eq_SmrKRGsymnext}-\ref{eq_SmrKRGvarnext}), semi-variograms (Eq.\ref{eq_Semivariogramfit}-\ref{eq_Semivariogramwindow}), and validation (Eq.\ref{eq_SmrKRGsymnextVAL}-\ref{eq_SmrKRGvarnextVAL}) are self-calibrated via the internal meta-optimization routines, whose settings are hard-coded since they generally hold (Sections \ref{sec_castudy} and \ref{app_Tuning}). Only intuitive termination criteria, initial sampling, and dimensional bounds are strictly required, and there are just two hyperparameters for fine-tuning: $v_{ratio}$ for errors and quality checks, and $g_{ratio}$ for balancing exploration and exploitation. They can calibrate performance on a case-by-case basis, but robust default settings exist. The algorithm returns the final surrogate configuration and its evolved history of surrogates during training, i.e., $\gamma$ models and observations at each layer, delta/direct mixture, and greedy subset at each iteration. 
 Nevertheless, Kriging is notoriously expensive to compute, especially for many observations, and decomposed Kriging calculates multiple surrogates at every iteration. A series of measures are therefore implemented for algorithmic efficiency, especially through shared databases and initialization boosting for meta-optimizers. Details on speed-up techniques are provided in~\ref{subapp_speedup}.

 \section{Analytical benchmark for numerical validation}
 \label{sec_castudy}
 

 The efficiency and effectiveness of MLIO via decomposed Kriging are proven on a testbed and compared to a state-of-the-art method for optimization under uncertainty. To demonstrate the generalization claims of this study and properly assess performance, the key features of the test landscape should be known beforehand and easy to quantify, at least numerically, if not analytically.
 Noteworthy analytical functions exist in the field of uncertainty quantification (e.g., Ishigami \cite{ishigami1990importance}), and a vast literature on test functions for global optimization is available \cite{jamil2013literature}, with some eminent benchmark problems used for international competitions \cite{kumar2021benchmark}. Nevertheless, to the best of our knowledge, no renowned testbeds combine the two.

 The analytical benchmark for the present study is then built by adapting standard functions for global minimization. Part of the problem variables $\mathbf{x}$ are treated as design variables $\mathbf{u}$ for the $OPT_u$ process, and part as uncertain parameters to characterize the $UQ_p(\mathbf{u})$ process. The well-known mathematical characteristics of the functions remain unaltered, alongside the analytical minima and maxima. Designing under uncertainty can then be conducted numerically through a dense sampling to explore the whole variability of $f(\mathbf{u},\mathbf{p})$, and later apply $UQ(\mathbf{u}) = UQ_p(f(\mathbf{u},\mathbf{p}))$ and $OPT_u = min_u UQ(\mathbf{u})$ in two steps, emulating a real-life scenario-based approach with a large number of scenarios.
 It is possible to carefully choose a set of analytical functions $f$ showing indeterminate dimensional scaling and variegated properties to represent the heterogeneous traits of real problems. Symmetry, separability, uni- and multi-modality, peaks, barriers, strong interdependence, and noise modulation effects, even non-differentiability and ill-conditioning, can be investigated. Tab.~\ref{tab_functions} summarizes the 6 functions chosen for this study to embody all the properties just mentioned. They are accompanied by name, 2D visualization centered around $\mathbf{x_0}=0^D$, mathematical formulation, and box-bounds $\mathbf{B}$ for this study. A random seed for translation $\mathbf{T}=\mathcal{U}([0,1])^D$ is also introduced (diagonal ($T_1=T_2=...=T_D$) for the symmetric "Step" and "Alpine" functions). All functions are normalized, $\bar{f} \in [0,1]^1$, between their global minimum ($minf(\mathbf{T},\mathbf{B})$, analytical) and global maximum ($\tilde{maxf}(\mathbf{T},\mathbf{B})$, conservatively analytical), depending on bounds and translation. As well,  their variables are normalized, $\mathbf{\bar{x}} \in [0,1]^D$. The overall benchmark structure is summarized in the following equations:
 \begin{equation}\label{eq_fnorm}
    \begin{gathered}    
        \mathbf{x} \in \mathbb{R}^D, B_{d,1} \leq x_i \leq B_{d,2} ~\forall i=1,...,D\\
        \mathbf{\bar{x}} = \left[ \frac{x_1-B_{1,1}}{B_{1,2}-B_{1,1}}+T_1,...,\frac{x_D-B_{D,1}}{B_{D,2}-B_{D,1}}+T_D \right] \\
        \bar{f}(\mathbf{\bar{x}}) = \frac{f(\mathbf{x})-minf(\mathbf{T},\mathbf{B})}{\tilde{maxf}(\mathbf{T},\mathbf{B})-minf(\mathbf{T},\mathbf{B})} 
    \end{gathered}
 \end{equation} 
 Thanks to $\mathbf{T}$, it is possible to generate a family of functions with the same shape to evaluate the statistical performance of different methods. For each function, 25 repetitions are adopted, in line with competition benchmarks. Also, three sizes of dimensionality are tested: $2D$ minimal, to compare on the easiest case; $20D$ small, representative of many realistic and complex applications limited in dimensionality; $200D$ medium to large, low-end side for aggregated big problems as in operational research. Half of the total $D$ dimensions, $D_u$, are used as variables to optimize, and half, $D_p$, as parameters to characterize. 
 For each function, each number of dimensions, and each repetition, both robust and stochastic optimizations under uncertainty are conducted, in the representative form of $UQ_p = max_p$ and $UQ_p = \mathbb{E}_p$, respectively. The performance of the tested methods is evaluated against a reference pool of one million samples, treated as ground truth, $true$, and generated by a factorial combination of 1e3 evenly distributed Halton set points on both variable and parametric spaces. Two error metrics are defined in the form of normalized absolute errors, given the best design found as $\bar{\mathbf{u}}^{mthd}_{min} = \argmin_{\bar{\mathbf{u}}} UQ^{mthd}(\bar{\mathbf{u}})$ for each method, $mthd$:
 \begin{equation}\label{eq_ErrMetrics}
    \begin{gathered}   
        IA^{mthd} = \frac{|UQ^{mthd}(\bar{\mathbf{u}}^{mthd}_{min})-UQ^{true}(\bar{\mathbf{u}}^{mthd}_{min})|}{\max_u UQ^{true}(\bar{\mathbf{u}})-\min_u UQ^{true}(\bar{\mathbf{u}})} \\
        SO^{mthd} = \frac{|UQ^{true}(\bar{\mathbf{u}}^{mthd}_{min})-UQ^{true}(\bar{\mathbf{u}}^{true}_{min})|}{\max_u UQ^{true}(\bar{\mathbf{u}})-\min_u UQ^{true}(\bar{\mathbf{u}})} \\
    \end{gathered}
 \end{equation} 
 Inaccuracy (IA) measures how much the uncertainty quantification for the best design, $\mathbf{u}^{mthd}_{min}$, deviates from the true uncertainty quantification, $UQ^{true}$. Suboptimality (SO) measures how much the true uncertainty quantification for the best method design, $UQ^{true}(\mathbf{u}^{mthd}_{min})$, differs from the one of the true best design, $UQ^{true}(\mathbf{u}^{true}_{min})$. For the sake of fairness and consistency, all methods can select samples only among the reference pool. So, $SO^{mthd}=0$ and $IA^{mthd}=0$ mean the method meets the same quality as the reference dense sampling of one million points. The denominator is a normalizer over the true envelope. 

 The proposed multi-level informed optimization via decomposed Kriging is developed in MATLAB\textregistered~ version 9.14 (R2023a). Its performance is calculated by sampling the same 1e6 reference points through surrogate. It is compared with one of the most advanced PCE available for uncertainty quantification, i.e. the sparse, truncated, degree, and q-norm adaptive Polynomial Chaos Expansion within the UQLab tool \cite{marelli2014uqlab}, coupled with the single-objective Genetic Algorithm (GA) implemented in MATLAB\textregistered.
 Despite its conventional two-step nature, the coupled PCE and GA have recently proven capable of addressing difficult black-box engineering cases of design under uncertainty \cite{coppitters2020robust}. It is herein labeled PCE+GA and serves as a cutting-edge exponent of traditional techniques to compare MLIO with. 
 GA will minimize $UQ_p(\tilde{\bar{f}}(\bar{\mathbf{u}},\bar{\mathbf{p}}))$ where $\tilde{\bar{f}}$ is the approximation returned by the normalized PCE, fitted for each candidate design $\bar{\mathbf{u}}$ over a subset of the 1e3 parametric reference samples $\bar{\mathbf{p}}$.

 
  \begin{table}  
    \caption{Features of the validation benchmark comprising $\sim$7e9 samples in total, mainly belonging to reference samples.}
    \centering
    \footnotesize
    \begin{tblr}{
      cells={valign=m,halign=c},
      cell{1}{1} = {r=2,c=2}{},
      cell{1}{3} = {c=3}{},
      cell{3}{1} = {r=6}{},
      cell{3}{3} = {r=6,c=3}{},
      row{1-2}={font=\bfseries},
      column{1-2}={font=\bfseries},
      colspec={|c|c|c|c|c|}
    }
    \hline
    & & Dimensionalities & & \\
    \cline{3-5}
    & & {$\mathbf{D=2}$ \\ $\mathbf{D_u,D_p=1}$} & {$\mathbf{D=20}$ \\ $\mathbf{D_u,D_p=10}$} & {$\mathbf{D=200}$ \\ $\mathbf{D_u,D_p=100}$} \\
    \hline
    \begin{sideways}Test functions\end{sideways} & Step & {Repetitions = 25 \\ Tuning sets \# = 6 for PCE+GA, 2 for MLIO \\ Max samples = 1e4 for PCE+GA, 1e3 for MLIO \\ Reference = 1e6 samples Halton set \\ Optimization runs = 2, robust and stochastic} & & \\
    \hline
    & Alpine & & & \\
    \hline
    & SumSquares & & & \\
    \hline
    & Levy & & & \\
    \hline
    & Rosenbrock & & & \\
    \hline
    & Ackley & & & \\
    \hline
    \end{tblr}
    \label{tab_testbed}
 \end{table}
 Properly setting the two methods is essential to express their potential for a meaningful comparison. Hyperparameters are then arranged as reported in \ref{app_Tuning}, distinguishing between major ones, which are directly tuned with a few configurations denoted with \# label, and secondary ones, fixed to standard values. Results (Fig.\ref{fig_MetricResults} and supplementary figures) will investigate the best tuning among \# configurations.
 PCE+GA tuning process consumed 30 times the $f$ evaluations compared to MLIO.
 Conclusively, Tab.~\ref{tab_testbed} summarizes the testbed settings for this study. 
 SO and IA errors are statistically tracked over repetitions for robust and stochastic optimization separately, as a function of the number of $f$ observations required by each method. This identifies the minimum number of samples needed to reach a given quality level for PCE+GA and MLIO, depending on the test function and dimensionality. IA and SO are set to 1 (theoretical maximum) until the first estimation of the best design under uncertainty is available after initialization.


 The computational burden is measured in terms of equivalent averaged execution time per sample. 
 This is calculated by timing each method on the test functions under the same load conditions of the running machine, which is a 12th Gen Intel\textregistered~ i7-12700K, 3.61 GHz, and 128 GB RAM workstation. 
 The time each method needs for an approximation level of 1\%, sufficient for most engineering problems, is compared as average among IA and SO metrics on robust and stochastic optimizations.


\section{Aggregated results}
\label{sec_results}
 \begin{figure*}
    \centering
    \subfloat[]{{\includegraphics[width=.6\textwidth]{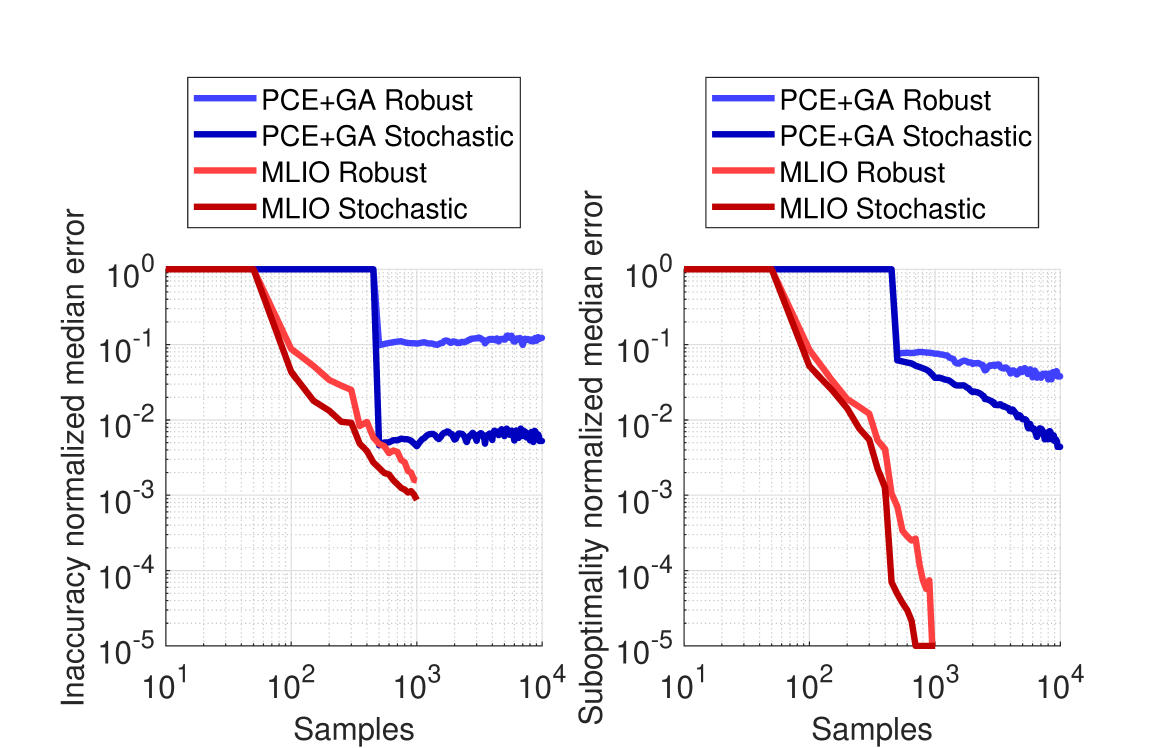}}\label{fig_MetricResultsB}}
    \quad
    \subfloat[]{{\includegraphics[width=.36\textwidth]{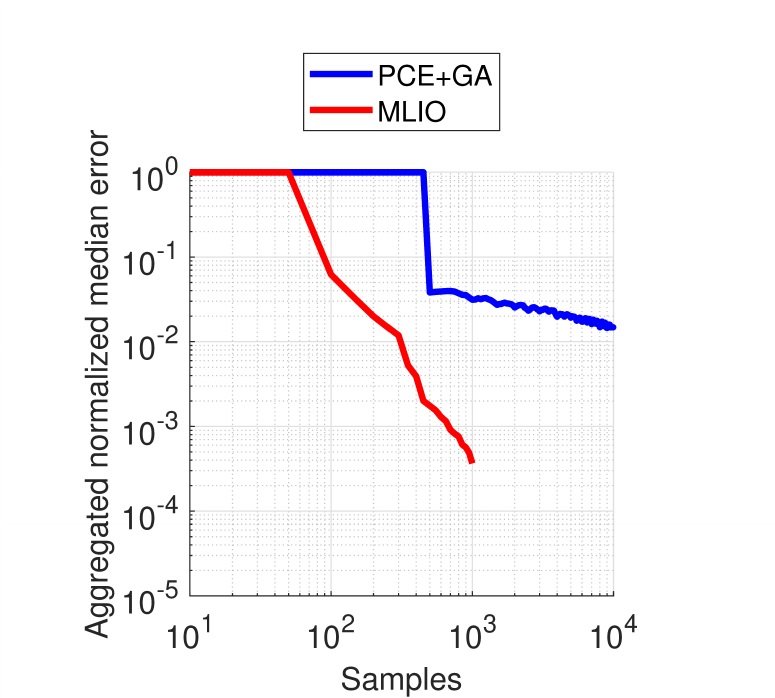}}\label{fig_MetricResultsA}}%
    \caption{Aggregated median errors over the testbed for tuned PCE+GA (setting \#2 over 6) and MLIO (setting \#1 over 2) vs. 1e6 Halton set. a) shows MLIO vs. PCE+GA inaccuracy and suboptimality for robust and stochastic optimizations. b) shows MLIO vs. PCE+GA statistical error merging all results.}
    \label{fig_MetricResults}
 \end{figure*}
 

Fig.\ref{fig_MetricResults} compares the IA and SO performance of the PCE+GA and MLIO methods for robust and stochastic optimization. Figure~\ref{fig_MetricResultsB} shows all four contributors distinctly, while Fig.~\ref{fig_MetricResultsA} represents the most aggregated form of them, i.e. IA and SO together for robust ans stochastic optimization, over the 25 repetitions, across the 6 test functions, and 3 dimensionalities. 
The best-performing tuning is found to be setting \#2 for PCE+GA and setting \#1 for MLIO (see \ref{subapp_hyper} and Fig.3-10 in the Supplementary material), which proves its strong self-adaptation. 
Thanks to normalization and known features of the test functions, the range $[1,0.1]$ for IA and SO errors can be considered poor, $[0.1,0.01]$ good, $[0.01,0.001]$ very good, and anything below is excellent. Errors are limited to 1e-5. 
The following distinguished traits emerge:
 \begin{itemize}[noitemsep,topsep=0pt]
     \item MLIO is significantly faster than PCE+GA to improve both IA and SO (logarithmic rates in \ref{tab_ResultsRates}), saving 1.5-100 times the resources for the same accuracy, or being 2-8000 times more accurate for the same resources.
     \item IA is constant as a function of samples for PCE+GA since UQ is performed in the same static way via PCE.
     \item The number of samples required by PCE+GA to produce the first estimation of the best design under uncertainty ($\sim$400, including one PCE training and the first generation of GA) is 10 times the MLIO counterpart ($\sim$50, minimal initialization only).
 \end{itemize}
 \begin{table}[!h]
    \caption{Tuned MLIO vs. PCE+GA  median performance metrics vs. 1e6 Halton set after 1e3 samples, for robust and stochastic optimization, per function, per dimensionality. The best between the two methods in each case is highlighted by cell color (blue for PCE+GA and red for MLIO), and particularly poor performance ($>$10\%) is highlighted in bold with the same color scheme.}
    \centering
    \tiny
    \begin{tblr}{
      cells={valign=m,halign=c},
      row{1-2}={font=\bfseries},
      column{1-3}={font=\bfseries},
      colspec={|c|c|c!{\vrule width 1.5pt}c|c!{\vrule width 1.5pt}c|c!{\vrule width 1.5pt}c|c!{\vrule width 1.5pt}},
      cell{1}{1} = {r=2,c=3}{},
      cell{1}{4} = {c=2}{},
      cell{1}{6} = {c=2}{},
      cell{1}{8} = {c=2}{},
      cell{3}{1} = {r=4}{},
      cell{7}{1} = {r=4}{},
      cell{11}{1} = {r=4}{},
      cell{15}{1} = {r=4}{},
      cell{19}{1} = {r=4}{},
      cell{23}{1} = {r=4}{},
      cell{3}{2} = {r=2}{},
      cell{5}{2} = {r=2}{},
      cell{7}{2} = {r=2}{},
      cell{9}{2} = {r=2}{},
      cell{11}{2} = {r=2}{},
      cell{13}{2} = {r=2}{},
      cell{15}{2} = {r=2}{},
      cell{17}{2} = {r=2}{},
      cell{19}{2} = {r=2}{},
      cell{21}{2} = {r=2}{},
      cell{23}{2} = {r=2}{},
      cell{25}{2} = {r=2}{},
    }
    \hline
    & & & D=2 & & D=20 & & D=200 & \\
    \cline{4-9}
    & & & PCE+GA & MLIO & PCE+GA & MLIO & PCE+GA & MLIO \\
    \hline[1.5pt]
    \begin{sideways}Step\end{sideways} & Rob. & IA & 1.81e-2 & \SetCell[c=1]{bg=red!25}4.87e-3 & 3.21e-2 & \SetCell[c=1]{bg=red!25}1.00e-5 & \color{blue}\textbf{3.46e-1} & \SetCell[c=1]{bg=red!25}1.00e-5 \\
    \hline
    & & SO & 3.46e-5 & \SetCell[c=1]{bg=red!25}1.00e-5 & \color{blue}\textbf{1.06e-1} & \SetCell[c=1]{bg=red!25}1.00e-5 & 7.10e-2 & \SetCell[c=1]{bg=red!25}1.00e-5 \\
    \hline
    & Stoch. & IA & 2.92e-3 & \SetCell[c=1]{bg=red!25}1.50e-3 & 7.61e-4 & \SetCell[c=1]{bg=red!25}5.35e-4 & 1.08e-2 & \SetCell[c=1]{bg=red!25}4.63e-4 \\
    \hline
    & & SO & \SetCell[c=1]{bg=blue!25}1.00e-5 & \SetCell[c=1]{bg=red!25}1.00e-5 & 8.05e-2 & \SetCell[c=1]{bg=red!25}1.00e-5 & 3.46e-2 & \SetCell[c=1]{bg=red!25}1.00e-5 \\
    \hline[1.5pt]
    \begin{sideways}Alpine\end{sideways} & Rob. & IA & 9.58e-2 & \SetCell[c=1]{bg=red!25}3.28e-3 & \color{blue}\textbf{2.65e-1} & \SetCell[c=1]{bg=red!25}1.00e-5 & \color{blue}\textbf{5.58e-1} & \SetCell[c=1]{bg=red!25}5.04e-3 \\
    \hline
    & & SO & 9.04e-2 & \SetCell[c=1]{bg=red!25}3.65e-4 & \color{blue}\textbf{1.31e-1} & \SetCell[c=1]{bg=red!25}1.00e-5 & 7.92e-2 & \SetCell[c=1]{bg=red!25}1.00e-5 \\
    \hline
    & Stoch. & IA & 1.57e-2 & \SetCell[c=1]{bg=red!25}3.10e-4 & 1.87e-2 & \SetCell[c=1]{bg=red!25}1.81e-3 & 1.41e-2 & \SetCell[c=1]{bg=red!25}7.20e-4 \\
    \hline
    & & SO & 7.81e-3 & \SetCell[c=1]{bg=red!25}1.94e-4 & \color{blue}\textbf{1.35e-1} & \SetCell[c=1]{bg=red!25}1.00e-5 & 9.52e-2 & \SetCell[c=1]{bg=red!25}1.00e-5 \\
    \hline[1.5pt]
    \begin{sideways}SumSquares\end{sideways} & Rob. & IA & \SetCell[c=1]{bg=blue!25}1.00e-5 & 2.94e-5 & \SetCell[c=1]{bg=blue!25}1.00e-5 & 8.85e-4 & \color{blue}\textbf{1.00e-0} & \SetCell[c=1]{bg=red!25}7.48e-2 \\
    \hline
    & & SO & 1.51e-3 & \SetCell[c=1]{bg=red!25}1.00e-5 & \color{blue}\textbf{1.05e-1} & \SetCell[c=1]{bg=red!25}1.00e-5 & \color{blue}\textbf{1.54e-1} & \SetCell[c=1]{bg=red!25}1.81e-2 \\
    \hline
    & Stoch. & IA & \SetCell[c=1]{bg=blue!25}1.00e-5 & 2.80e-5 & \SetCell[c=1]{bg=blue!25}1.00e-5 & 5.81e-4 & 3.70e-2 & \SetCell[c=1]{bg=red!25}8.60e-3 \\
    \hline
    & & SO & 8.28e-5 & \SetCell[c=1]{bg=red!25}1.00e-5 & \color{blue}\textbf{1.15e-1} & \SetCell[c=1]{bg=red!25}1.00e-5 & 5.65e-2 & \SetCell[c=1]{bg=red!25}8.07e-3 \\
    \hline[1.5pt]
    \begin{sideways}Levy\end{sideways} & Rob. & IA & 3.63e-2 & \SetCell[c=1]{bg=red!25}5.48e-4 & 8.77e-2 & \SetCell[c=1]{bg=red!25}2.00e-2 & \color{blue}\textbf{1.06e-1} & \SetCell[c=1]{bg=red!25}3.49e-3 \\
    \hline
    & & SO & 7.75e-3 & \SetCell[c=1]{bg=red!25}2.45e-4 & 4.72e-2 & \SetCell[c=1]{bg=red!25}7.54e-3 & 2.98e-2 & \SetCell[c=1]{bg=red!25}6.52e-3 \\
    \hline
    & Stoch. & IA & 2.14e-3 & \SetCell[c=1]{bg=red!25}2.23e-5 & 3.789-3 & \SetCell[c=1]{bg=red!25}3.67e-3 & 3.96e-3 & \SetCell[c=1]{bg=red!25}1.17e-3 \\
    \hline
    & & SO & 2.54e-3 & \SetCell[c=1]{bg=red!25}1.30e-5 & 3.66e-2 & \SetCell[c=1]{bg=red!25}5.03e-3 & 2.22e-2 & \SetCell[c=1]{bg=red!25}3.40e-3 \\
    \hline[1.5pt]
    \begin{sideways}Rosenbrock\end{sideways} & Rob. & IA & \SetCell[c=1]{bg=blue!25}1.00e-5 & 6.05e-5 & \color{blue}\textbf{3.35e-1} & \SetCell[c=1]{bg=red!25}6.98e-3 & \color{blue}\textbf{7.58e-1} & \SetCell[c=1]{bg=red!25}2.02e-4 \\
    \hline
    & & SO & 3.28e-3 & \SetCell[c=1]{bg=red!25}1.00e-5 & 9.15e-2 & \SetCell[c=1]{bg=red!25}3.86e-3 & 7.10e-2 & \SetCell[c=1]{bg=red!25}1.00e-5 \\
    \hline
    & Stoch. & IA & \SetCell[c=1]{bg=blue!25}1.00e-5 & 1.89e-4 & 9.52e-3 & \SetCell[c=1]{bg=red!25}2.91e-3 & 1.14e-2 & \SetCell[c=1]{bg=red!25}1.85e-3 \\
    \hline
    & & SO & 2.19e-3 & \SetCell[c=1]{bg=red!25}5.94e-5 & 5.43e-2 & \SetCell[c=1]{bg=red!25}1.00e-5 & 3.34e-2 & \SetCell[c=1]{bg=red!25}1.00e-5 \\
    \hline[1.5pt]
    \begin{sideways}Ackley\end{sideways} & Rob. & IA & \color{blue}\textbf{1.23e-1} & \SetCell[c=1]{bg=red!25}2.63e-3 & 9.78e-2 & \SetCell[c=1]{bg=red!25}9.27e-3 & \color{blue}\textbf{7.26e-1} & \SetCell[c=1]{bg=red!25}3.02e-3 \\
    \hline
    & & SO & \color{blue}\textbf{1.10e-1} & \SetCell[c=1]{bg=red!25}7.76e-3 & \color{blue}\textbf{3.17e-1} & \SetCell[c=1]{bg=red!25}7.60e-3 & \color{blue}\textbf{1.60e-1} & \SetCell[c=1]{bg=red!25}1.00e-5 \\
    \hline
    & Stoch. & IA & 5.01e-3 & \SetCell[c=1]{bg=red!25}9.78e-5 & \SetCell[c=1]{bg=blue!25}1.13e-2 & 1.29e-2 & 1.01e-2 & \SetCell[c=1]{bg=red!25}1.22e-3 \\
    \hline
    & & SO & 4.90e-2 & \SetCell[c=1]{bg=red!25}1.20e-3 & \color{blue}\textbf{2.00e-1} & \SetCell[c=1]{bg=red!25}7.24e-3 & \color{blue}\textbf{1.19e-1} & \SetCell[c=1]{bg=red!25}4.71e-5 \\
    \hline[1.5pt]
    \end{tblr}
    \label{tab_MetricResults}
\end{table}
 MLIO reaches approximation errors around 0.1\% and 0.001\% for inaccuracy and suboptimality, respectively, within 1e3 samples. The lower SO error reflects the effectiveness of the greedy Layer 3. PCE+GA instead presents a worse 10\%-1\% approximation within a larger number of 1e4 samples. Conducting robust optimization is more difficult for both methods since approximating a statistical moment is easier than finding the maxima envelope. Nevertheless, MLIO exhibits a similar performance, while PCE+GA shows a difference of around one order of magnitude. Consequently, PCE+GA performance for robust optimization is almost poor, while it is at least good for stochastic optimization.

 Tab.\ref{tab_MetricResults} adds performance details based on the specific shape and dimensionality of the problem. Given a budget of 1e3 samples for both methods, IA and SO results are reported for each test function and each dimension under robust and stochastic optimization. MLIO clearly supersedes PCE+GA in the vast majority of the cases, especially $D=200$, ensuring an error stably below 1\% even for the most complex non-separable functions (Rosenbrock and Ackley). Higher errors are found only for the robust optimization of the ill-conditioned SumSquares (error $\sim$2-7\%) in 200D. PCE+GA is better only for the uncertainty quantification (IA error) on the smoothest functions (SumSquares and Rosenbrock) in low dimensionality (D=2 and D=20).
 Poor performance is registered for PCE+GA alone, in many $D=200$ cases, some $D=20$, and only $D=2$ Ackley, mainly for robust optimization. 

 \section{Discussion}
 \label{sec_discussion}

\begin{figure}
   \begin{center}
         \includegraphics[trim={6.5cm 0 6.5cm 0},clip,width=.45\textwidth]{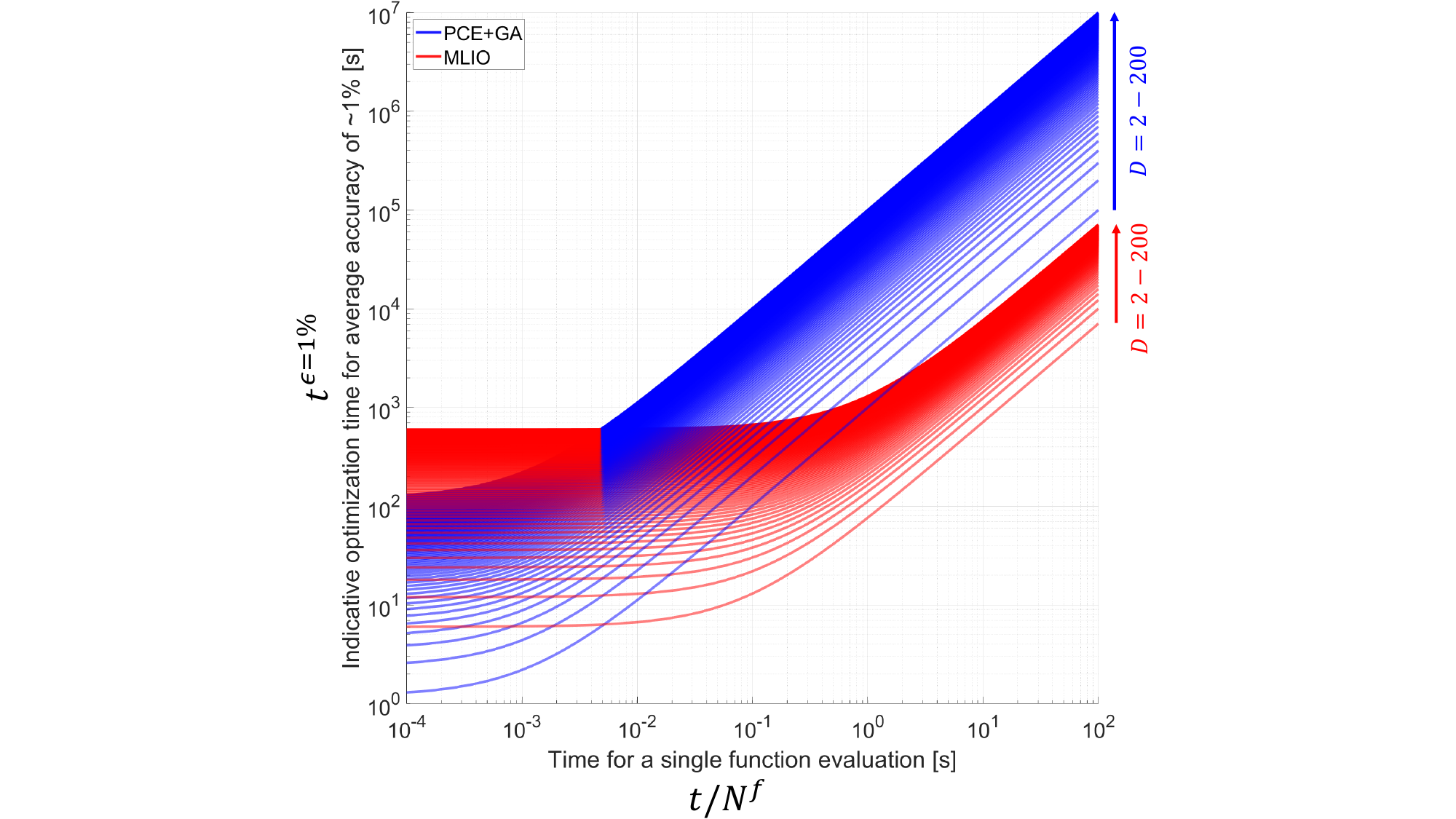}\\
         \caption{Complexity of MLIO and PCE+GA methods for 1\% accuracy as a function of $f$ dimensionality ($D$) and evaluation cost.}\label{fig_TimeComplexity}
   \end{center}
 \end{figure}

 Although MLIO takes much fewer samples under the same level of approximation for optimizing under uncertainty, it is also much more algorithmically complex than PCE+GA, in terms of the number of operations per function evaluation. Therefore, there is a threshold on the single $f$ evaluation time for which MLIO requires fewer total computational resources than PCE+GA to perform the optimization as a function of dimensionality. The aggregation of the results along the dimensionality $D$ (Fig.4 and 8 of the Supplementary material) makes it possible to identify the indicative number of samples $N^{\epsilon=1\%}(D)$ needed to reach, on average, the 1\% error level. 
 The regression laws are identified to be $N^{\epsilon=1\%}_{PCE+GA}(D) \sim 500 D$ (linear) and $N^{\epsilon=1\%}_{MLIO}(D) \sim 50 D^{0.5}$ (sublinear) for PCE+GA and MLIO, respectively. The correspondent time per iteration returns an almost constant $t/N^{\epsilon=1\%}_{PCE+GA}(D) \sim 0.0012$ seconds for PCE+GA and a sublinear $t/N^{\epsilon=1\%}_{MLIO}(D) \sim 0.06 D^{0.5}$ seconds for MLIO. It is possible to compute the indicative optimization time in seconds, $t^{\epsilon=1\%}_{mthd}(t/N^{f},D) \sim \left(t/N^{f}+t/N^{\epsilon=1\%}_{mthd}(D)\right)N^{\epsilon=1\%}_{mthd}(D)$, required for the two methods to reach 1\% accuracy on a $D$ dimensional problem depending on the function evaluation time $t/N^f$ without parallelization. The envelopes of the computational complexity in Fig.\ref{fig_TimeComplexity} are: 
 \begin{equation}\label{eq_TimeComplexity}
    \begin{gathered} 
            t^{\epsilon=1\%}_{PCE+GA}(t/N^{f},D) \sim (t/N^{f} + 0.0012)500D [s]\\
            t^{\epsilon=1\%}_{MLIO}(t/N^{f},D) \sim (t/N^{f} + 0.06D^{0.5})50D^{0.5} [s]\\
    \end{gathered}
 \end{equation} 
 Evidently, a function evaluation in the millisecond range, $\sim$5e-3 seconds (more precisely $(0.06 \times 50-0.0012 \times 500)/(500D-50 D^{0.5})$ seconds) renders MLIO convenient over PCE+GA in terms of computational complexity. If $f$ evaluations require seconds to be carried out, as expected for most real-world applications, the MLIO gain over PCE+GA already stably reaches at least one order of magnitude. For larger timing $t/N^{\epsilon=1\%}_{mthd}$ becomes negligible, and the difference between the methods is asymptotically driven by the difference in the number of samples only, $(N^{\epsilon=1\%}_{MLIO})/(N^{\epsilon=1\%}_{PCE+GA}) \sim 0.1 D^{-0.5}$. This means that an MLIO optimization takes around $\sim$50-700 times the single $f$ evaluation to tackle a D=2-200 problem, while a PCE+GA optimization takes $\sim$1e3-1e5 times for the same. In this regime, an MLIO optimization for a 200D problem is even faster than the equivalent PCE+GA on a much smaller 2D problem. 
 Despite that the absolute timing is machine-specific, the relative features of MLIO vs. PCE+GA and optimization time vs. single evaluation time hold regardless.


 The collected results emphasize how PCE works well for low-dimensional smooth problems but struggles with higher-dimensional and/or more irregular landscapes, even if clear patterns of symmetry and separability are present. In such conditions, the meta-heuristic step by GA is misled by the large errors associated with the uncertainty quantification step via PCE.
 Advanced PCE approaches preceded by an order reduction have recently been developed to enhance scalability \cite{zhou2020surrogate}, but errors are anyway poor on high-dimensional problems.
 Furthermore, looking at performance stability (statistical ranges in the Supplementary file, especially Figure 1), MLIO presents a much lower variability than PCE+GA in absolute terms: MLIO 75\% quantile is still better or equivalent to PCE+GA 25\% quantile; MLIO worst case is comparable to PCE+GA 75\% quantile. Lastly, this remarkable performance is in practice tuning-free for MLIO, while a significant amount of additional resources must be spent on PCE+GA to identify the best settings. 


 Despite the statistical performance being assessed, the proposed method is suitable for hundreds of variables and parameters, which may be insufficient in real-world applications, even after the application of reduction techniques. Extending the MLIO to thousands and more dimensions is possible but would require additional developments to mitigate Kriging's computational complexity. This drawback becomes especially penalizing when dealing with thousands of dimensions, thousands of observations, quick-to-evaluate functions, or a combination of these factors.

 \section{Conclusions}
 \label{sec_conclusions}

 This paper introduces a new point of view in optimization under uncertainty, i.e., mapping design and parameter interactions via Multi-Level Informed Optimization (MLIO). The method leverages an ensemble of orthogonal and hierarchical decomposed Kriging surrogates, to tackle large, complex, and resource-consuming problems. MLIO is formally described and statistically compared to a state-of-the-art two-step approach, PCE+GA, on a heterogeneous analytical testbed up to 200 dimensions (100 design variables and 100 uncertain parameters). Both MLIO and the competitor PCE+GA, are suited for robust and stochastic optimization under uncertainty of black-box problems in engineering and applied sciences. 

 According to the statistical results, MLIO can stably reach $<$1\% error with respect to a dense sampling of 1e6 points within less than 1e3 samples, scaling sub-linearly with the problem's dimensionality. In contrast, PCE+GA cannot guarantee the same even within 1e4 samples, scaling linearly with the number of dimensions. Based on the numerical evidence, MLIO via decomposed Kriging proves accurate and scalable for complex decision under uncertainty tasks, and it is at least one order of magnitude better than conventional two-step approaches. Promisingly, it bridges the gap towards a deeper understanding of uncertain large systems. 

 MLIO can be applied to problems beyond design under uncertainties, such as global optimization, uncertainty quantification, sensitivity analysis, quantile estimation, reliability and risk assessment, and many others. To further extend its applicability potential, future work will include parallel speed-up, preliminary reduction, trust-region refinements, separability aggregation, gradient enhancements, multi-fidelity and multi-objective operations. The method enables addressing problems previously oversimplified within accessible resources, such as optimizing the net-zero transition of the European energy system under realistic climate and weather uncertainties. 

 \section*{Declaration of competing interest}
 \label{sec_Interests}
 The authors declare that they have no known competing financial interests or personal relationships that could have appeared to influence the work reported in this paper.

 \section*{Acknowledgements}
 \label{sec_Acknowledgements}
 Dr. Enrico Ampellio and Prof. Dr. Giovanni Sansavini are part of the SPEED2ZERO, a Joint Initiative co-financed by the ETH Board. In addition, the research published in this publication was carried out with the support of the Swiss Federal Office of Energy as part of the SWEET consortium RECIPE. The authors bear sole responsibility for the conclusions and results presented in this publication. The authors are grateful to Lorenzo Zapparoli, Alfredo Oneto, and Christoph Funke, PhD students of the RRE Lab at ETH Zurich, for their suggestions, in particular regarding the usability of mathematical formulations. Special thanks also go to Dr. Paolo Gabrielli for the insightful initial discussions. 

 \section*{Author contributions}
 \label{sec_Contributions}
 \textbf{Enrico Ampellio}: Conceptualization, Methodology, Software, Validation, Formal analysis, Investigation, Data Curation, Writing - Original Draft, Visualization. \textbf{Blazhe Gjorgiev}: Conceptualization, Writing - Review \& Editing, Supervision. \textbf{Giovanni Sansavini}: Conceptualization, Resources, Writing - Review \& Editing, Supervision, Project administration, Funding acquisition.

 \section*{Supplementary material}
 \label{sec_Supplementary}
 The supplementary file reports the full set of statistical results for both PCE+GA and MLIO on the analytical benchmark presented. Two composed figures, one for robust and one for stochastic optimization, are produced in each setting of the two algorithms, for a total of 16 pictures. They describe the statistical performance (min, 25\% quantile, median, 75\% quantile, and max) of the two methods in terms of IA and SO error metrics for every test problem and dimensionality as a function of the number of samples. Aggregated performance over test problems and dimensions separately are also embedded. 

\appendix

\newpage
\section{Kriging} 
 \label{app_KRG}
  
Kriging assumes the underlying function $z(\mathbf{x})$ to be random and its value at any unobserved location $\mathbf{x_0} \in \mathbb{R}^D$ to be predicted as $\tilde{z}(\mathbf{x_0})$ by the weighted sum of $N$ observations $z(\mathbf{x_n})$ \cite{lichtenstern2013kriging}:
 \begin{equation}\label{eq_OrdKRGweight}
     \begin{gathered}
         z(\mathbf{x_0}) \approx \tilde{z}(\mathbf{x_0}) = \sum_{n=1}^{N}{w_{n,0}z(\mathbf{x_n})} = \mathbf{w_0}^T\mathbf{z} \\
         \mathbf{x} \in \mathbb{R}^D, z : \mathbb{R}^D \rightarrow \mathbb{R}^1, \mathbf{w_0} = [w_{1,0},...,w_{N,0}]^T \in \mathbb{R}^N \\
         \mathbf{x_0} = [x_{1,0},...,x_{D,0}], \mathbf{z} = [z(\mathbf{x_1}),...,z(\mathbf{x_n})]^T \in \mathbb{R}^N  \\
     \end{gathered}
 \end{equation}  
 Building the Kriging surrogate entails finding the set of weights $w_{n,0}$ leading to a minimal unbiased prediction variance $\tilde{\sigma}^2(\mathbf{x_0}) = Var(\tilde{z}(\mathbf{x_0})-z(\mathbf{x_0}))$, where unbiasedness and variance express as:
 \begin{equation}\label{eq_OrdKRGvar}
     \begin{gathered}
         \sum_{n=1}^{N}{w_{n,0}}=1 \Leftrightarrow \mathbf{w_0}^T\mathds{1} = 1 \\
         \tilde{\sigma}^2(\mathbf{x_0}) = \mathbb{E}\left[\left(\sum_{n=1}^{N}{w_{n,0}z(\mathbf{x_i})}-z(\mathbf{x_0})\right)^2\right] = -\mathbf{w_0}^T\mathbf{\Gamma}\mathbf{w_0}+2\mathbf{w_0}^T\mathbf{\gamma_0}
     \end{gathered}
 \end{equation}   
 where $\mathds{1}$ is the unitary vector of length $N$, the symmetric semivariogram matrix $\mathbf{\Gamma}$ is composed by the semivariances $\gamma_{i,j}=\gamma(||\mathbf{x_i}-\mathbf{x_j}||)= \frac{1}{2}\left(z(\mathbf{x_i}) - z(\mathbf{x_j})\right)^2,i,j=1,...,N$ of $z(\mathbf{x})$ between two observations, and $\mathbf{\gamma_0}=\gamma(||\mathbf{x_i}-\mathbf{x_0}||)=[||\gamma(\mathbf{x_1}-\mathbf{x_0})||,...,||\gamma(\mathbf{x_N}-\mathbf{x_0})||]^T \in \mathbb{R}^N$ between each sample and the new point $\mathbf{x_0}$ to predict.   
 The variance in Eq.\ref{eq_OrdKRGvar} is minimized subject to $\mathbf{w_0}^T\mathds{1} = 1$ trough the Lagrangian multiplier $\lambda_0$ in Eq.\ref{eq_OrdKRGmin}, leading to the final linear system in Eq.\ref{eq_OrdKRGsys} to be solved for the Kriging weights~\cite{webster2007geostatistics}:
 \begin{equation}\label{eq_OrdKRGmin}
     \begin{gathered}        
        \varphi(\mathbf{w_0},\lambda_0) = -\mathbf{w_0}^T\mathbf{\Gamma}\mathbf{w_0} + 2\mathbf{w_0}^T\mathbf{\gamma_0} - 2\lambda_0(\mathbf{w_0}^T\mathds{1} - 1) \\
        \frac{\partial\varphi(\mathbf{w_0},\lambda_0)}{\partial\mathbf{w_0}} = -2\Gamma\mathbf{w_0} + 2\mathbf{\gamma_0} - 2\lambda_0\mathds{1} = 0  \\
     \end{gathered}
 \end{equation} 
 \begin{equation}\label{eq_OrdKRGsys}
     \begin{gathered}        
        \Gamma\mathbf{w_0} + \lambda_0\mathds{1} = \mathbf{\gamma_0},~ \mathbf{w_0}^T\mathds{1} = 1 \Rightarrow \mathbf{\alpha} \mathbf{\xi} = \mathbf{\beta} 
        \\  
        \mathbf{\alpha} \in \mathbb{R}^{(N+1)\times(N+1)} =
        \begin{bmatrix}
            \mathbf{\Gamma} & \mathds{1} \\
            \mathds{1}^T & 0 \\
        \end{bmatrix}
        \\
        \mathbf{\xi} \in \mathbb{R}^{N+1} = 
        \begin{bmatrix}
            \mathbf{w_0} \\
            \lambda_0 \\
        \end{bmatrix}
        \\
        \mathbf{\beta} \in \mathbb{R}^{N+1} = 
        \begin{bmatrix}
            \mathbf{\gamma_{0}} \\
            1 \\
        \end{bmatrix}
        \\        
     \end{gathered}
 \end{equation} 

 The corresponding minimal Kriging variance $\tilde{\sigma}^2(\mathbf{x_0})$ allows to define a confidence interval of the prediction in addition to the expected value $\tilde{z}(\mathbf{x_0})$, paving the way to sound quality metrics and adaptive feedback. Such variance can be written in matrix form as per Eq.\ref{eq_OrdKRGsigmasys}, directly related to the linear system in Eq.\ref{eq_OrdKRGsys}, from the plain form developed in Eq.\ref{eq_OrdKRGsigma} by introducing the minimization of Eq.\ref{eq_OrdKRGmin} in Eq.\ref{eq_OrdKRGvar} : 
 \begin{equation}\label{eq_OrdKRGsigma}
     \begin{gathered}
         \tilde{\sigma}^2(\mathbf{x_0}) = \mathbf{w_0}^T\mathbf{\gamma_0} -\mathbf{w_0}^T(\mathbf{\Gamma}\mathbf{w_0} - \mathbf{\gamma_0}) = \mathbf{w_0}^T\mathbf{\gamma_0} + \lambda_0\mathbf{w_0}^T\mathds{1} \\
     \end{gathered}
 \end{equation}   
 \begin{equation}\label{eq_OrdKRGsigmasys}
     \begin{gathered}     
         \tilde{\sigma}^2(\mathbf{x_0}) = \mathbf{w_0}^T\mathbf{\gamma_0} + \lambda_0 = 
         \begin{bmatrix}
            \mathbf{w_0}^T & \lambda_0\\
         \end{bmatrix}
         \begin{bmatrix}
            \mathbf{\gamma_0} \\
            1 \\
         \end{bmatrix}
     \end{gathered}
 \end{equation}  
   
 A different set of weights $[\mathbf{w_0},...,\mathbf{w_M}]$ is needed for each new M-th prediction $[\mathbf{x_0},...,\mathbf{x_M}]$, so that predicting them all together means solving the following matrix form, extension of the linear system of Eq.\ref{eq_OrdKRGsys}:
 \begin{equation}\label{eq_OrdKRGsysM}
     \begin{gathered}        
        \begin{bmatrix}
            w_{1,0} & \hdots & w_{1,M} \\
            \vdots & \vdots & \vdots \\
            w_{N,0} &  \hdots & w_{N,M} \\
            \lambda_{0} & \hdots & \lambda_{M} \\
        \end{bmatrix}
        =
        \begin{bmatrix}
            \mathbf{\Gamma} & \mathds{1} \\
            \mathds{1}^T & 0 \\
        \end{bmatrix}
        ^{-1}
        \begin{bmatrix}
            \gamma_{1,0} & \hdots & \gamma_{1,M} \\
            \vdots & \hdots & \vdots \\
            \gamma_{N,0} & \hdots & \gamma_{N,M} \\
            1 & \hdots & 1 \\
        \end{bmatrix}
     \end{gathered}
 \end{equation}  
 In order for the Kriging to a best linear unbiased predictor, $\gamma$ needs to respect specific global properties and is practice represented by a semivariance model of auto-correlation fitted on the $z(\mathbf{x_n})$ observations (see \ref{subapp_variogram} for further details). The weights then depend on the features of the $\gamma$ model used. 
 Equation~\ref{eq_OrdKRGsysM} is called ordinary Kriging, the most frequently used form in practice.
 
 The extension to universal Kriging assumes that the function $z(\mathbf{x})$ can be decomposed in a nonrandom trend or drift function $\mu(\mathbf{x})$, as linear combination of $L$ basis $f$ by coefficients $a_l$, plus a real-valued residual random function $Y$ without the drift:
 \begin{equation}\label{eq_UnvKRGweight}
 	\begin{gathered}
 		z(\mathbf{x_i}) = \mu(\mathbf{x_i}) + Y(\mathbf{x_i}) \rightarrow  \mathbf{z}=\mathbf{F}\mathbf{a} + \mathbf{Y} \\
 		\mu(\mathbf{x}) = \sum_{l=0}^{L} a_l f_l(\mathbf{x}) \\
            \mathbf{F} =
            \begin{bmatrix}
                1 & f_1(\mathbf{x_1}) & \dots & f_L(\mathbf{x_1}) \\
                \vdots & \vdots & \hdots & \vdots & \\
                1 & f_1(\mathbf{x_N}) & \dots & f_L(\mathbf{x_N}) \\
            \end{bmatrix}
            \in \mathbb{R}^{N \times (L+1)} \\
            f_i : \mathbb{R}^D \rightarrow \mathbb{R}^1 ~\forall i=1,...,L, \mathbf{a} = [a_0,...,a_L]^T \in \mathbb{R}^{L+1} \\
            \mathbf{z} = [z(\mathbf{x_1}),...,z(\mathbf{x_N})]^T, \mathbf{Y} = [Y(\mathbf{x_1}),...,Y(\mathbf{x_N})]^T \in \mathbb{R}^N \\
 	\end{gathered}
 \end{equation}  
 Eq.\ref{eq_OrdKRGweight} still holds, so the minimal prediction variance transforms Eq.\ref{eq_OrdKRGsys} in:
 \begin{equation}\label{eq_UnvKRGsys}
     \begin{gathered}       
        \mathbf{\alpha} \mathbf{\xi} = \mathbf{\beta}
        \\
        \mathbf{\alpha} \in \mathbb{R}^{(N+L+1)\times(N+L+1)} = 
        \begin{bmatrix}
            \mathbf{\Gamma^Y} & \mathbf{F} \\
            \mathbf{F}^T & \mathbf{0} \\
        \end{bmatrix}
        \\
        \mathbf{\xi} \in \mathbb{R}^{N+L+1} = 
        \begin{bmatrix}
            \mathbf{w_0} \\
            \mathbf{\lambda_0} \\
        \end{bmatrix}
        \\
        \mathbf{\beta} \in \mathbb{R}^{N+L+1} =
        \begin{bmatrix}
            \mathbf{\gamma^Y_{0}} \\
            \mathbf{f_0} \\
        \end{bmatrix}
        \\
        \mathbf{\lambda_0} = [\lambda_{0,0},...,\lambda_{L,0}]^T, \mathbf{f_0} = [1,f_1(\mathbf{x_0}),...,f_L(\mathbf{x_0})]^T \in \mathbb{R}^{L+1} \\
     \end{gathered}
 \end{equation} 
 where $\mathbf{F}$ is taken from Eq.\ref{eq_UnvKRGweight}, $\mathbf{\Gamma^Y}$ and $\mathbf{\gamma^Y_{0}}$ have the same form of ordinary Kriging equations \ref{eq_OrdKRGmin} and \ref{eq_OrdKRGsys}, but they now they refer to the residual variogram with respect to the drift, obtained as $Y = z - \mu$.
 The corresponding variance is:
 \begin{equation}\label{eq_UnvKRGsigmasys}
     \begin{gathered}     
         \tilde{\sigma}^2(\mathbf{x_0}) =  
         \begin{bmatrix}
            \mathbf{w_0}^T & \mathbf{\lambda_0}^T \\
         \end{bmatrix}
         \begin{bmatrix}
            \mathbf{\gamma^Y_{0}} \\
            \mathbf{f_0} \\
         \end{bmatrix}
     \end{gathered}
 \end{equation}  
 If $L=0$, the drift $\mu(\mathbf{x})$ is reduced to just a constant term $a_0$, and universal Kriging collapses to ordinary Kriging. For comprehensive details about Kriging fundamentals, refer to \cite{cressie2023spatial}.

\section{Details about decomposed Kriging algorithm}
\label{app_SmrKRG}

\subsection{Variogram and auto-correlation models}
\label{subapp_variogram}

 \begin{figure}
    \begin{center}
         \includegraphics[trim={4cm 0 3cm 0},clip,width=0.5\textwidth]{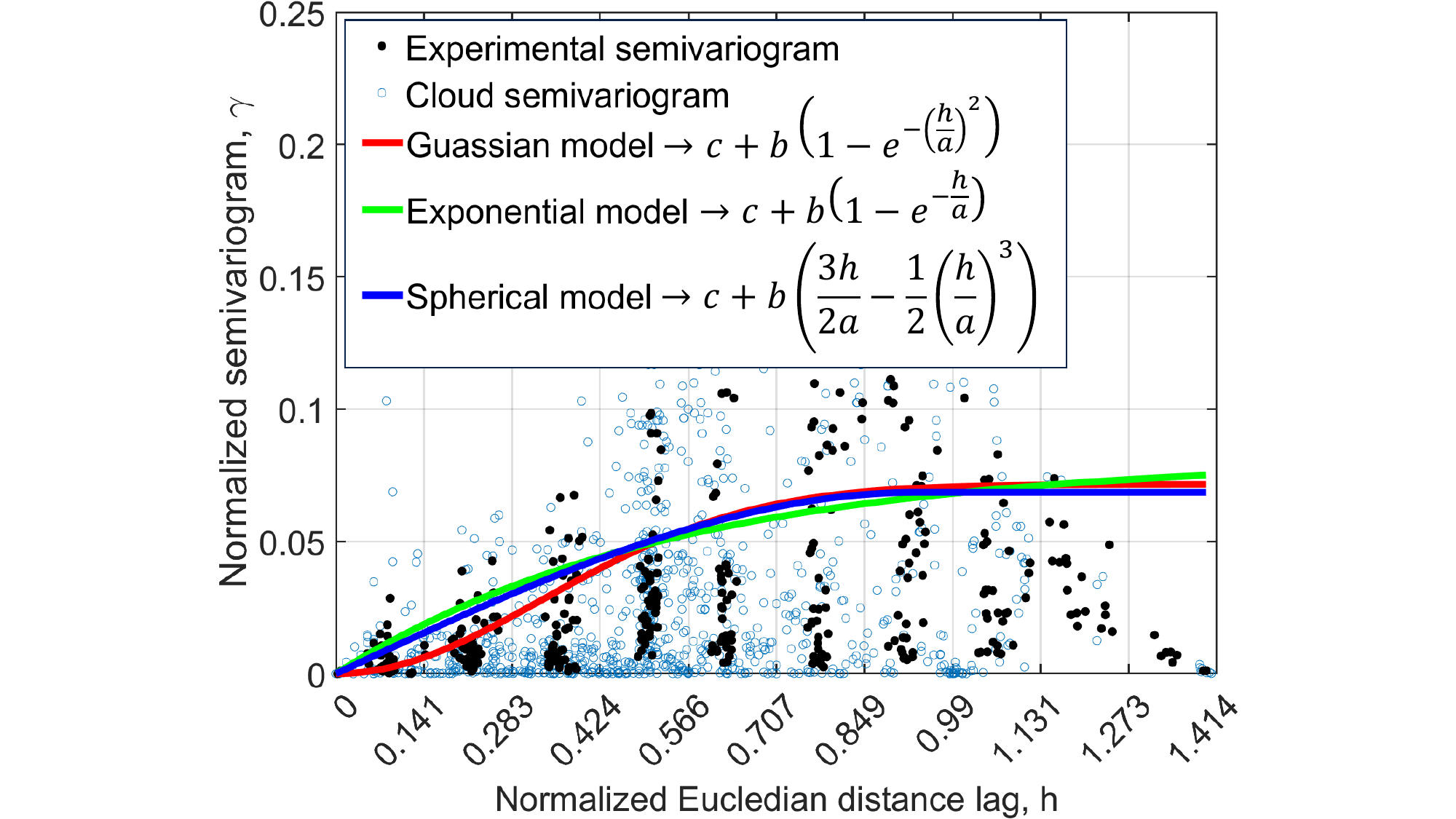}\\
         \caption{Normalized example of a cloud and related experimental semivariogram ($N^H=10$ windows), fitted with three popular parametric models}\label{fig_Variogram}
    \end{center}
 \end{figure} 
 In Kriging, $\gamma$ is basically the auto-correlation that approximates $z$'s covariance properties on the space. This kernel must be a conditionally negative semidefinite function to depict a theoretical semivariogram, which is a fitting of the experimental semivariogram. The latter is a windowing of the cloud semivariogram, obtained by plotting half the squared $z$ difference of each pair of observations $k$ as a function of their distance lag $h_{k}$. Fig.\ref{fig_Variogram} depicts an example: empty blue dots display the cloud semivariogram; the black dots are a windowing of the cloud semivariogram with 10 windows along the normalized lag axis, and display the experimental semivariogram; colored lines represent three $\gamma$ models fitted on experimental semivariogram.
 
 Many mathematical forms for $\gamma$ have been proposed \cite{chiles2012geostatistics}, and their parametric fitting on the experimental semivariogram is a meta-optimization to solve several times within the decomposed Kriging loops:
  \begin{equation}\label{eq_Semivariogramfit}    
    \begin{gathered}
     \argmin_{m} \left[\min_{a,b,c} \sum^{N^h}_{k=1} \left(\gamma^*(h^*_k)-\gamma_{m,a,b,c}(h^*_k)\right)^2\right]_{m=1,...,N^{\gamma}} \\ \rightarrow \gamma_{m,a,b,c}(h), ~\forall h_{k=[i,j]}=||\mathbf{x_i} - \mathbf{x_j}|| \\      
     [a,b,c] \geq 0, [b,c] \leq 1, a \leq D^{0.5} (normalized) \\ 
     0 \leq  \gamma \leq 1,  0 \leq  x_d \leq 1 ~\forall d = 1,..,D ~ (normalization)
     \end{gathered}
 \end{equation}     
 For the sake of effectiveness while limiting computational resources, the following is needed: i) a small $N^{\gamma}$ number of parametric semivariogram models $\gamma_{m}$; ii) a limited $N^h$ number of Euclidean distance lags $h^*_k$; iii) an appropriate algorithm to solve the least squares minimization of the experimental semivariogram $\gamma^*(h^*_k)$ for the best-fitted $\gamma^{\gamma^*}_{m,a,b,c}(h)$ model, as in Eq.\ref{eq_Semivariogramfit}. This problem is non-trivial, small but constrained, non-linear, and potentially multi-modal.
 Each $\gamma_m$ model features three parametric effects, $a$, $b$, and $c$: the nugget $c$, i.e., a jump discontinuity at the origin; the sill $b$, i.e., the asymptotic value $\lim_{h \to \infty} \gamma(h)$ kept constant after the $\gamma(h)$ exceeds it for the first time, for a given distance alg $h$ called range, $a$. $c$ offers the possibility of "passing-through" noisy or ill-conditioned regions valued around the same order of magnitude as the nugget. Instead, $b$ and $a$ are a way to control the influence of far away, almost unrelated samples and regulate the $\gamma$ curve slope in between. All this is essential to reconstruct the shape of Kriging predictions for a functional approximation. 
 
 The number of $N^h$ samples in the experimental semivariogram $\gamma^*$ (black dots in Fig.\ref{fig_Variogram}) depends, in turn, on the $N^H$ number of windows of wideness $H$ adopted along the lag space (10 in the example figure) to derive it from the cloud semivariogram with $N$ observations in total (empty blue dots in Fig.\ref{fig_Variogram}): 
 \begin{equation}\label{eq_Semivariogramwindow}
     \begin{gathered}    
        \gamma^*_{i,n} = \frac{1}{2N_{i,n}} \sum_{j \in J_{i,n}} \left(z(\mathbf{x_i}) - z(\mathbf{x_j})\right)^2,  h^*_{i,n} = \frac{1}{N_{i,n}} \sum_{j \in J_{i,n}}||\mathbf{x_i} - \mathbf{x_j}|| \\ 
        J_{i,n} \in \mathbb{Z}^{+N_{i,n}} = \left\{ j: H_n \leq ||\mathbf{x_i} - \mathbf{x_j}|| \leq H_{n+1} ~ for ~ j=1,...,N \right\}
        \\
        \gamma^* \in \mathbb{R}^{N^h = N \cdot N^H} : h^*_{i,n} \to \gamma^*_{i,n} ~\forall i=1,...N ~\forall n=1,...N^H \\
     \end{gathered}
 \end{equation} 
 Note that the experimental semivariogram for the decomposed Kriging is calculated point-wise $\forall i$ with respect to the $N^H$ consecutive distance windows $[H_n,H_{n+1}]$. This is why there is a group of black dots for each window in Fig.\ref{fig_Variogram}, instead of only one. The goal is to retain more diversity in both $\gamma$ and $h$ from the cloud semivariogram with respect to the standard total aggregated average \cite{matheron1962traite}, and evolve it in a dynamic way. Indeed, $\gamma^*$ increases linearly with the number of observations, progressively populated by the adaptive training.  
 Refer to \ref{subapp_hyper} for the list of $\gamma$ models and related hyperparameters chosen for the purposes of this paper, generalizable to any other use of decomposed Kriging. 
 
\subsection{Error definitions for exit criteria}
\label{subapp_error}

 The error exit criteria for decomposed Kriging are Normalized Root Mean Square Error (NRMSE) on validation points, and normalized maximum predicted deviation for confidence:
 \begin{equation}\label{eq_SmrKRGsymexitVAL}
     \begin{gathered}    
        \epsilon^{SYM}_{VAL} = \sqrt{ \frac{1}{V^{SYM}} \sum^{V^{SYM}}_{n=1} \left( \frac{z(\mathbf{v^{SYM}_n}) - \tilde{z}^{SYM}(\mathbf{v^{SYM}_n})}{\Delta^{SYM}} \right)^2 } \leq \tau^{SEP}_{VAL} \\
     \end{gathered}
 \end{equation}  
 \begin{equation}\label{eq_SmrKRGsymexitCI}
     \begin{gathered}    
         \epsilon^{SYM}_{CI} = \frac{\max_{\mathbf{x^{SYM}_0}} \left(CI_{\mathbf{x^{SYM}_0}}(0.95)-\tilde{z}^{SYM}(\mathbf{x^{SYM}_0})\right)}{\Delta^{SYM}} \leq \tau^{SEP}_{CI}
     \end{gathered}
 \end{equation}  
 \begin{equation}\label{eq_SmrKRGsepexitVAL}
     \begin{gathered}    
        \epsilon^{SEP}_{d,VAL} = \sqrt{ \frac{1}{V^{SEP}_d} \sum^{V^{SEP}_d}_{n=1} \left( \frac{z(\mathbf{v^{SEP}_{d,n}}) - \tilde{z}^{SEP}(\mathbf{v^{SEP}_{d,n}})}{\Delta^{SEP}} \right)^2 } \leq \tau^{SEP}_{VAL} \\
        \forall d=2,...,D \\
     \end{gathered}
 \end{equation} 
 \begin{equation}\label{eq_SmrKRGsepexitCI}
     \begin{gathered}    
        \epsilon^{SEP}_{d,CI} = \frac{\max_{\mathbf{x^{SEP}_{d,0}}} \left((CI_{\mathbf{x^{SEP}_{d,0}}}(0.95)-\tilde{z}^{SEP}(\mathbf{x^{SEP}_{d,0}})\right)}{\Delta^{SEP}} \leq \tau^{SEP}_{CI}\\ 
        \forall d=2,...,D \\
     \end{gathered}
 \end{equation} 
 \begin{equation}\label{eq_SmrKRGvarexitVAL}
     \begin{gathered}    
        \epsilon^{FRE}_{VAL} = \sqrt{ \frac{1}{V^{DKG}} \sum^{V^{DKG}}_{n=1} \left( \frac{z(\mathbf{v^{DKG}_n}) - \tilde{z}^{DKG}(\mathbf{v^{DKG}_n})}{\Delta^{TOT}} \right)^2 } \leq \tau^{FRE}_{VAL} \\
     \end{gathered}
 \end{equation} 
 \begin{equation}\label{eq_SmrKRGvarexitCI}
     \begin{gathered}    
        \epsilon^{FRE}_{CI} = \frac{\max_{\mathbf{x_0}} \left(CI_{\mathbf{x_0}}(0.95)-\tilde{z}^{DKG}(\mathbf{x_0})\right)}{\Delta^{TOT}} \leq \tau^{FRE}_{CI}\\ 
     \end{gathered}
 \end{equation} 
 To facilitate the selection of thresholds and their interpretation in relative terms, errors are normalized by a measure of the min/max value range, $\Delta$, computed from all the observations:
 \begin{equation}\label{eq_SmrKRGexitsdelta}
     \begin{gathered}    
        \Delta^{SYM} = \max(z([\mathbf{x^{SYM}};\mathbf{v^{SYM}}]))-\min(z([\mathbf{x^{SYM}};\mathbf{v^{SYM}}])) \\
        \Delta^{SEP} = \max(z([\mathbf{x^{SEP}};\mathbf{v^{SEP}}]))-\min(z([\mathbf{x^{SEP}};\mathbf{v^{SEP}}])) \\
        \Delta^{TOT} = \max(z([\mathbf{x^{DKG}};\mathbf{v^{DKG}}]))-\min(z([\mathbf{x^{DKG}};\mathbf{v^{DKG}}])) \\
     \end{gathered}
 \end{equation}  

\subsection{Acceleration techniques}
\label{subapp_speedup}

The following measures are taken, and implemented in MATLAB\textregistered, to significantly alleviate the computational burden of the adaptive Kriging surrogate:
 \begin{itemize}[noitemsep,topsep=0pt]
     \item Linear algebra is extensively used for all the key calculations, expressed in matrix form. Kriging $\alpha$ matrices (Eq.\ref{eq_SmrKRGsymweights}-\ref{eq_SmrKRGvarweights}) are computed once and then stored as inverted to accelerate the subsequent calculation of the predictors (Eq.\ref{eq_SmrKRGsymsplit}-\ref{eq_SmrKRGvarsplit}). Indeed, predictions are called more times with respect to the matrix size, especially by next sample meta-optimizations (Eq.\ref{eq_SmrKRGsymnext}-\ref{eq_SmrKRGvarnext}) and greedy process (Eq.\ref{eq_SmrKRGvargreedy}).
     \item Each Kriging surrogate is updated, stored in full, and evaluated only if strictly necessary (pseudo-code \ref{alg_SmrKRG}). Until updating, a prediction database where only eventual new points are computed is passed throughout the algorithm.
     \item Depending on the delta/direct mix of Kriging surrogates valid for the current iteration, unused options are utterly step-wise disregarded when reconstructing the prediction for each Level (Eq.\ref{eq_SmrKRGsymsplit}-\ref{eq_SmrKRGvarsplit} and \ref{eq_SmrKRGsepsplitdirect}-\ref{eq_SmrKRGvarsplitdirect} for Eq.\ref{eq_SmrKRG}).
     \item Best fit parameters for each semivariogram model are stored and used to initialize the next instance of the fitting meta-optimization (Eq.\ref{eq_Semivariogramfit}). This has a strong acceleration effect because it leverages on the experimental semivariogram convergence.
     \item Similarly to the previous item, the greedy subset (Eq.\ref{eq_SmrKRGvargreedy}) is passed over among the initial population of the next exploitative meta-optimizer step.
     \item If given a dense pool of possible samples to choose from, the meta-optimization for the validation samples (Eq.\ref{eq_SmrKRGsymnextVAL}-\ref{eq_SmrKRGvarnextVAL}) can be substituted by a distance calculation among all pool's points, predetermined only once.
     \item Euclidean distance among observations and possible new samples is the most repeated operation in decomposed Kriging. A common database accessible at all Levels is then created to inhibit wasteful recalculation.
 \end{itemize}

 \subsection{Pseudo-code}
 \label{subapp_pseudocode}

 The pseudo-code of Kriging corresponding to the scheme in Fig.\ref{fig_SmrKRGflow} is reported in the Alg.\ref{alg_SmrKRG} below.
  \begin{algorithm*}
     \caption{Decomposed Kriging}\label{alg_SmrKRG}
     \footnotesize
     \begin{algorithmic}[1]

         \Statex \LeftComment{0}{Initialization phase} 
         \State Define the reference point, symmetric, separable and assumption-free sampling pools, including validation     
         \State Confidence and validation errors = $\infty$, iter=0, next = 0, greedy = 0

         \Statex \LeftComment{0}{Adaptive training phase} 
         \While{[Any error $>$ tol (Eq.\ref{eq_SmrKRGsymexitVAL}-\ref{eq_SmrKRGvarexitCI}) $\lor V^{DKG} < V^{DKG}_{min}] \land [N^{TOT} < N^{TOT}_{max}$]}
            \State iter = iter + 1
            \Statex \LeftComment{1}{Symmetric surrogate update}
            \If{next=0 $\lor$ next=2}
                \State Update symmetric experimental semivariogram (Eq.\ref{eq_Semivariogramwindow}) and fit symmetric model (Eq.\ref{eq_Semivariogramfit})
                \State Update symmetric surrogate (Eq.\ref{eq_SmrKRGsymweights}) and update symmetric errors
                (Eq.\ref{eq_SmrKRGsymexitVAL}),\ref{eq_SmrKRGsymexitCI})
            \EndIf
            \Statex \LeftComment{1}{Separable surrogate update}
            \If{next=0 $\lor$ next=2 $\lor$ next=3}
                \State Update delta and direct separable experimental semi-variograms (Eq.\ref{eq_Semivariogramwindow})
                \State Update the fitting of delta and direct separable semivariogram models (Eq.\ref{eq_Semivariogramfit} from delta and direct variograms)
                \State Update delta and direct separable surrogates (Eq.\ref{eq_SmrKRGsepweights} through Eq.\ref{eq_SmrKRGsepsplit} or Eq.\ref{eq_SmrKRGsepsplitdirect} predictor)
                \State Update separable errors (Eq.\ref{eq_SmrKRGsepexitVAL},\ref{eq_SmrKRGsepexitCI}) and choose delta or direct surrogate based on validation error
            \EndIf
            \Statex \LeftComment{1}{Assumption-free surrogate update}
            \State Update delta and direct assumption-free experimental semi-variograms (Eq.\ref{eq_Semivariogramwindow})
            \State Update the fitting of delta and direct assumption-free semivariogram models (Eq.\ref{eq_Semivariogramfit})
            \State Update delta and direct assumption-free surrogates (Eq.\ref{eq_SmrKRGvarweights} through Eq.\ref{eq_SmrKRGvarsplit} or Eq.\ref{eq_SmrKRGvarsplitdirect} predictor)
            \State Update assumption-free errors (Eq.\ref{eq_SmrKRGvarexitVAL},\ref{eq_SmrKRGvarexitCI}) and choose delta or direct surrogate based on validation error

            \Statex \LeftComment{1}{Eventual symmetric next sample}
            \IIf{next=0} next=1 \EndIIf
            \If{next=1} 
                \If{[Symmetric errors $>$ tol (Eq.\ref{eq_SmrKRGsymexitVAL}),\ref{eq_SmrKRGsymexitCI}) $\lor V^{DKG} < V^{DKG}_{min}] \land [N^{SYM}+V^{SYM} < N^{SS}_{d,max}]$}
                    \State Add new training sample to separable pool to maximize symmetric surrogate confidence (Eq.\ref{eq_SmrKRGsymnext})
                    \If{$mod(N^{SYM},\lceil 1/v_{ratio} \rceil)=0$} \LeftComment{0}{Eventual symmetric next validation point}
                        \State Add new validation sample to separable pool to maximize symmetric samples' diversification
                        (Eq.\ref{eq_SmrKRGsymnextVAL})
                    \EndIf
                \Else{} next = next + 1
                \EndIf
            \EndIf
            \Statex \LeftComment{1}{Eventual separable next sample}
            \If{next=2}
                \If{[Separable errors $>$ tol (Eq.\ref{eq_SmrKRGsepexitVAL}),\ref{eq_SmrKRGsepexitCI}) $\lor V^{DKG} < V^{DKG}_{min}] \land [N^{SEP}_d+V^{SEP}_d < N^{SS}_{d,max}]$}
                    \State Add new training sample to separable pool to maximize separable surrogate confidence (Eq.\ref{eq_SmrKRGsepnext})
                    \If{$mod(\sum^{D}_{d=2} N^{SEP}_d,\lceil 1/v_{ratio} \rceil)=0$} \LeftComment{0}{Eventual separable next validation point}
                        \State Add new validation sample to separable pool to maximize separable samples' diversification
                        (Eq.\ref{eq_SmrKRGsepnextVAL})
                    \EndIf
                \Else{} next = next + 1 \EndIf
            \EndIf
            \Statex \LeftComment{1}{Eventual Assumption-free next sample}
            \If{next=3}
                \If{Any assumption-free error $>$ tol (Eq.\ref{eq_SmrKRGvarexitVAL}),\ref{eq_SmrKRGvarexitCI}) $\lor V^{DKG} < V^{DKG}_{min}$}
                    \If{$greedy/(N^{FRE}-greedy)<g_{ratio}$} \LeftComment{0}{Eventual greedy subset}
                        \State Define a new training sample subset according to the acquisition function $g$ (Eq.\ref{eq_SmrKRGvargreedy})
                        \State greedy = greedy + 1
                    \EndIf
                    \State Add new training sample to assumption-free pool to maximize assumption-free surrogate confidence (Eq.\ref{eq_SmrKRGvarnext})                                   
                    \If{$mod(N^{FRE},\lceil 1/v_{ratio} \rceil)=0$} \LeftComment{0}{Eventual assumption-free next validation point}
                        \State Add new validation sample to assumption-free pool to maximize total samples' diversification
                        (Eq.\ref{eq_SmrKRGvarnextVAL})
                    \EndIf
                \EndIf
            \EndIf
            \Statex \LeftComment{1}{Managing recursive multi-layer looping}
            \IIf{next $<$ 3} next = next + 1 \ElseIIf{} next = 1 \EndIIf

         \EndWhile
     \end{algorithmic}
 \end{algorithm*}

\section{Tuning on the benchmark}
\label{app_Tuning}

\subsection{PCE+GA and MLIO hyperparameters}
\label{subapp_hyper}

\begin{table}[H]
    \centering
    \caption{Tuned hyper-parametric configurations for the PCE+GA method to meet a total of 1e4 samples}
    \footnotesize
    \begin{tabular}{|c|c|c|c|}
        \hline
        \textbf{Tuning}  & \textbf{GA} & \textbf{PCE} & \textbf{GA} \\
        \textbf{setting} & \textbf{population} & \textbf{samples} & \textbf{generations} \\
        \hline
        \textbf{\#1} & 10 & 25 & 40 \\
        \hline 
        \textbf{\#2} & 10 & 50 & 20 \\
        \hline 
        \textbf{\#3} & 10 & 100 & 10 \\
        \hline
        \textbf{\#4} & 25 & 25 & 16 \\
        \hline
        \textbf{\#5} & 25 & 50 & 8 \\
        \hline
        \textbf{\#6} & 50 & 25 & 8 \\
        \hline
    \end{tabular}
    \label{tab_PCE+GAtuning}
 \end{table}
 Concerning PCE in PCE+GA, the number of parametric samples for the PCE, the population size, and max samples for the meta-heuristics impact performance the most. Given a total budget of 1e4 samples, 6 options \# are tested (Tab.~\ref{tab_PCE+GAtuning}), sweeping a wide range of balance between the resources dedicated to UQ and those dedicated to OPT, within reasonable limits.  When both GA population and PCE sampling are selected, the number of maximum generations is consequently determined to fit within the 1e4 sample cap. The initial population is randomly picked among the reference 1e3 samples. The best-performing configuration overall will be selected to compare versus MLIO.
 Among secondary hyperparameters, uncertainty is considered uniformly distributed $\mathcal{U}([0,1])$ to explore the whole variability, experiments are picked randomly among the 1e3 reference parametric samples for each design, and Least Angle Regression (LARS) is adopted for sparse compressing sensing. The most relevant hyperparameters to decide upon are indeed the degree and the q-norm truncation for the polynomials. Ideally, they could be set both to high values, up to 50 for the degree (large enough to cope with highly multi-modal problems) and 1 for the q-norm (full retention) and let the PCE implementation in UQlab to find the best settings in terms of LOO error, but this would require a calculation time diverging with the number of dimensions. Instead, notable proprieties of analytical functions are leveraged to limit the degree, and sensitivity on q-norm showed a quality threshold for high-dimensional cases where orthogonal projections are difficult to discern in any way. Tab.~\ref{tab_PCEset} reports the setting of PCE for degree and q-norm adopted on the present benchmark.
 \begin{table}
    \centering
    \caption{UQLab PCE settings for the present benchmark}
    \footnotesize
    \begin{tabular}{cc}
        \begin{tabular}{|c|c|}
            \hline
            \textbf{Function} & \textbf{Poly degree} \\
            \hline
            Step & 1-20 \\
            \hline 
            Alpine & 1-40 \\
            \hline 
            SumSquares & 1-5 \\
            \hline 
            Levy & 1-40 \\
            \hline 
            Rosembrock & 1-5 \\
            \hline 
            Alpine & 1-20 \\
            \hline
        \end{tabular}
        \begin{tabular}{|c|c|}
            \hline
            {$\mathbf{D} \rightarrow \mathbf{D_u}=\mathbf{D_p}$} & \textbf{q-norm} \\
            \hline
            2$\rightarrow$1 & 0.75(default) \\
            \hline 
            20$\rightarrow$10 & 0.5 \\
            \hline 
            200$\rightarrow$100 & 0.1 \\
            \hline
        \end{tabular}
    \end{tabular}
    \label{tab_PCEset}
 \end{table}

 A long series of adjustable parameters characterize GAs, mainly related to crossover and mutation. Fine-tuning them is a demanding but potentially high-reward process \cite{mosayebi2020tuning}, exceeding the scope of this study, as is the selection of alternatives to GA among the several meta-heuristics belonging to whether evolutionary, swarm intelligence, or hybrid strategies. Even the best algorithm with the best tuning, which will usually cost many additional $f$ evaluations to discover, will need, at the very least, hundreds of samples to cope with complex optimization problems featuring dozens of dimensions \cite{hansen2010comparing}. A similar amount is needed for UQ with PCE, leading to a total number of samples for a canonical two-step approach on generalized high-dimensional problems anyway likely $>\mathcal{O}(1e4)$. For this reason, a GA with standard optional settings \cite{MATLABga} is employed in this paper.  Together with the debatable selection of the meta-heuristics and its tuning, the needed insights about degree, q-norm, and probability distribution for the PCE, the balancing between the population size and UQ observations competing for resources, PCE+GA and akin methods are much more challenging to set up and tune with respect to MLIO, and much less flexible.

 Concerning MLIO via decomposed Kriging initialization is relevant. A minimum number of samples is required to initiate the process, depending on dimensionality: 1 reference point, 1 additional training point for each dimension (symmetric and separable), and 1 assumption-free, 1 symmetric, 1 separable, and 1 assumption-free validation points, for a total count of $1+D+1+1+1+1=5+D$ points. The actual number depends on how many samples are dedicated to i) each dimension of the separable pool $N^{SS}_d=N^{SYM}=N^{SEP}_d$, ii) the assumption-free pool $N^{FRE}$, and iii) $v_{ratio}$. In total, there will be $1+N_dD+N^{FRE}$ training points and $\lceil N_d v_{ratio} \rceil + \lceil N^{SS}_d(D-1) v_{ratio} \rceil + \lceil N^{FRE} v_{ratio} \rceil$ validation points. 
 2 initialization settings, \#1 and \#2, are then tested for the tuning (Tab.~\ref{tab_SmrKRGtuning}), namely the minimal one respecting $v_{ratio}$ and the smallest to enable a quadratic estimation on the separable space.
 Tab.~\ref{tab_SmrKRGtuning} reports the number used for the two settings of MLIO. 
 The former starts adding points autonomously as soon as possible, while the second includes points for the cheapest second-order approximation on the separable pool. Initial validation points already follow the diversity rules in Eq.\ref{eq_SmrKRGsymnextVAL}-Eq.\ref{eq_SmrKRGvarnextVAL}.
 \begin{table}[!ht]
    \centering
    \caption{Tuned hyper-parametric configurations for decomposed Kriging in multi-level informed optimization with $v_{ratio}=50\%$}
    \footnotesize
    \begin{tabular}{|c|c|c|c|c|c|}
        \hline
        \textbf{Tuning} & \multirow{2}{*}{$\mathbf{N^{SS}_d}$} & \multirow{2}{*}{$\mathbf{N^{FRE}}$} & \multicolumn{3}{c|}{\textbf{Initial total samples}} \\
        \cline{4-6}
        \textbf{setting} & & & \textbf{D=2} & \textbf{D=20} & \textbf{D=200} \\
        \hline
        \textbf{\#1} & 1 & 1 & 7 & 34 & 304 \\
        \hline 
        \textbf{\#2} & 2 & 1 & 9 & 63 & 603 \\
        \hline
    \end{tabular}
    \label{tab_SmrKRGtuning}
 \end{table}
 Due to its computational complexity, disproportionate to statistical repetitions of instantaneous to compute analytical functions, the total number of samples for MLIO is limited to $N^{TOT}_{max}=1e3$.
 Coming to secondary hyperparameters, the ones belonging to internal meta-optimizations can be hard-coded thanks to self-adaptation. Next sample search in Eq.\ref{eq_SmrKRGsymnext}-\ref{eq_SmrKRGsepnext} is solved employing the same GA introduced before, but this time with generous population and generations (both 100), being at least 1e4 samples needed to approach the global optimum \cite{razali2011genetic} and given that Kriging surrogates are very fast to compute. Another crucial meta-optimization for decomposed Kriging regards the experimental semivariogram in Eq.\ref{eq_Semivariogramwindow} and its fitting in Eq.\ref{eq_Semivariogramfit}. Since the former is already dynamically adjusted with the number of samples, a windowing $N^H=10$ is sufficient to guarantee appropriate $\gamma$ fitting while limiting the process complexity. For similar reasons, only 3 parametric variogram models among the many in literature are fitted, for each Kriging Layer at each iteration, namely spherical, exponential, and Gaussian models, chosen to cover the panorama of slope variations from 0 to $a$. They are all equal to 0 if $h=0$; refer back to Fig.\ref{fig_Variogram} for a visual representation of these specific settings. The fitting is solved via the interior point method coded in "fmicon" of MATLAB\textregistered~ optimization toolbox. $a$ and $b$ are initialized as $a=1/N^h \sum^{N^h}_{k=1} h^*_k$ and $b=1/N^h \sum^{N^h}_{k=1} \gamma^*(h^*_k)/ \gamma_{m,1,1,0}(1)$ in the first run on all the three models, and set as the solution from the previous run for the following ones. Instead, the nugget $c$ is primarily used to avoid ill-conditioning of $\alpha$ matrices due to eventual very close points in the training set. If the conditioning number is above $1e8$, then $c=1e-8$ is imposed to avoid numerical issues. A similar principle applies to managing particularly noisy landscapes.
 $v_{ratio}$ and $g_{ratio}$ are another pair of important parameters, and robust default settings exist: $v_{ratio} = 50\% \in [20\%-100\%]$; $g_{ratio} = 100\% \in [25\%-200\%]$. In this study first is set to 50\%, meaning 1/3 of the total observations are used just for validation (66\%/33\%). This is a larger number than the usual 80\%/20\% by Pareto principle, but not uncommon in machine learning \cite{xu2018splitting} and justified to support the exit criteria based on validation tolerances conservatively. Furthermore, sensitivities showed it is a good compromise to return trustworthy error metrics. $g_{ratio}$ is also set to 50\% to favor global optimum convergence; since balancing exploration and exploitation, its effect depends on the function but is not as large as initialization, which is the only tuned hyperparameter.
 Separable training points are initialized on the edges of the box-bounds to use orthogonal Kriging surrogates via interpolation, while assumption-free points are chosen randomly among the reference sampling. Initial validation points already follow the diversity rules in Eq.\ref{eq_SmrKRGsymnextVAL}-Eq.\ref{eq_SmrKRGvarnextVAL}.
 $N^{TOT}_{max}=1e3$ is reached unless validation and confidence errors do not go below $\tau_{VAL}=0.1\%$ and $\tau_{CI}=1\%$ sooner. Low threshold values are selected to privilege step-be-step progression and fully compare the MLIO potential with PCE+GA, as a function of increasing samples. Symmetric and separable Layers are stopped by quality or by a budget of $N^{SS}_{d,max}=100$ samples per dimension $d$. A minimum number of $V^{DKG}_{min}=D$ validation samples is also imposed. The greedy function simply evaluates the whole reference sample basis to minimize $UQ_p(\tilde{\bar{f}}(\bar{\mathbf{u}},\bar{\mathbf{p}}))$ through decomposed Kriging and returns the corresponding $\mathbf{X^{FRE}_0}=[\bar{\mathbf{u}}_{min},\bar{\mathbf{p}}]$ subset.
 
\subsection{Tuned results}
\label{subapp_tuning}
 
 Results depend on the tuning settings, as evident from Figures 2-9 in the Supplementary file. Actually, MLIO tuning is exploited primarily to check the algorithm's adaptive capabilities. Indeed, there is not much difference between MLIO setting \#1 and \#2: the latter is marginally better within the first hundreds of samples, but the former is a little better afterward. This, combined with the first returning results before the second, makes setting \#1 preferable, which means the adaptive process is more effective than a simply larger initialization.
 
 More marked differences are instead found with regard to the tuning of the PCE+GA; in particular, the AI metric is better for a larger sampling dedicated to uncertainty quantification.
 However, this improvement does not progress linearly with PCE samples, especially for the complex high-dimensional functions. This leads to preferring sampling \#2 over \#3 to privilege lower SO errors thanks to a quicker initialization phase, with respect to a slightly better IA. The very small population size of 10 for GA is anyway favored to maximize generations, given the overall limit of 1e4 samples. Nonetheless, based on result projections, around 1e5 samples are estimated necessary for PCE+GA in 200D (Section \ref{sec_results} and Eq.\ref{eq_TimeComplexity}) to reach the 1\% accuracy level, given that the method cannot guarantee such accuracy up to the 1e4 samples of this benchmark. This is highlighted by the logarithmic rate of error decrease in Tab.~\ref{tab_ResultsRates}.
 \begin{table}[!ht]
    \centering
    \caption{Average improving rates of median results for PCE+GA and MLIO on IA and SO metrics in terms of order of magnitudes}
    \footnotesize
    \begin{tblr}{
      cells={valign=m,halign=c},
      cell{1}{2} = {c=2}{},
      cell{1}{4} = {c=2}{},
      colspec={|c|c|c|c|c|}
    }
    \hline
    & Robust & & Stochastic & \\
    \cline{2-5}
    & $\frac{\mathcal{O}(IA)}{\mathcal{O}(Samples)}$ & $\frac{\mathcal{O}(SO)}{\mathcal{O}(Samples)}$ & $\frac{\mathcal{O}(IA)}{\mathcal{O}(Samples)}$ & $\frac{\mathcal{O}(SO)}{\mathcal{O}(Samples)}$ \\
    \hline
    PCE+GA & $\sim 0$ & $\sim 0.2$ & $\sim 0$ & $\sim 0.9$ \\
    \hline
    MLIO & $\sim 2.1$ & $\sim 3.8$ & $\sim 2.3$ & $\sim 3.8$ \\
    \hline
    \end{tblr}
    \label{tab_ResultsRates}
\end{table}

\subsection{Analytical testbed}
\label{subapp_testbed}

\begin{table*}
    \caption{The 6 variegated analytical functions used in this paper as a benchmark for the numerical validation of design under uncertainty methods}
    \centering
    \scriptsize
    \begin{tblr}{
      cells={valign=m,halign=c},
      cell{1}{1} = {r=2,c=3}{},
      cell{1}{4} = {c=2}{},
      cell{3}{1} = {r=18}{},
      cell{3,9,15}{2} = {r=6}{},
      row{1-2}={font=\bfseries},
      column{1-2}={font=\bfseries},
      colspec={|c|c|c|c|c|}       
    }
    \hline
    & & & Modality & \\
    \cline{4-5}
    & & & Uni-modal & Multi-modal \\
    \hline
    \begin{sideways}Separability\end{sideways} & \begin{sideways}Symmetric\end{sideways} & ID (features) & \textbf{Step} (non-differentiable sphere) & \textbf{Alpine} (peak effects) \\ 
    & & 2D view & \includegraphics[width=.25\textwidth]{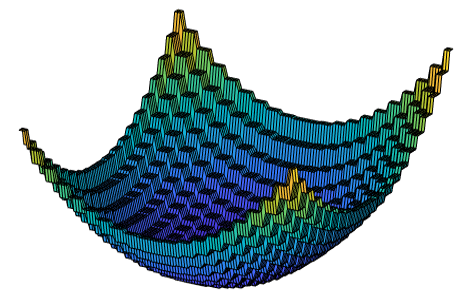} & \includegraphics[width=.25\textwidth]{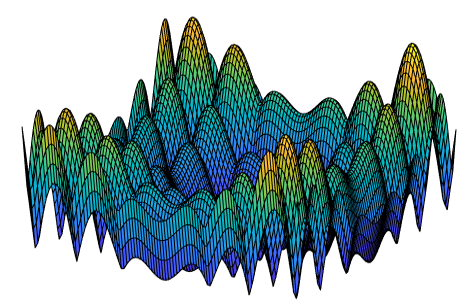} \\
    & & $f(\mathbf{x})$ & $\sum^{D}_{d=1} \lfloor x_d+0.5 \rfloor ^2$ & $\sum^{D}_{d=1} |x_d sin(x_d)+0.1x_d|$ \\ 
    & & $B_d \forall d$ & [0,20] & [0,20] \\ 
    & & $minf(\mathbf{T},\mathbf{B})$ & 0 & 0 \\  
    & & $\tilde{maxf}(\mathbf{T},\mathbf{B})$ & $f(\mathbf{x}^{\tilde{max}})$  & $1.1 \sum^{D}_{d=1} |x^{\tilde{max}}_d|$ \\
    \hline
    & \begin{sideways}Separable\end{sideways} & ID (features) & \textbf{SumSquares} (ill-conditioned ellipsoid) & \textbf{Levy} (barrier effects) \\ 
    & & 2D view & \includegraphics[width=.25\textwidth]{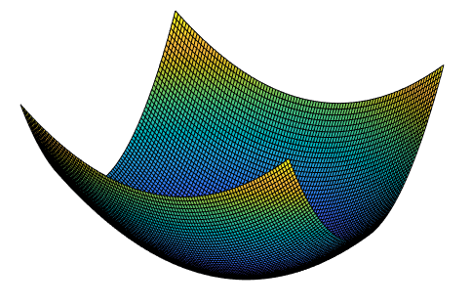} & \includegraphics[width=.25\textwidth]{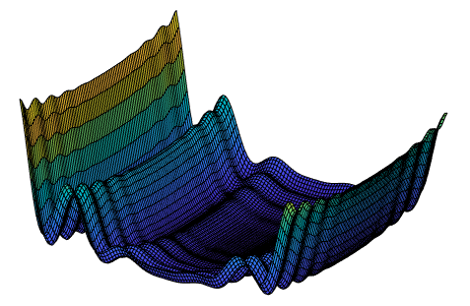} \\
    & & $f(\mathbf{x})$ & $\sum^{D}_{d=1} dx_d^2$ & {$sin^2(\pi \omega_1) + (\omega_D-1)^2[1+sin^2(2 \pi \omega_D)]+$ \\ $+\sum^{D-1}_{d=2} (\omega_d-1)^2[1+10sin^2(2 \pi \omega_d + 1)]$ \\ where $\omega_d = 1 + (x_d-1)/4$} \\ 
    & & $B_d \forall d$ & [0,20] & [0,20] \\   
    & & $minf(\mathbf{T},\mathbf{B})$ & 0 & 0 \\  
    & & $\tilde{maxf}(\mathbf{T},\mathbf{B})$ & $f(\mathbf{x}^{\tilde{max}})$ & {$1+11 \sum^{D-1}_{d=1} (\omega^{\tilde{max}}_d-1)^2 + 2 (\omega^{\tilde{max}}_D-1)^2$ \\ where $\omega^{\tilde{max}}_d = 1 + (x^{\tilde{max}}_d-1)/4$} \\ 
    \hline
    & \begin{sideways}Assumption-free\end{sideways} & ID (features) & \textbf{Rosenbrock} (correlated valley) & \textbf{Ackley} (noise modulation effects) \\ 
    & & 2D view & \includegraphics[width=.25\textwidth]{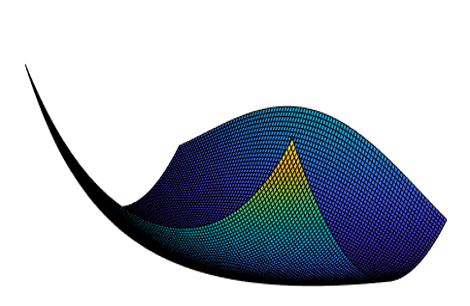} & \includegraphics[width=.25\textwidth]{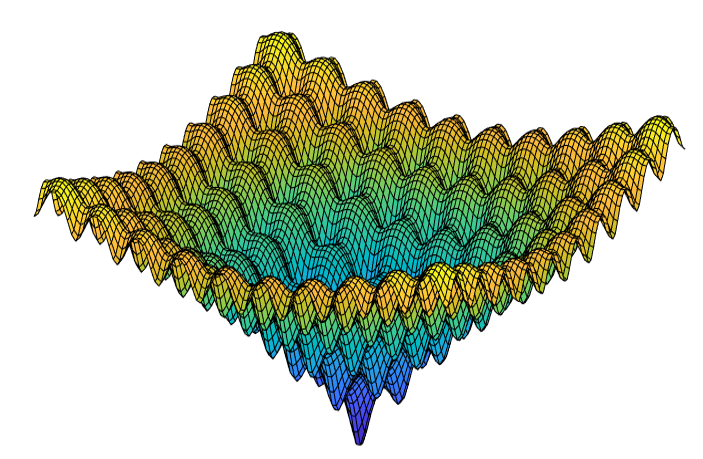} \\
    & & $f(\mathbf{x})$ & $\sum^{D-1}_{d=1} 100(x_d^2-x_{d+1})^2+(x_d-1)^2$ & {$-20exp(-0.2\sqrt{1/D \sum^{D}_{d=1} x_d^2}) + 20 +$ \\ $-exp(1/D \sum^{D}_{d=1} cos(2 \pi x_d))+exp(1)$}\\ 
    & & $B_d \forall d$ & [0,1] & [0,10] \\  
    & & $minf(\mathbf{T},\mathbf{B})$ & 0 & 0 \\  
    & & $\tilde{maxf}(\mathbf{T},\mathbf{B})$ & {$max(\sum^{D-1}_{d=1} [100({x^{n_i}_d}^2-x^{n_j}_{d+1})^2+(x^{n_i}_d-1)^2])$ \\ $\forall n_i,n_j=\{0,0.5,1\}$} & {$-20exp(-0.2\sqrt{1/D \sum^{D}_{d=1} {x^{\tilde{max}}_d}^2})+$ \\ $+20-exp(-1) + exp(1)$}\\ 
    \hline
    \end{tblr}
    \begin{tablenotes}
        \centering
        \item $\mathbf{x}^{\tilde{max}}=[x^{\tilde{max}}_1,...,x^{\tilde{max}}_], x^{\tilde{max}}_d=\bar{x}^{\tilde{max}}_d(B_{d,2}-B_{d,1})+B_{d,1} ~\forall d=1,...,D$
        \item $\bar{x}^{\tilde{max}}_d=max(|0-T_d|,|1-T_d|) ~\forall d=1,...,D$
        \item $ x^{n}_d=(n-T_d)(B_{d,2}-B_{d,1})+B_{d,1} ~\forall d=1,...,D$  
    \end{tablenotes}
    \label{tab_functions}
\end{table*}

\newpage
\bibliographystyle{elsarticle-num-names}



\section*{Supplementary material}

# Full statistical results for the analytical benchmark

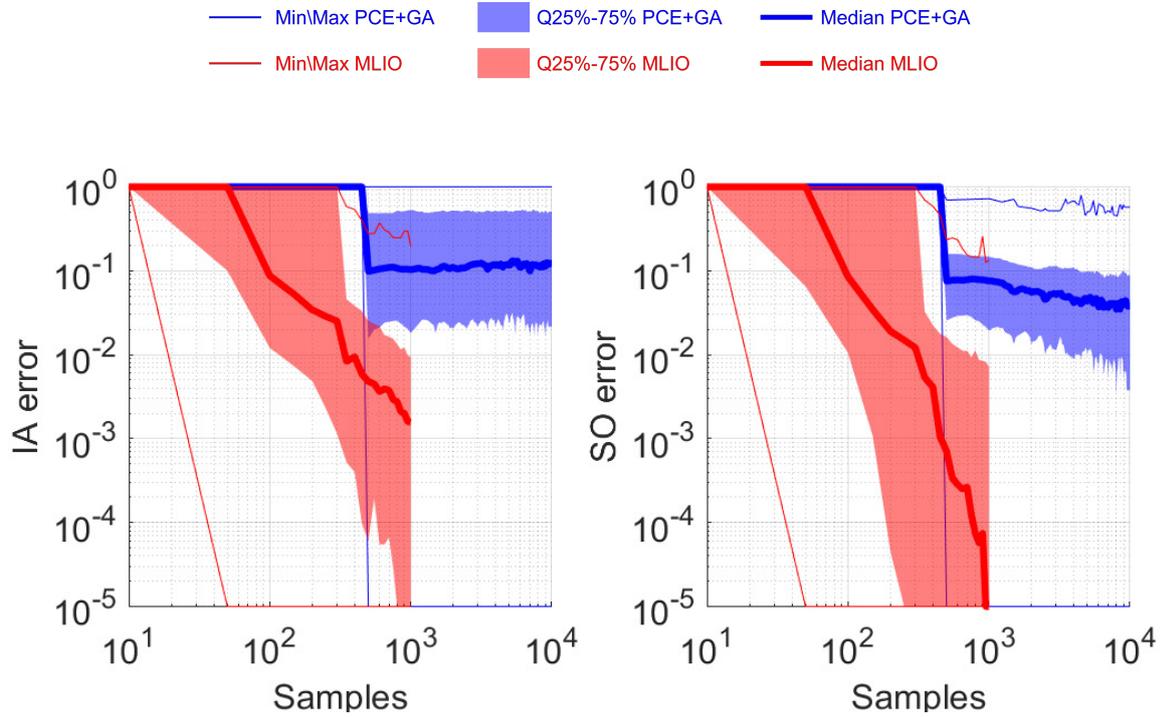

(a) MLIO vs. PCE+GA for robust optimization

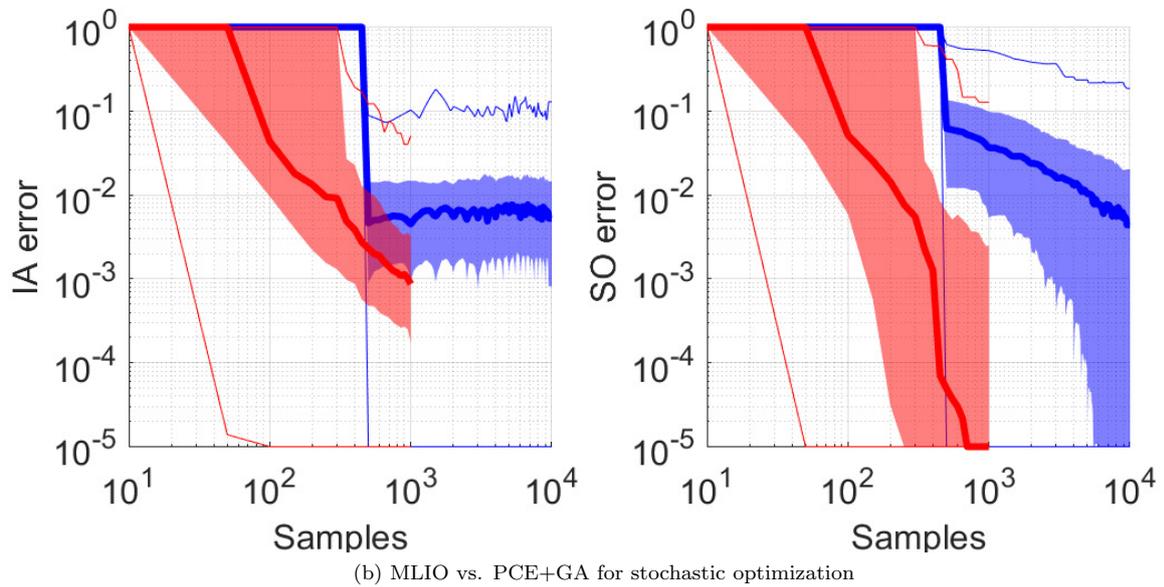

(b) MLIO vs. PCE+GA for stochastic optimization

Figure 1: Aggregated statistical results over the testbed for tuned MLIO and PCA+GA

1

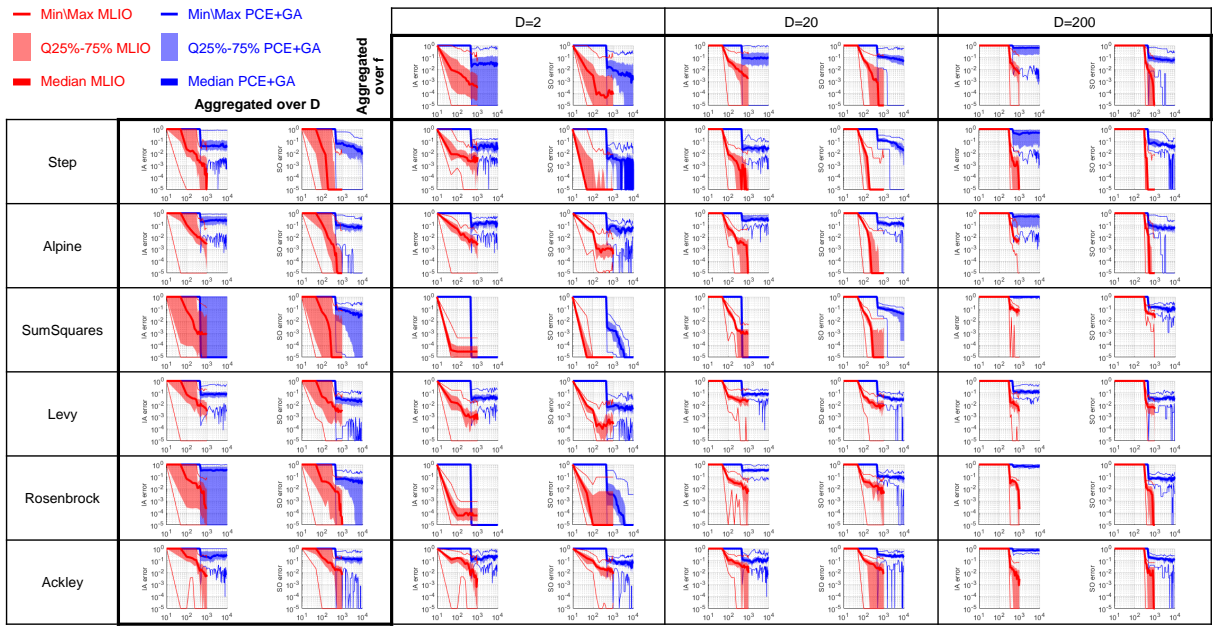

(a) Statistical results of PCE+GA vs. MLIO for robust optimization

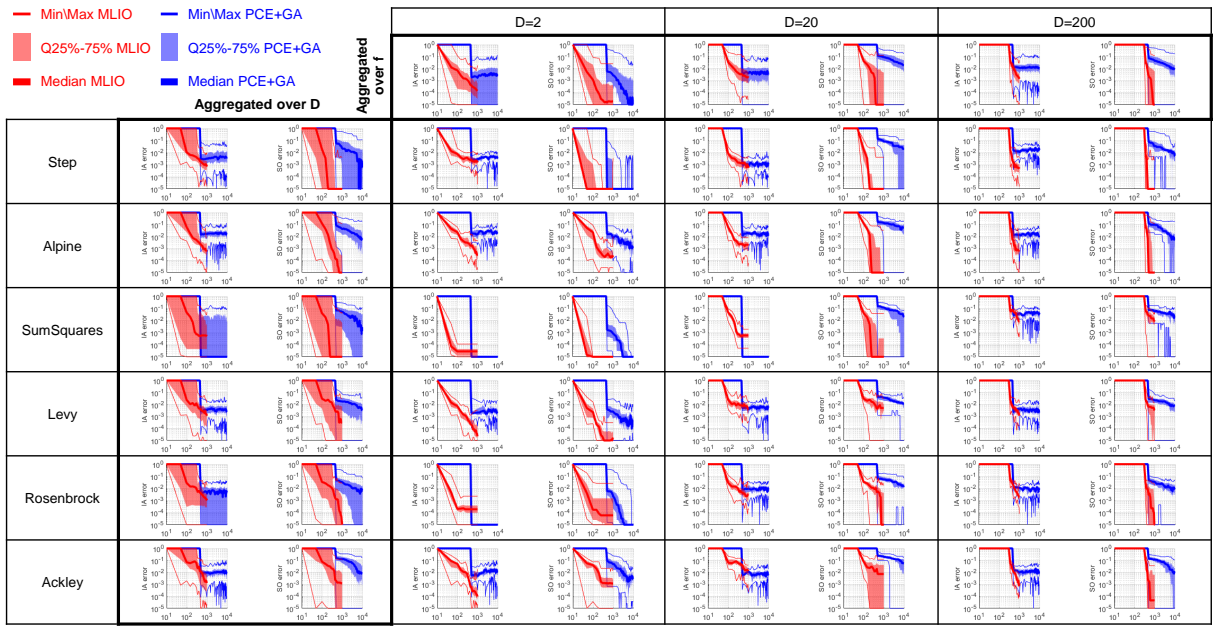

(b) Statistical results of PCE+GA vs. MLIO for stochastic optimization

Figure 2: Full statistical results of tuned PCE+GA vs. MLIO, per function, per dimension over the testbed

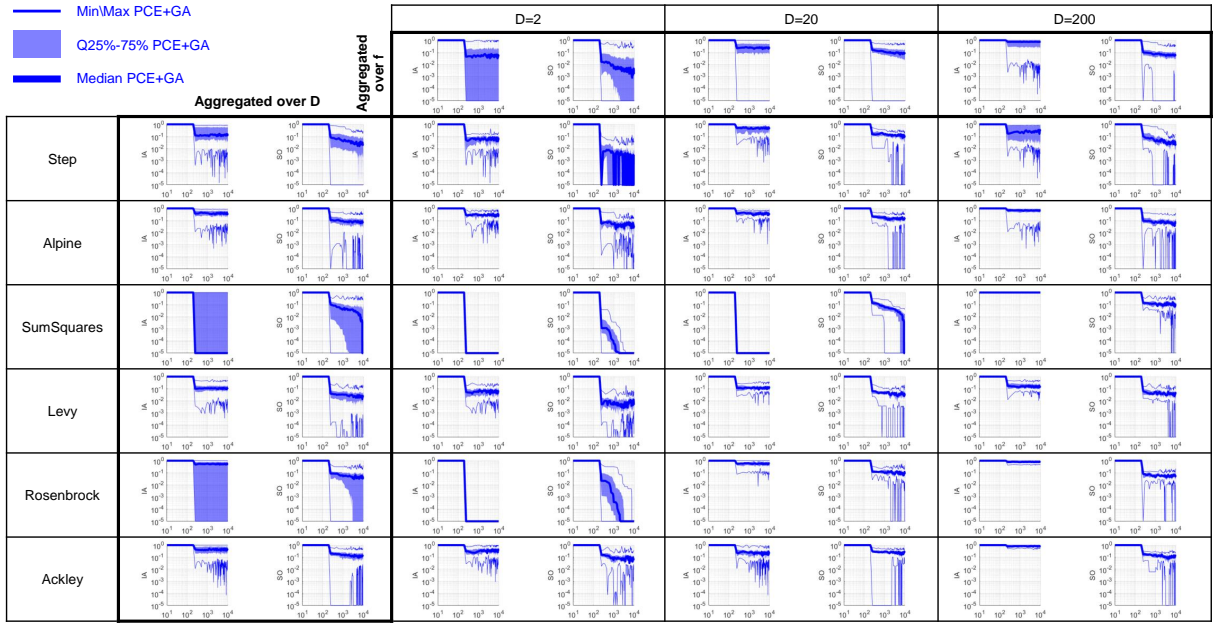

(a) Statistical results of PCE+GA for robust optimization

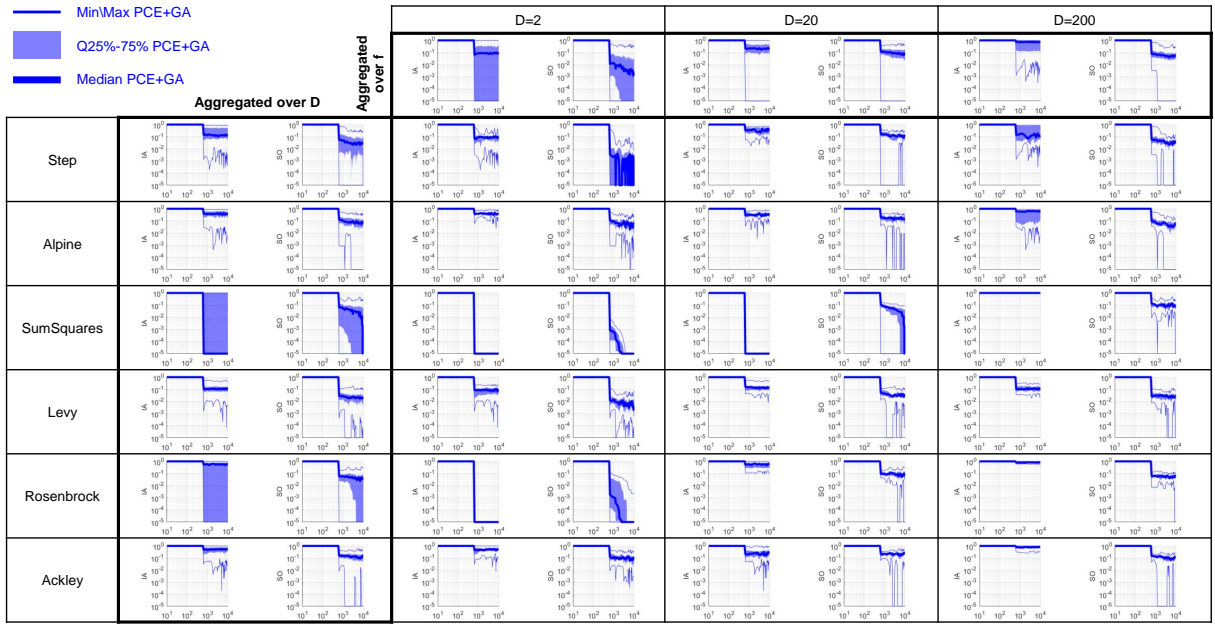

(b) Statistical results of PCE+GA for stochastic optimization

Figure 3: Full statistical results of PCE+GA tuning setting #1, per function, per dimension over the testbed

3

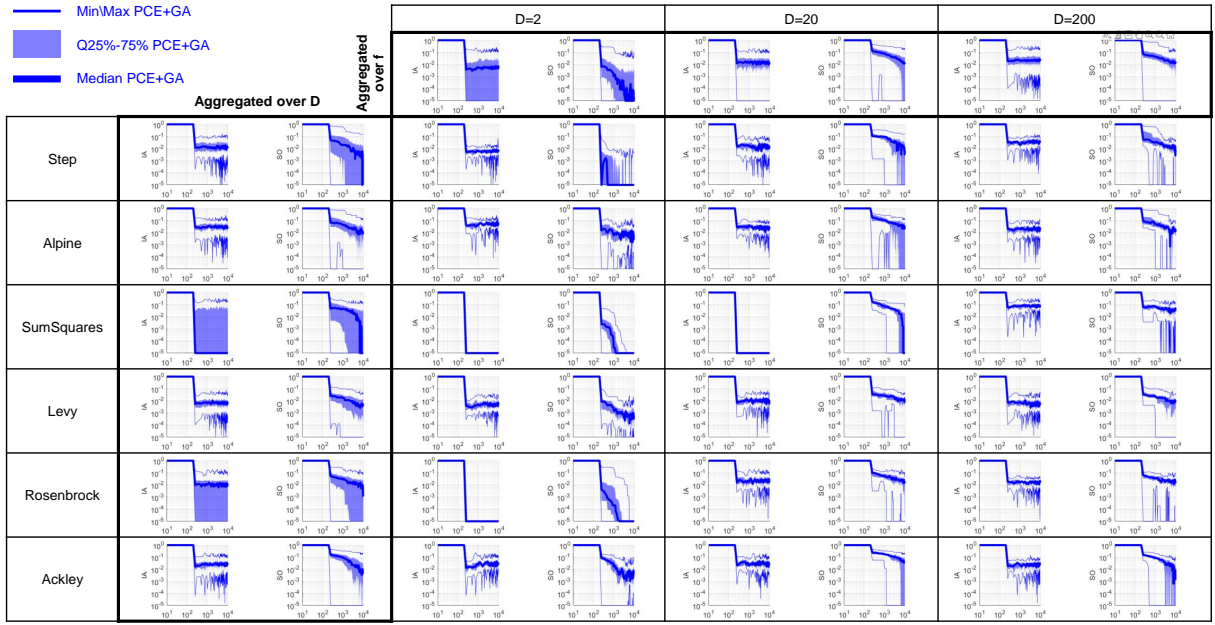

(a) Statistical results of PCE+GA for robust optimization

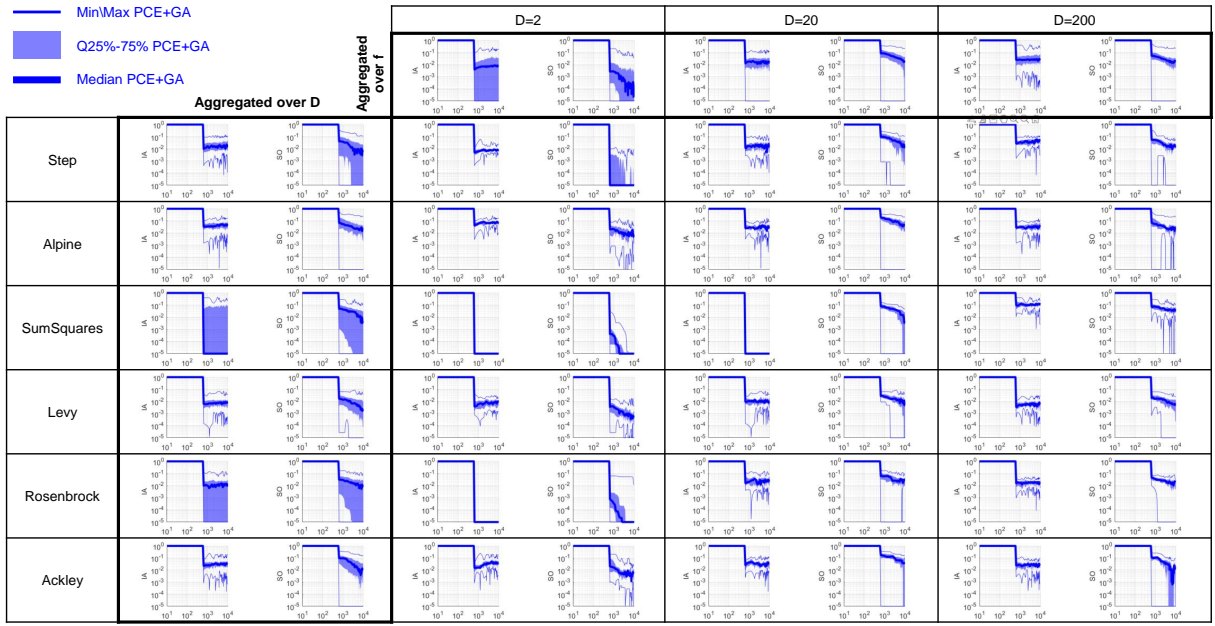

(b) Statistical results of PCE+GA for stochastic optimization

Figure 4: Full statistical results of PCE+GA tuning setting #2, per function, per dimension over the testbed

4

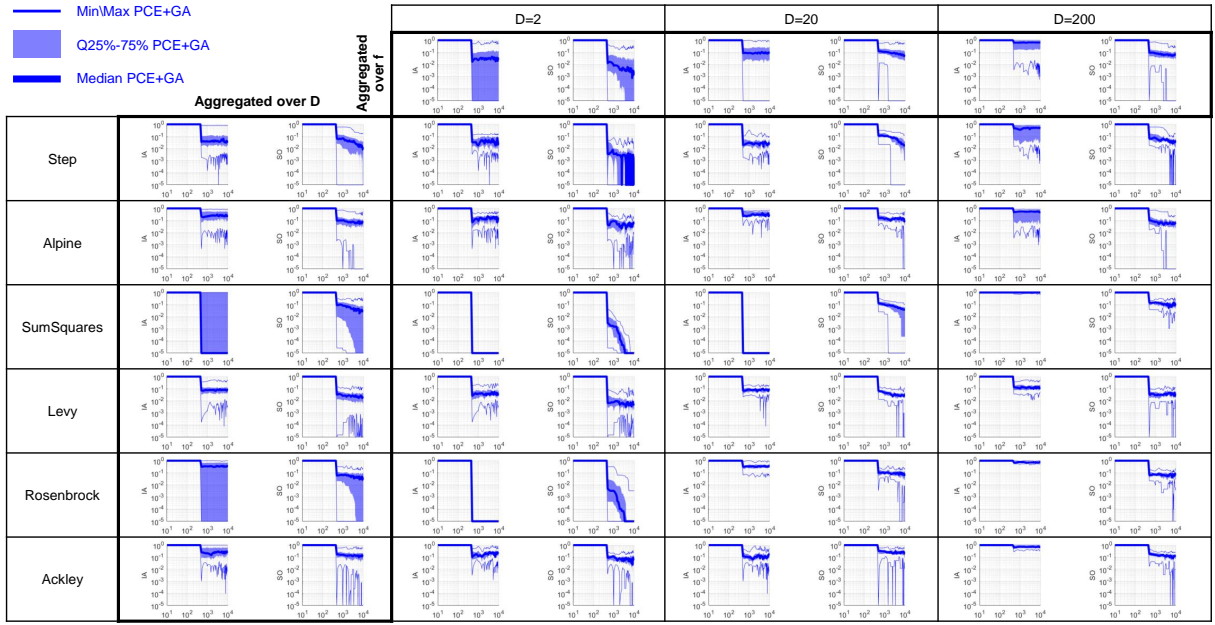

(a) Statistical results of PCE+GA for robust optimization

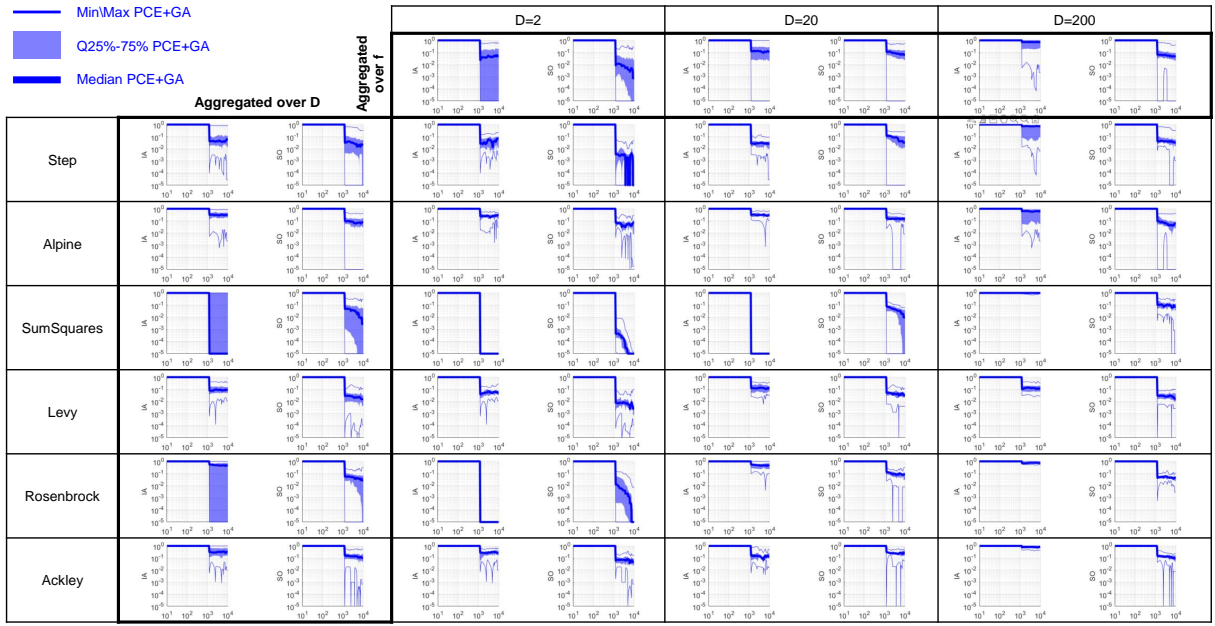

(b) Statistical results of PCE+GA for stochastic optimization

Figure 5: Full statistical results of PCE+GA tuning setting #3, per function, per dimension over the testbed

5

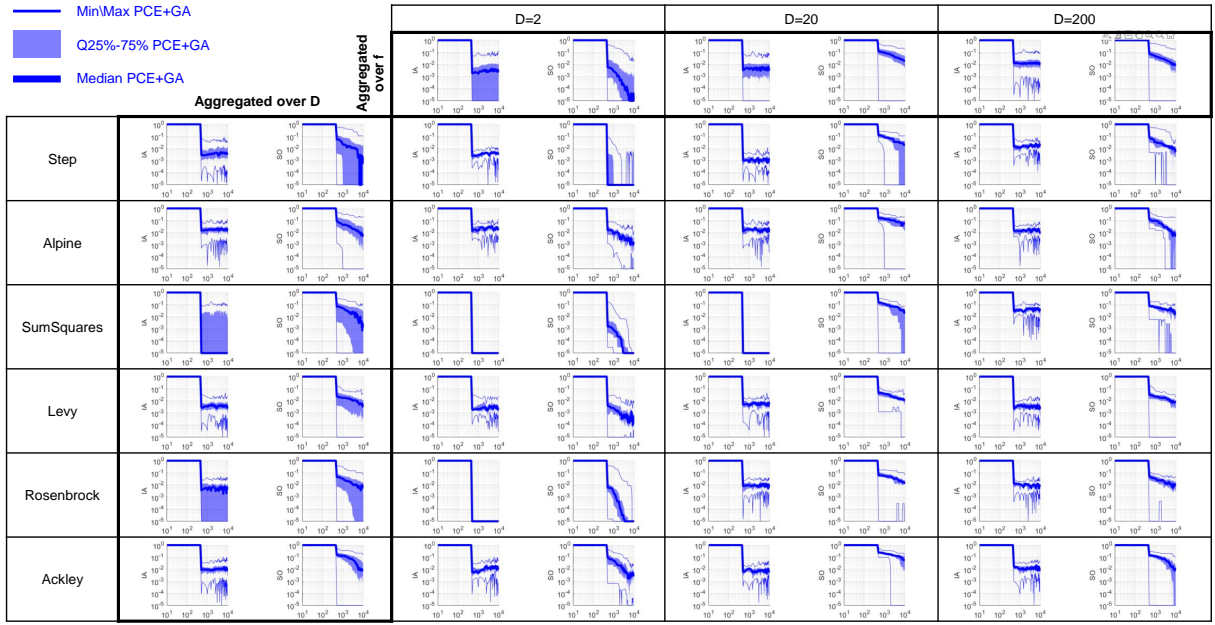

(a) Statistical results of PCE+GA for robust optimization

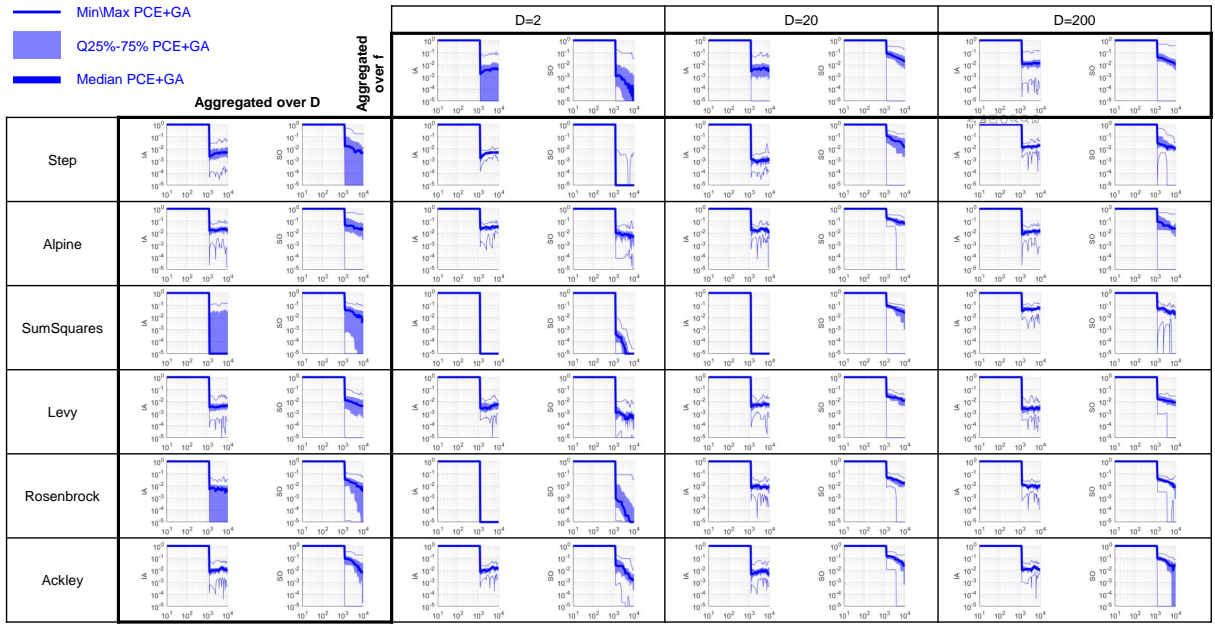

(b) Statistical results of PCE+GA for stochastic optimization

Figure 6: Full statistical results of PCE+GA tuning setting #4, per function, per dimension over the testbed

6

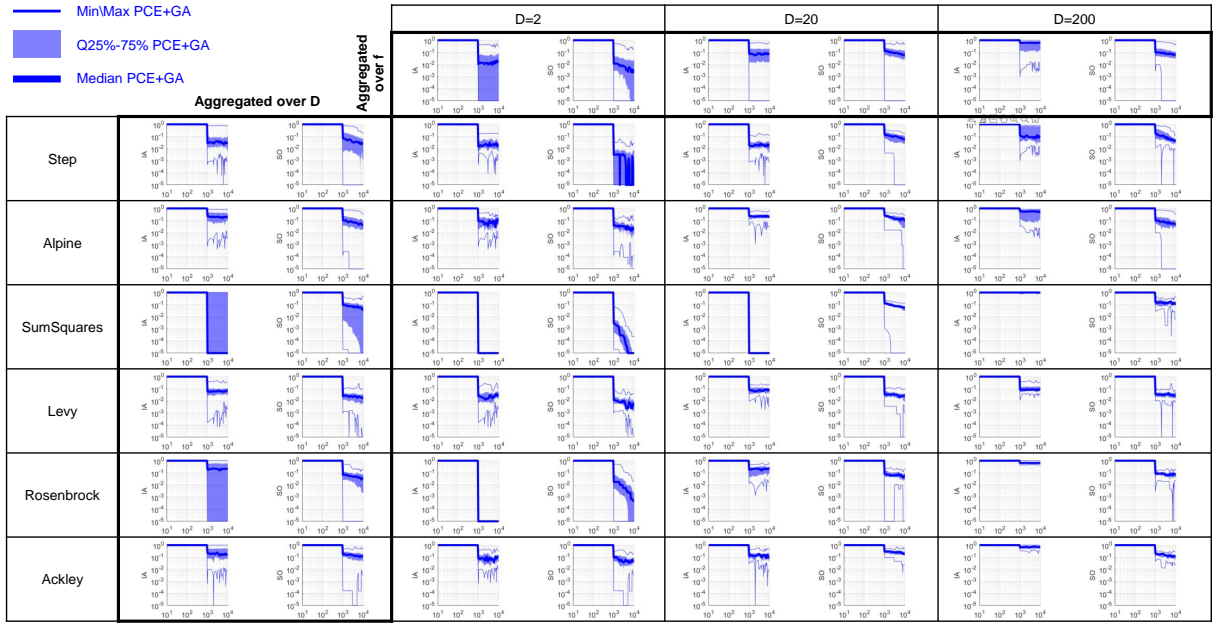

(a) Statistical results of PCE+GA for robust optimization

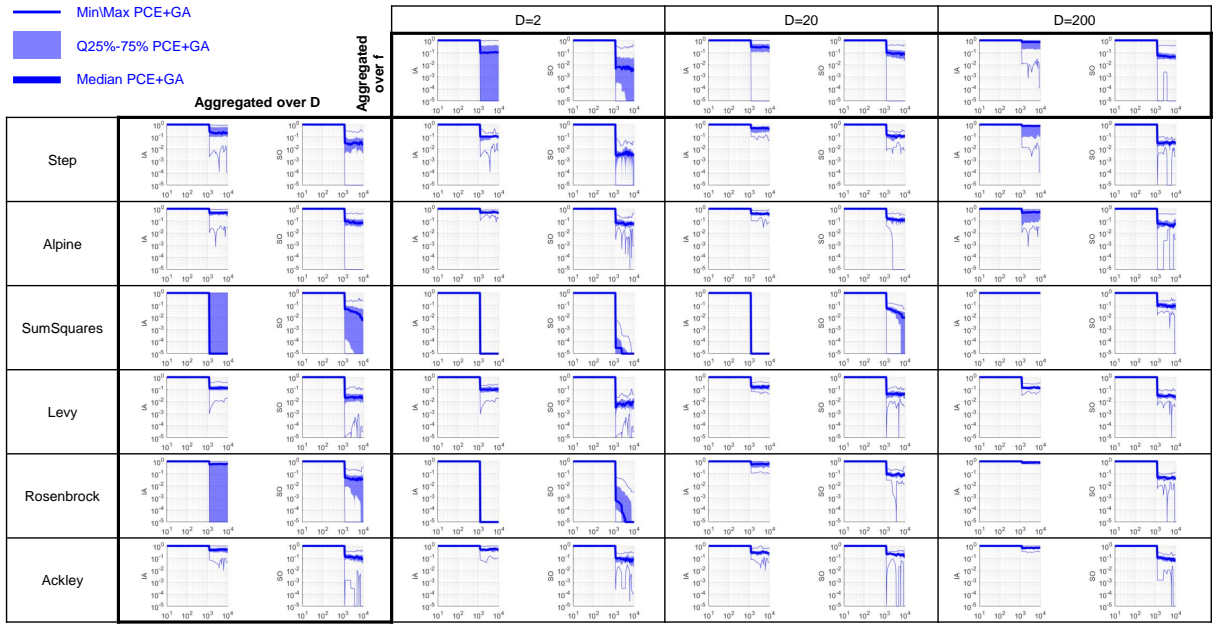

(b) Statistical results of PCE+GA for stochastic optimization

Figure 7: Full statistical results of PCE+GA tuning setting #5, per function, per dimension over the testbed

7

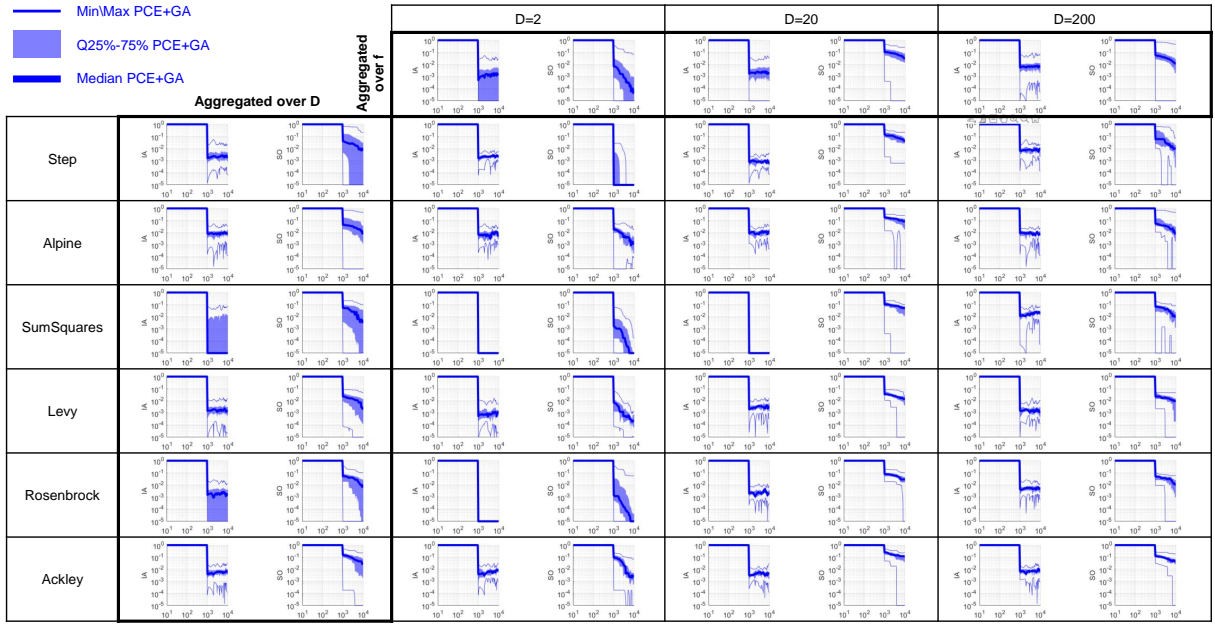

(a) Statistical results of PCE+GA for robust optimization

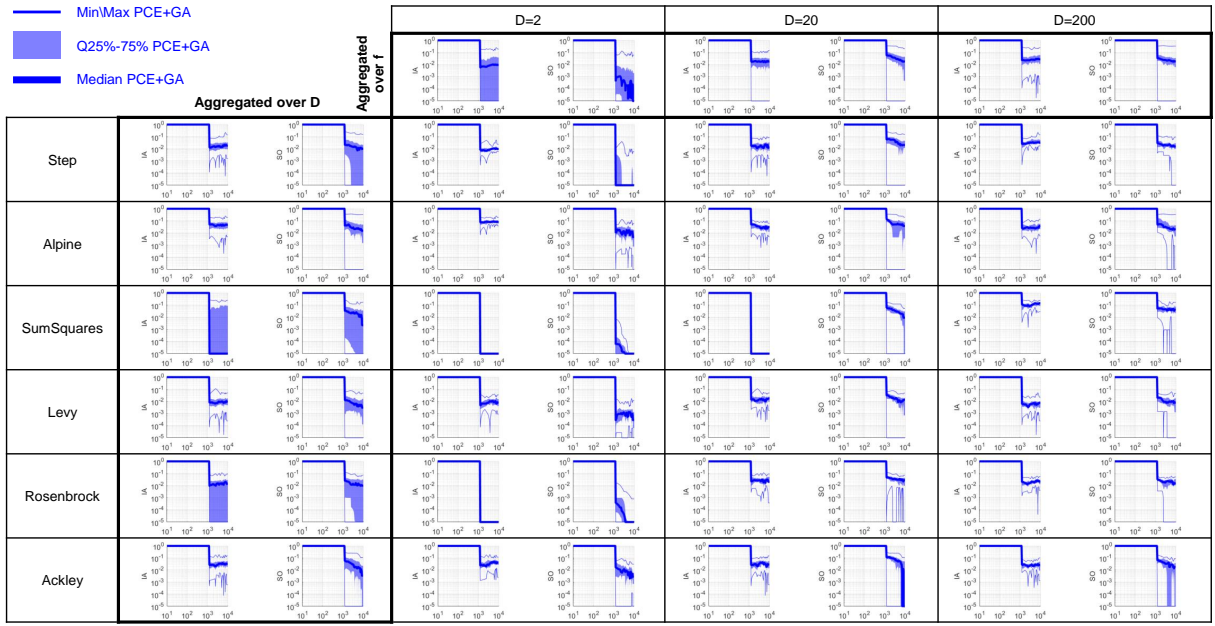

(b) Statistical results of PCE+GA for stochastic optimization

Figure 8: Full statistical results of PCE+GA tuning setting #6, per function, per dimension over the testbed

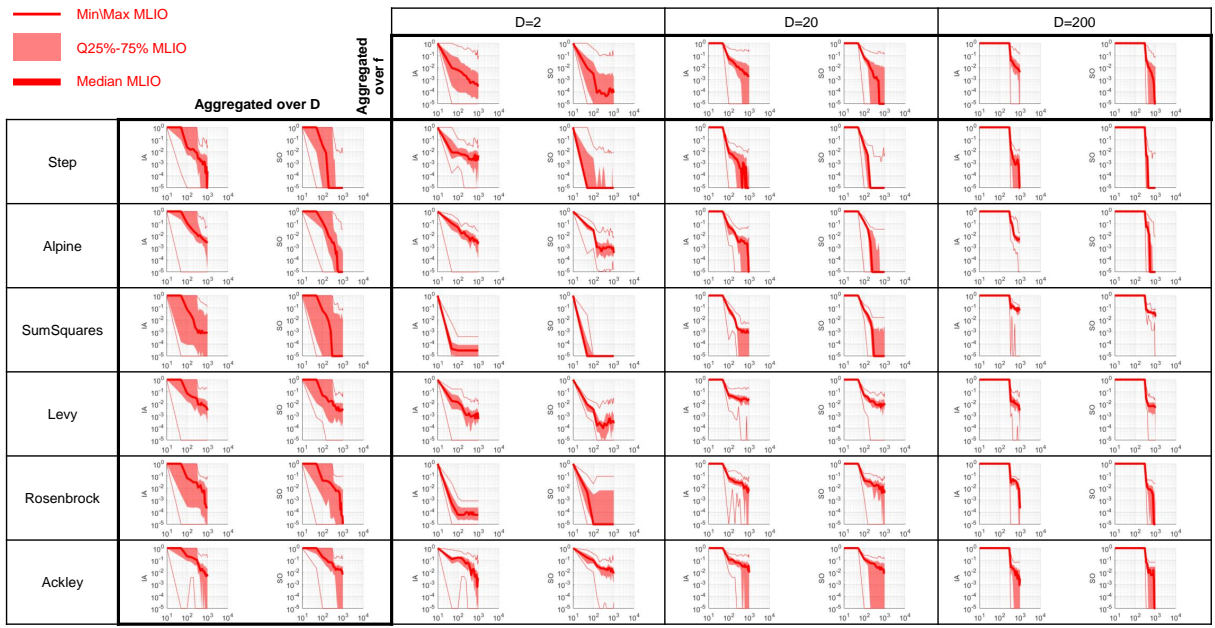

(a) Statistical results of MLIO for robust optimization

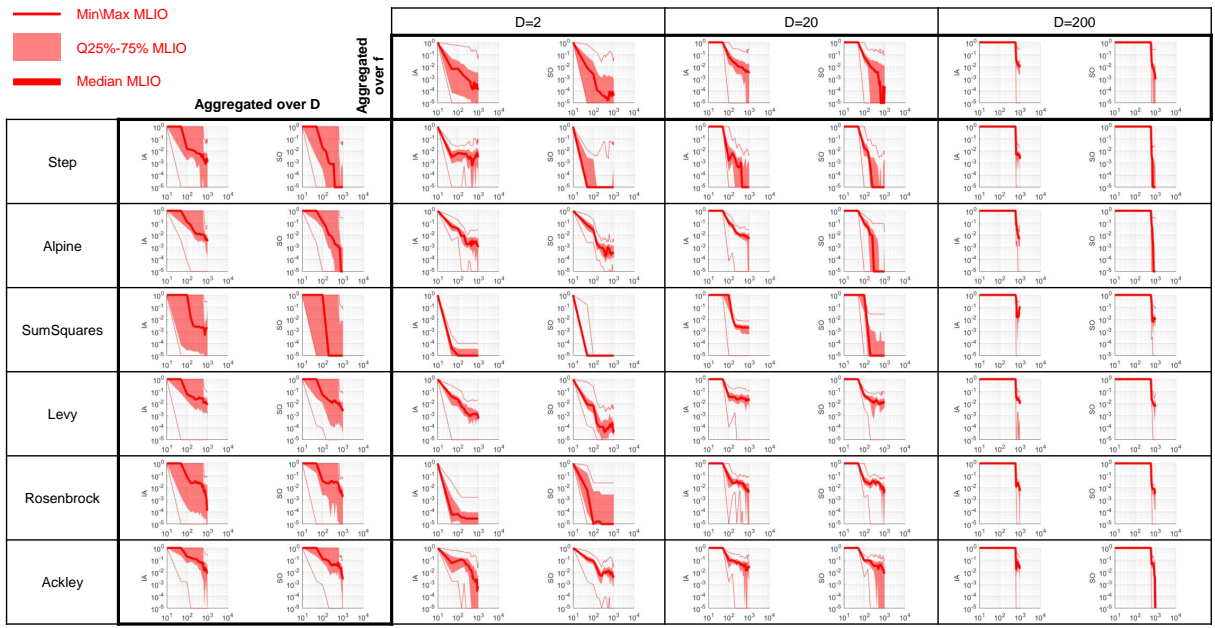

(b) Statistical results of MLIO for stochastic optimization

Figure 9: Full statistical results of MLIO tuning setting #1, per function, per dimension over the testbed

9

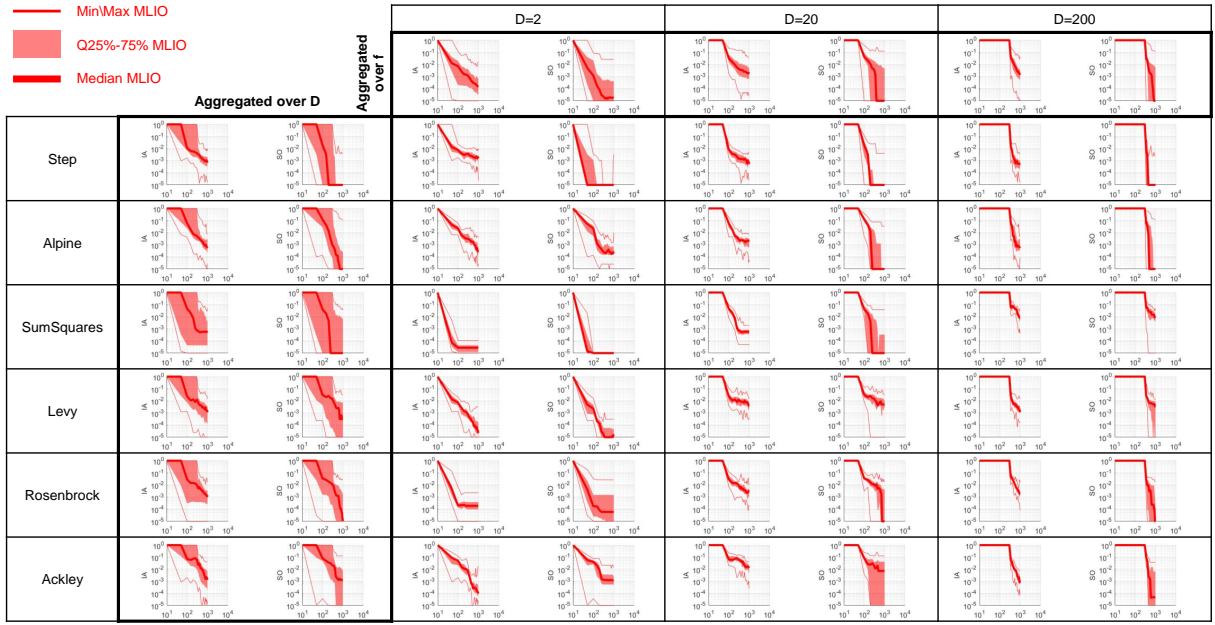

(a) Statistical results of MLIO for robust optimization

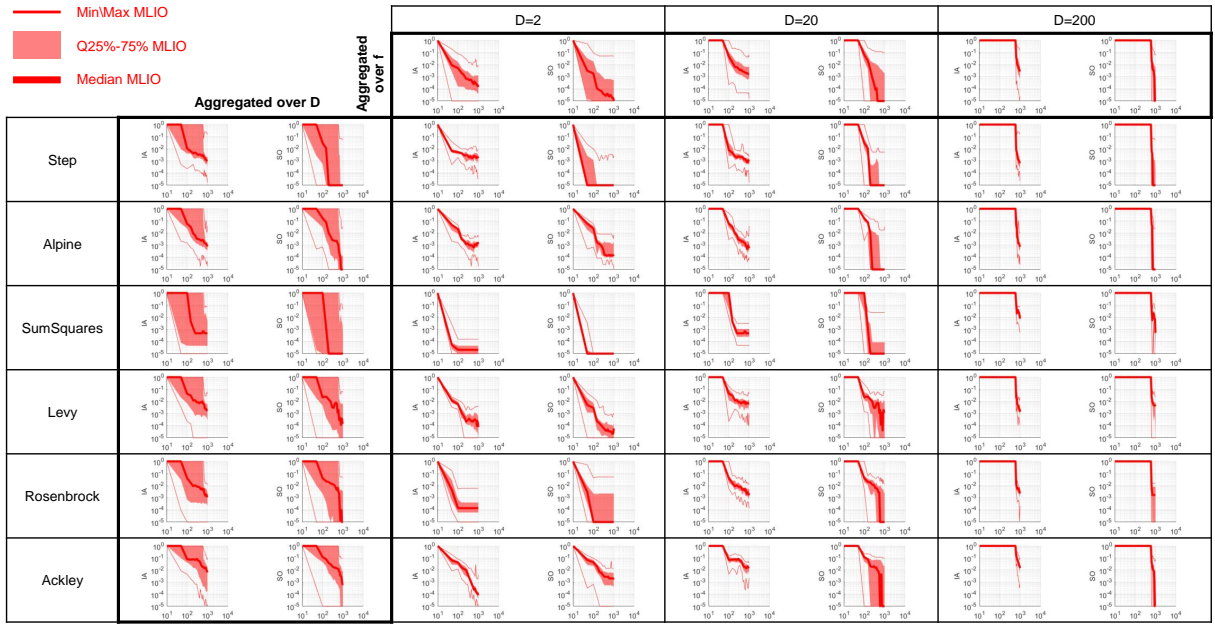

(b) Statistical results of MLIO for stochastic optimization

Figure 10: Full statistical results of MLIO tuning setting #2, per function, per dimension over the testbed

10



\end{document}